\documentclass[prd,aps,nofootinbib,tightenlines]{revtex4}
\usepackage{mathrsfs}
\usepackage{amsmath}
\usepackage{amssymb}
\usepackage{epsfig}
\usepackage{graphicx}
\usepackage{booktabs}
\usepackage{multirow}
\usepackage{siunitx}
\usepackage{subfigure}
\usepackage[section]{placeins}
\usepackage{rotating}
\usepackage{float}
\usepackage{slashed}
\usepackage{longtable}
\usepackage[colorlinks, citecolor=blue, linkcolor=red, urlcolor=blue]{hyperref}

\begin{document}
\newcommand{\psl}{ p \hspace{-1.8truemm}/ }
\newcommand{\nsl}{ n \hspace{-2.2truemm}/ }
\newcommand{\vsl}{ v \hspace{-2.2truemm}/ }
\newcommand{\epsl}{\epsilon \hspace{-1.8truemm}/\,  }
\title{ Semileptonic neutral current decays of $\Xi_b$ with dileptons or dineutrinos in the final state}
\author{Zhou Rui$^1$ }\email{jindui1127@126.com}
\author{Zhi-Tian Zou$^1$ }\email{zouzt@ytu.edu.cn}
\author{Ya Li$^2$}\email{liyakelly@163.com}
\author{Ying Li$^1$ }\email{liying@ytu.edu.cn}
\affiliation{Department of Physics, Yantai University, Yantai 264005, China}
\affiliation{Department of Physics and Institute of Theoretical Physics, Nanjing Normal University, Nanjing 210023, China}
\date{\today}
\begin{abstract}
We perform a detailed analysis of semileptonic $\Xi_b$ decays mediated by flavor-changing neutral currents ($b\to s$ and $b\to d$) with dilepton or dineutrino final states within the perturbative quantum chromodynamics (PQCD) framework. All independent form factors including vector, axial-vector, tensor, and pseudotensor currents are calculated and are  used to analyze the decay branching fractions and angular distributions. Our numerical results for the branching fractions of $\Xi_b\to \Xi \ell^+\ell^-$ decays suggest they are within measurable reach for the LHCb experiment in the near future. 
For the case of unpolarized $\Xi_b$ baryons, we derive several angular observables, which can provide new and complementary constraints on Wilson coefficients in semileptonic flavor-changing neutral current transitions compared to those from mesonic decays. Finally, we present a combined analysis of dilepton and dineutrino channels, comparing various observables in detail. Our results offer further insights into the long-standing anomalies observed in $B$ meson decays.
\end{abstract}



\maketitle
\section{Introduction}
Flavor-changing neutral current (FCNC) transitions involving $b$-quark decays have attracted significant attention over the past several decades. These processes are highly suppressed in the Standard Model (SM), as they occur only through penguin or box loop diagrams, making them exceptionally sensitive probes of potential new physics (NP) at high energy scales~\cite{Bobeth:2007dw, Blake:2016olu}. Most existing investigations have focused on FCNC-induced $B$-meson decays, such as $B \to K^{(*)}\ell^+\ell^-$ and $B_s \to \phi \ell^+\ell^-$~\cite{Bouchard:2013mia, Horgan:2013pva, Straub:2015ica, Du:2015tda, Zhang:2024rmb,Ali:2025xkw,Rajeev:2021ntt}, where experimental precision has reached an advanced level~\cite{LHCb:2014cxe, Aaij:2013iag, Aaij:2013aln, Aaij:2015esa, LHCb:2016ykl}. A broad array of observables has been measured, including differential branching fractions~\cite{LHCb:2014cxe, LHCb:2016ykl, LHCb:2021zwz}, angular distributions~\cite{LHCb:2015svh, CMS:2017rzx, ATLAS:2018gqc, LHCb:2020lmf, LHCb:2020gog, LHCb:2021xxq, LHCb:2023gel, LHCb:2023gpo, LHCb:2024onj, CMS:2024atz}, and tests of lepton-flavor universality (LFU)~\cite{LHCb:2021lvy, LHCb:2022qnv, LHCb:2022vje, CMS:2024syx, LHCb:2024rto}.  Intriguingly, several measurements of angular observables and branching fractions exhibit tensions with SM predictions (see Refs.~\cite{Matias:2012xw, Descotes-Genon:2013vna, Aaij:2013aln, Aaij:2015oid, Abdesselam:2016llu, ATLAS:2017dlm, CMS:2017ivg, Parrott:2022zte} and references therein). In particular, the measured ratios of the branching fractions between the electron and muon modes hint at a possible LFU violation in $b\to s \ell^+\ell^-$ transition~\cite{Aaij:2014ora, Aaij:2017vbb, Aaij:2015dea, Wehle:2016yoi, LHCb:2021trn}. However, with the full LHCb dataset~\cite{ LHCb:2022qnv, LHCb:2022vje}, the anomalies in  $\mathcal{R}(K^{(*)}) $ have significantly weakened and are now largely consistent with SM predictions. Furthermore, a recent Belle-II measurement~\cite{Belle-II:2023esi} indicates that the branching fraction of the dineutrino mode $B \to K \nu \bar{\nu}$ exceeds the SM expectation by approximately $2.7\sigma$~\cite{Altmannshofer:2009ma, Buras:2014fpa}. Although these deviations are not yet conclusive, they collectively reinforce the possibility of physics beyond the SM.

In this context, it is particularly compelling to investigate $b$-baryon decays that proceed through the same underlying quark-level transitions, as they offer valuable cross-checks of the long-standing anomalies observed in mesonic channels. Owing to their half-integer spin, $b$-baryons retain essential information about the helicity structure of the effective Hamiltonian in their weak decays~\cite{Boer:2014kda, Blake:2017une, Das:2018sms, Yan:2019tgn}. Consequently, the study of baryonic modes provides complementary phenomenological insight into the nature of heavy baryons and sheds light on the dynamics governing their weak decays. Moreover, angular observables in baryonic transitions can also illuminate the anomalies reported in $B$-meson decays. For these reasons, it is timely and necessary to place baryonic decay modes on equal footing with mesonic ones, as probing FCNC transitions in the $b$-baryon sector is of particular interest.

Experimental studies of $b$-baryon decays are not yet as mature as those of their mesonic counterparts, but significant progress has been achieved in recent years. Following the first observation of $\Lambda_b \to \Lambda \mu^+\mu^-$ by the CDF Collaboration~\cite{CDF:2011buy}, the LHCb Collaboration confirmed this decay mode in~\cite{LHCb:2013uqx} and subsequently performed detailed angular analyses in~\cite{LHCb:2015tgy, LHCb:2018jna}. From the theoretical perspective, one of the central challenges in describing $\Lambda_b \to \Lambda \mu^+\mu^-$ is the calculation of the hadronic $\Lambda_b \to \Lambda$ form factors. These form factors have been extensively studied using a variety of approaches, including quark models, QCD sum rules, lattice QCD, and covariant quark models, as documented in~\cite{Mannel:1990vg, Hussain:1990uu, Hussain:1992rb, Cheng:1994kp, Cheng:1995fe, Huang:1998ek, Mohanta:1999id, He:2006ud, Wang:2008sm, Wang:2009hra, Aliev:2010uy, Mott:2011cx, Feldmann:2011xf, Mannel:2011xg, Wang:2011uv, Gutsche:2013pp, Liu:2015qfa, Mott:2015zma, Wang:2015ndk, Detmold:2016pkz, Lu:2025gjt, Yang:2025yaw, Blake:2022vfl}. The dineutrino modes are experimentally very challenging due to the two undetected neutrinos in the final state and have not yet been observed. They could become accessible at future high-luminosity $e^+ e^-$ machines operating at the $Z$-pole, such as FCC-ee and CEPC~\cite{FCC:2018evy, Bernardi:2022hny, CEPCStudyGroup:2018ghi, CEPCPhysicsStudyGroup:2022uwl}. Theoretically, these channels are exceptionally clean because they do not suffer from hadronic uncertainties beyond the form factors such as nonfactorizable effects and photon-penguin contributions. Several analyses of $\Lambda_b \to \Lambda \nu\bar{\nu}$ have been performed~\cite{Chen:2000mr, Aliev:2007rm, Sirvanli:2007yq, Hiller:2021zth, Das:2023kch, Amhis:2023mpj, Altmannshofer:2025eor, Das:2025zrn}. These developments naturally motivate the study of other baryonic FCNC $b \to s$ transitions, such as $\Xi_b \to \Xi$, which in principle could be observed at LHCb in the near future. For the $b \to d$ transition, the corresponding semileptonic FCNC decays of the $\Xi_b$ are even richer, encompassing $\Xi_b \to (\Sigma,\Lambda)(\ell^+\ell^-,\nu\bar\nu)$. Given ongoing experimental advancements, these modes may become measurable soon, making rare $\Xi_b$ decays a compelling subject of study. Although the family nonuniversal $Z^\prime$ model~\cite{Nayek:2020hna} and QCD sum-rule analyses~\cite{Azizi:2011mw} have offered useful insights, a comprehensive and systematic investigation of the dineutrino channels remains lacking to the best of our knowledge.

The purpose of this paper is to present a systematic analysis of the semileptonic decays $\Xi_b\to (\Xi,\Sigma,\Lambda)(\ell^+\ell^-,\nu\bar\nu)$ which proceed via the FCNC transitions $b\to s(d)$. The transition form factors governing the hadronic matrix elements of the vector, axial-vector, tensor, and pseudotensor currents are calculated using the perturbative QCD (PQCD) approach in the large-recoil region, corresponding to low momentum transfer $q^2$. To obtain reliable form-factor shapes over the entire kinematic domain, we further adopt the $z$-expansion parametrization to extrapolate the PQCD results and derive complete $q^2$ distributions. We analyze the decay distributions within the helicity-amplitude formalism for unpolarized $\Xi_b$ baryons, retaining finite lepton-mass effects ($m_\ell\neq0$). The resulting angular observables are expressed through ten transversity amplitudes constructed from the baryonic form factors. In addition, we present numerical predictions for the branching fractions, LFU ratios, and several angular observables, including the longitudinal polarization fraction of the dilepton system, as well as the leptonic, baryonic, and mixed forward-backward asymmetries for both dilepton and dineutrino channels.

The remainder of this paper is organized as follows. In Sec.~\ref{sec:framework}, we introduce the effective Hamiltonian governing the relevant quark-level FCNC transitions. Section~\ref{sec:FFs} summarizes the definitions of the baryonic transition form factors, while Sec.~\ref{sec:obe} presents the full set of formulas for the angular distributions and associated observables. Section~\ref{sec:results} contains the PQCD analysis of the various physical quantities of interest. In Sec.~\ref{sec:FFsv}, we provide numerical results for the Weinberg and helicity form factors at zero and maximum recoils, together with their $q^2$ dependence. Section~\ref{sec:ratio} reports both differential and integrated branching fractions and the corresponding LFU ratios. The differential and integrated angular observables for all lepton flavors are presented in Sec.~\ref{sec:ang}, where we also discuss long-distance effects on the decay rates. In Sec.~\ref{sec:dineu}, we evaluate the branching fractions and angular observables for the dineutrino channels. Finally, Sec.~\ref{sec:sum} offers a brief summary of our findings. Several Appendices are included to provide supplementary technical details of the calculations.

\section{Formalism}\label{sec:framework}
\subsection{Baryonic form factors } \label{sec:FFs}
The analysis of weak decays in the SM begins with an effective Hamiltonian constructed from four-fermion operators multiplied by their corresponding Wilson coefficients. At low energies, the general effective Hamiltonian governing both  $b\to Q \ell^+\ell^- $ and $b\to Q \nu\bar\nu $ transitions can be written as~\cite{Buchalla:2000sk}
\begin{eqnarray}\label{eq:effH}
\mathcal{H}_{eff}=-\frac{4G_F}{\sqrt{2}}V^*_{tQ}V_{tb}\frac{\alpha_e}{8\pi}\sum_{i}C_i(\mu)O_i(\mu),
\end{eqnarray}
where $Q$ denotes a down-type quark, $s$ or $d$. Here, $\alpha_e$ is the electromagnetic coupling evaluated at the $b$-quark mass scale $m_b$,  $V_{ij}$ are the Cabibbo-Kobayashi-Maskawa (CKM) matrix elements~\cite{ParticleDataGroup:2024cfk}, and $G_F$ is the Fermi constant~\cite{ParticleDataGroup:2024cfk}. The quantities $C_i(\mu)$ are the short-distance Wilson coefficients at the renormalization scale $\mu$, while $O_i(\mu)$ denote the corresponding local operators encoding long-distance dynamics.
Contributions proportional to the CKM combinations $V_{uQ}^*V_{ub}$ and $V_{uQ}^*V_{ub}$
are strongly canceled due to the   Glashow-Iliopoulos-Maiani (GIM) mechanism~\cite{Glashow:1970gm}   and can therefore be safely neglected in an FCNC process.
For $\ell^+\ell^-$ final states, the dominant SM contributions arise from the electromagnetic dipole operator $O_7$ and the semileptonic operators $O_{9\ell}$ and $O_{10\ell}$. The decay amplitude for the exclusive process $\mathcal{B}_i\to \mathcal{B}_f \ell^+\ell^-$ is obtained by sandwiching the effective Hamiltonian between the initial- and final-baryon states, yielding
\begin{eqnarray}\label{eq:amplitude}
\mathcal{M}(\mathcal{B}_i\to \mathcal{B}_f  \ell^+ \ell^-) & = &
\frac{ G_F\alpha_e V^*_{tQ}V_{tb}}{2\sqrt{2}\,\pi}
\biggl\{
 C_9^{\rm eff}\,
  \langle\mathcal{B}_f\,|\,\bar{Q}\,\gamma^\mu(1-\gamma_5)\, b\,|\,\mathcal{B}_i\rangle \,\bar\ell\gamma_\mu \ell \nonumber\\&+&
C_{10}^{\rm eff}\, \langle\mathcal{B}_f\,|\,\bar{Q}\,\gamma^\mu(1-\gamma_5)\, b\,|\,\mathcal{B}_i\rangle
\, \bar\ell\gamma_\mu \gamma_5 \ell
\nonumber\\&-&
\frac{2m_b}{q^2}\,C_7^{\rm eff}\,
\langle\mathcal{B}_f\,|\,  \bar{Q}\,i\sigma^{\mu \nu}q_\nu\,(1+\gamma^5)\,\,b\,|\,\mathcal{B}_i\rangle
\,  \bar\ell\gamma_\mu \ell
\biggr\} .
\end{eqnarray}
where $\sigma^{\mu\nu}=\frac{i}{2}(\gamma^\mu\gamma^\nu-\gamma^\nu\gamma^\mu)$. The effective Wilson coefficients $C_{7,9,10}^{\rm eff}$ include the one-loop matrix elements of the four-quark operators $O_{1,2,\ldots,6}$. In the numerical analysis presented below, we adopt the next-to-leading logarithmic results for the Wilson coefficients evaluated at the scale $\mu=m_b$, namely $C_7^{\rm eff}=-0.2923$, $C_9^{\rm eff}=4.0749$, and $C_{10}^{\rm eff}=-4.3085$~\cite{Descotes-Genon:2013vna}. It is important to emphasize that, in addition to short-distance contributions, $C_9^{\rm eff}$ receives sizable long-distance effects from real $c\bar c$ resonant states belonging to the $J/\psi$ family~\cite{Gutsche:2013pp}. The impact of these long-distance contributions will be discussed in detail in the next section. Since the operator $O_{10\ell}$ does not renormalize under QCD, its Wilson coefficient is scale independent at $\mu=m_b$. Any residual scheme dependence must cancel in the physical decay amplitude once all contributions are consistently taken into account.

For the $b\to Q\bar\nu\nu$ transitions only the single operator, $(\bar{Q} \gamma_\mu (1-\gamma_5) b)(\bar{\nu} \gamma^\mu (1-\gamma_5) \nu)$,  is relevant in the SM. The corresponding decay amplitude reads as
\begin{eqnarray}\label{eq:amplitude2}
\mathcal{M}(\mathcal{B}_i\to \mathcal{B}_f  \nu\bar\nu) & = &
\frac{ G_F\alpha_e V^*_{tQ}V_{tb}}{2\sqrt{2}\,\pi}
\biggl\{
 C_L\,
  \langle\mathcal{B}_f\,|\,\bar{Q}\,\gamma^\mu(1-\gamma_5)\, b\,|\,\mathcal{B}_i\rangle \,\bar\nu\gamma_\mu \nu \nonumber\\&-&
C_L\, \langle\mathcal{B}_f\,|\,\bar{Q}\,\gamma^\mu(1-\gamma_5)\, b\,|\,\mathcal{B}_i\rangle
\, \bar\nu\gamma_\mu \gamma_5 \nu \biggl\},
\end{eqnarray}
with Wilson coefficient $C_L=-6.322$~\cite{Altmannshofer:2025eor}.

To evaluate the amplitudes above, the hadronic matrix elements on the right-hand side must be expressed in terms of form factors defined in full QCD. For the purposes of the PQCD analysis, it is convenient to introduce twelve dimensionless form factors, defined as in Ref.~\cite{Rui:2025iwa}.
\begin{align}
\label{eq:invariant}
 \langle \mathcal{B}_f(p^\prime,\lambda^\prime) | \overline{Q} \,\gamma^\mu\, b | \mathcal{B}_i(p,\lambda) \rangle &= \overline{u}_{\mathcal{B}_f} \left[ F_1(q^2)\: \gamma^\mu + F_2(q^2)\frac{ p^\mu}{M} + F_3(q^2)\frac{p'^\mu}{M} \right] u_{\mathcal{B}_i},   \nonumber\\
 \langle \mathcal{B}_f(p^\prime,\lambda^\prime) | \overline{Q} \,\gamma^\mu\gamma_5\, b | \mathcal{B}_i(p,\lambda) \rangle &=\overline{u}_{\mathcal{B}_f} \left[ G_1(q^2)\: \gamma^\mu + G_2(q^2)\frac{ p^\mu}{M} + G_3(q^2)\frac{p'^\mu}{M} \right]\gamma_5\: u_{\mathcal{B}_i},    \nonumber\\
 \langle \mathcal{B}_f(p^\prime,\lambda^\prime) | \overline{Q} \,i\sigma^{\mu\nu}q_\nu\, b | \mathcal{B}_i(p,\lambda) \rangle &= \overline{u}_{\mathcal{B}_f} \left[ F^T_1(q^2)\: \gamma^\mu + F^T_2(q^2)\frac{ p^\mu}{M} + F^T_3(q^2)\frac{p'^\mu}{M} \right] u_{\mathcal{B}_i},  \nonumber\\
 \langle \mathcal{B}_f(p^\prime,\lambda^\prime) | \overline{Q} \,i\sigma^{\mu\nu}q_\nu\,\gamma_5\, b | \mathcal{B}_i(p,\lambda) \rangle &= \overline{u}_{\mathcal{B}_f} \left[ G^T_1(q^2)\: \gamma^\mu + G^T_2(q^2)\frac{ p^\mu}{M} + G^T_3(q^2)\frac{p'^\mu}{M} \right]\gamma_5\: u_{\mathcal{B}_i},
\end{align}
where $u_{\mathcal{B}_i}$ and $u_{\mathcal{B}_f}$  denote the Dirac spinors of the initial and final baryons, respectively, while $p^{(\prime)}$ and $\lambda^{(\prime)}$ represent their corresponding momenta and helicities. Equation~(\ref{eq:invariant}) presents, from top to bottom, the matrix elements of the vector, axial-vector, tensor, and pseudotensor currents. Each of the vector and axial-vector currents is parametrized by three independent form factors. For the tensor and pseudotensor currents, the contraction with $q_\nu$ allows the number of independent form factors to be reduced by applying the equations of motion, which leads to the following relations,
\begin{eqnarray}
F_1^T(M-m)+\frac{M^2-p\cdot p'}{M}F_2^T+\frac{p\cdot p'-m^2}{M}F_3^T&=&0, \\
-G_1^T(M+m)+\frac{M^2-p\cdot p'}{M}G_2^T+\frac{p\cdot p'-m^2}{M}G_3^T&=&0,
\end{eqnarray}
with $M(m)$ being the mass of $\mathcal{B}_{i(f)}$. Consequently, a total of ten independent form factors is required to describe an FCNC baryonic transition. The vector ($F_i$) and axial-vector ($G_i$) form factors have been calculated within the PQCD framework in Ref.~\cite{Rui:2025bsu}. Analogously, the factorization formulas for the tensor ($F_i^T$) and pseudotensor ($G_i^T$) form factors can be written as
\begin{eqnarray}\label{eq:FG}
F^T_i(G^T_i)=\frac{4f_{\mathcal{B}_i}  \pi^2 G_F}{27\sqrt{2}}\sum_\xi
\int\mathcal{D}x\mathcal{D}b
\alpha_s^2(t_\xi)e^{-S_{\mathcal{B}_i}-S_{\mathcal{B}_f}}\Omega_\xi(b_i,b^\prime_i) H^{F^T_i(G^T_i)}_\xi(x_i,x^\prime_i),
\end{eqnarray}
with the integration measures;
\begin{eqnarray}
\mathcal{D}x&=&dx_1dx_2dx_3\delta(1-x_1-x_2-x_3)dx^\prime_1dx^\prime_2dx^\prime_3\delta(1-x^\prime_1-x^\prime_2-x^\prime_3),\nonumber\\
\mathcal{D}b&=& d^2\textbf{b}_2d^2\textbf{b}_3d^2\textbf{b}^\prime_2d^2\textbf{b}^\prime_3.
\end{eqnarray}
Here, $x_i^{(\prime)}$ ($i=1,2,3$) denote the longitudinal momentum fractions carried by the valence quarks inside the initial (final) baryon, while $\mathbf{b}_i^{(\prime)}$ are the conjugate variables to the corresponding transverse momenta $\mathbf{k}_{iT}^{(\prime)}$ in impact-parameter space.
 The introduction of parton transverse momenta $\mathbf{k}_{iT}^{(\prime)}$
  eliminates end point divergences, thereby enabling the perturbative calculation of many heavy-to-light form factors, including their higher-twist corrections.
 The summation runs over all contributing Feynman diagrams, labeled by $\xi$. The Sudakov factors $S_{\mathcal{B}_{i,f}}$, the hard scales $t_\xi$, and the functions $\Omega_\xi(b_i,b'_i)$ are given in Refs.~\cite{Rui:2025iwa,Rui:2025bsu}.  As an illustration, the explicit expressions for $H^{F_i^T}_a(x_i,x'_i)$  and $H^{G_i^T}_a(x_i,x'_i)$ are collected in Appendix~\ref{sec:for}, while the various twist light-cone distribution amplitudes (LCDAs) of the initial and final baryons---universal functions defined on the light cone that describe the longitudinal momentum fraction distributions of partons within a hadron---are summarized in Appendix~\ref{sec:LCDAsp}. The PQCD predictions for these form factors are reliable in the low-$q^2$ range ($q^2 \in [0,m_\tau^2]$). To obtain their behavior over the entire kinematically allowed range, we employ a $z$-expansion parametrization to extrapolate the perturbative results to the high $q^2$region ($q^2\sim (M-m)^2$). Further details of this procedure are provided in Appendix~\ref{sec:LCDAs}.

Beyond the form in Eq.~\eqref{eq:invariant}, two other common parametrizations of the hadronic matrix elements appear in the literature. The first is based on Weinberg's classification~\cite{Detmold:2016pkz,Gutsche:2013pp}:
\begin{eqnarray} \label{eq:Weinberg}
 \langle \mathcal{B}_f(p^\prime,\lambda^\prime) | \overline{Q} \,\gamma^\mu\, b | \mathcal{B}_i(p,\lambda) \rangle &=& \overline{u}_{\mathcal{B}_f} \left[ f_1(q^2)\: \gamma^\mu - \frac{f_2(q^2)}{M} i\sigma^{\mu\nu}q_\nu + \frac{f_3(q^2)}{M} q^\mu \right] u_{\mathcal{B}_i},    \nonumber\\
 \langle \mathcal{B}_f(p^\prime,\lambda^\prime) | \overline{Q} \,\gamma^\mu\gamma_5\, b | \mathcal{B}_i(p,\lambda) \rangle &=& \overline{u}_{\mathcal{B}_f} \left[ g_1(q^2)\: \gamma^\mu - \frac{g_2(q^2)}{M} i\sigma^{\mu\nu}q_\nu + \frac{g_3(q^2)}{M} q^\mu \right]\gamma_5\: u_{\mathcal{B}_i}, \nonumber\\
\langle \mathcal{B}_f(p^\prime,\lambda^\prime) | \overline{Q} \,i\sigma^{\mu\nu}q_\nu\, b | \mathcal{B}_i(p,\lambda) \rangle &=& \overline{u}_{\mathcal{B}_f} \left[f_1^{T}(q^2)  \frac{\gamma^\mu q^2 - q^\mu \slashed{q}}{M}  - f_2^{T}(q^2) i\sigma^{\mu\nu}q_\nu  \right] u_{\mathcal{B}_i},  \nonumber\\
 \langle \mathcal{B}_f(p^\prime,\lambda^\prime) | \overline{Q} \,i\sigma^{\mu\nu}q_\nu\,\gamma_5\, b | \mathcal{B}_i(p,\lambda) \rangle &=& \overline{u}_{\mathcal{B}_f}\left[ g_1^{T}(q^2) \frac{\gamma^\mu q^2 - q^\mu \slashed{q} }{M}   - g_2^{T}(q^2) i\sigma^{\mu\nu}q_\nu  \right]\gamma_5\: u_{\mathcal{B}_i},
\end{eqnarray}
which are related to our definitions in Eq.~(\ref{eq:invariant}) as follows
\begin{eqnarray}\label{eq:Weinberg1}
f_1&=&F_1+\frac{1}{2}(F_2+F_3)(1+r), \quad f_2=-\frac{1}{2}(F_2+F_3), \quad f_3=\frac{1}{2}(F_2-F_3),  \nonumber\\
g_1&=&G_1-\frac{1}{2}(G_2+G_3)(1-r), \quad g_2=-\frac{1}{2}(G_2+G_3), \quad g_3=\frac{1}{2}(G_2-G_3),\nonumber\\
f^T_1&=&\frac{1}{2(M-m)}(F_3^T-F_2^T), \quad f^T_2=-\frac{1}{2M}(F_3^T+F_2^T),\nonumber\\
g^T_1&=&\frac{1}{2(M+m)}(G_2^T-G_3^T), \quad g^T_2=-\frac{1}{2M}(G_3^T+G_2^T).
\end{eqnarray}
Here $r=m/M$ is the mass ratio between the final and initial baryons. It has been demonstrated that $f^T_2=g^T_2$ holds at $q^2=0$  because they differ only by a term linear in $q^2$~\cite{Gutsche:2013pp}.

Another one is the helicity-based definition from Ref.~\cite{Feldmann:2011xf}, defined as
\footnote{The labeling convention for form factors in~\cite{Feldmann:2011xf} is distinct from that of this work. The equivalences are: $f_{t,0,\perp}=f_{0,+,\perp}$, $g_{t,0,\perp}=g_{0,+,\perp}$, $f_{0,\perp}^T=h_{+,\perp}$, and $g_{0,\perp}^T=\widetilde{h}_{+,\perp}$.}
\begin{align}\label{eq:helicity0}
 \langle \mathcal{B}_f(p',\lambda')| \overline{Q}  \, \gamma_{\mu} \, b|\mathcal{B}_i(p,\lambda)\rangle &= \overline u_{\mathcal{B}_f}
 \left[
  f_t(q^2) \, (M-m) \, \frac{q_\mu}{q^2}
\right.
\cr
& + \left.
  f_0(q^2) \, \frac{M+m}{s_+} \,
 \left(p_\mu + p_\mu' - \frac{q_\mu}{q^2} \, (M^2-m^2) \right)
\right.
\cr
&+ \left.
 f_\perp(q^2) \left( \gamma_\mu - \frac{2 m}{s_+} \, p_\mu - \frac{2  M}{s_+} \, p '_\mu
  \right)
 \right] u_{\mathcal{B}_i} \,,  \nonumber\\
 \langle \mathcal{B}_f(p',\lambda')| \overline{Q} \, \gamma_{\mu} \gamma_5 \, b|\mathcal{B}_i(p,\lambda)\rangle &= -\overline u_{\mathcal{B}_f}\gamma_5
 \left[
  g_t(q^2) \, (M+m) \, \frac{q_\mu}{q^2}
\right.
\cr
& + \left.
  g_0(q^2) \, \frac{M-m}{s_-} \,
 \left(p_\mu + p_\mu' - \frac{q_\mu}{q^2} \, (M^2-m^2) \right)
\right.
\cr
&+ \left.
 g_\perp(q^2) \left( \gamma_\mu + \frac{2 m}{s_-} \, p_\mu - \frac{2  M}{s_-} \, p '_\mu
  \right)
 \right] u_{\mathcal{B}_i} \,, \nonumber\\
 \langle \mathcal{B}_f(p',\lambda')| \overline{Q} \, i \sigma^{\mu\nu}q_{\nu} \, b|\mathcal{B}_i(p,\lambda)\rangle &= -\overline u_{\mathcal{B}_f}
\left[
  f_0^T(q^2) \, \frac{q^2}{s_+} \,
 \left(p_\mu + p_\mu' - \frac{q_\mu}{q^2} \, (M^2-m^2) \right)
\right.
\cr
&
+ \left.f^T_\perp(q^2)(M+m) \left( \gamma_\mu - \frac{2 m}{s_+} \, p_\mu - \frac{2  M}{s_+} \, p '_\mu
  \right)
 \right] u_{\mathcal{B}_i} \,, \nonumber\\
 \langle \mathcal{B}_f(p',\lambda')| \overline{Q} \, i \sigma^{\mu\nu}q_{\nu}\gamma_5 \, b|\mathcal{B}_i(p,\lambda)\rangle &= -\overline u_{\mathcal{B}_f} \gamma_5
 \left[
g_0^T(q^2) \, \frac{q^2}{s_-} \,
 \left(p_\mu + p_\mu' - \frac{q_\mu}{q^2} \, (M^2-m^2) \right)
\right.
\cr
&
+ \left.g^T_\perp(q^2)(M-m) \left( \gamma_\mu + \frac{2 m}{s_-} \, p_\mu - \frac{2  M}{s_-} \, p '_\mu
  \right)
 \right] u_{\mathcal{B}_i} ,
\end{align}
with $s_\pm = (M\pm m)^2-q^2$. The subscripts $t$, $0$, and $\perp$ correspond to timelike (scalar), longitudinal, and transverse polarizations with respect to the momentum transfer $q^\mu$, respectively.
The   helicity-based form factors are related to our definitions in Eq.~(\ref{eq:invariant}) as follows
\begin{eqnarray}\label{eq:helicity}
f_0&=&F_1+\frac{s_+}{2M^2(1+r)}(F_2+F_3), \quad f_t=F_1+\frac{q^2+M^2-m^2}{2M^2(1-r)}F_2+\frac{-q^2+M^2-m^2}{2M^2(1-r)}F_3,  \nonumber\\
g_0&=&G_1-\frac{s_-}{2M^2(1-r)}(G_2+G_3), \quad g_t=G_1-\frac{q^2+M^2-m^2}{2M^2(1+r)}G_2-\frac{-q^2+M^2-m^2}{2M^2(1+r)}G_3,  \nonumber\\
 f_{\perp}&=&F_1, \quad g_{\perp}=G_1,  \nonumber\\
f_0^T&=&\frac{F_2^T-F_3^Tr}{M-m}, \quad f^T_\perp=-\frac{F_1^T}{M+m},\quad g_0^T=\frac{G_2^T+G_3^Tr}{M+m}, \quad g^T_\perp=\frac{G_1^T}{M-m}.
\end{eqnarray}
We can also write the relations between the ``Weinberg form factors" and helicity form factors as
\begin{eqnarray}\label{eq:HW}
f_0&=&f_1+f_2\frac{q^2}{M(M+m)}, \quad f_\perp=f_1+f_2(1+r), \quad f_t=f_1+f_3\frac{q^2}{M(M-m)},\nonumber\\
g_0&=&g_1-g_2\frac{q^2}{M(M-m)}, \quad g_\perp=g_1-g_2(1-r), \quad g_t=g_1-g_3\frac{q^2}{M(M+m)},\nonumber\\
f_0^T&=&-f_2^T-f_1^T(1+r), \quad f^T_\perp=-f_2^T-f_1^T\frac{q^2}{M(M+m)}, \nonumber\\
g_0^T&=&-g_2^T+g_1^T(1-r), \quad g^T_\perp=-g_2^T+g_1^T\frac{q^2}{M(M-m)},
\end{eqnarray}
which can be inverted to give
\begin{eqnarray}\label{eq:HW1}
f_1&=&f_\perp+(f_0-f_\perp)\frac{(M+m)^2}{s_+}, \quad f_2=-(f_0-f_\perp)\frac{(M+m)M}{s_+}, \nonumber\\
f_3&=&-\frac{M(M-m)}{q^2s_+}(f_0(M+m)^2-f_\perp q^2-f_ts_+),\nonumber\\
g_1&=&g_\perp+(g_0-g_\perp)\frac{(M-m)^2}{s_-}, \quad g_2=(g_0-g_\perp)\frac{(M-m)M}{s_-},\nonumber\\ g_3&=&\frac{M(M+m)}{q^2s_-}(g_0(M-m)^2-g_\perp q^2-g_ts_-),\nonumber\\
f_1^T&=&-(f^T_0-f^T_\perp)\frac{(M+m)M}{s_+}, \quad f^T_2=\frac{q^2f_0^T-(M+m)^2f^T_\perp}{s_+}, \nonumber\\
g_1^T&=&(g^T_0-g^T_\perp)\frac{(M-m)M}{s_-}, \quad g^T_2=\frac{q^2g_0^T-(M-m)^2g^T_\perp}{s_-}.
\end{eqnarray}
As can be seen from Eqs.~(\ref{eq:Weinberg1}), ~(\ref{eq:helicity}), ~(\ref{eq:HW}),  and~(\ref{eq:HW1}), one obtain some useful end point relations~\cite{Golz:2021imq}
\begin{eqnarray}\label{eq:FFC1}
 f_0(0) &=& f_t(0)=f_1(0), \quad  g_0(0) = g_t(0)=g_1(0), \nonumber\\ f^T_\perp(0)&=&-f_2^T(0)=g^T_\perp(0)=-g_2^T(0),
\end{eqnarray}
for $q^2=0$, and
\begin{eqnarray}\label{eq:FFC2}
 g_\perp(q^2_{\rm max}) = g_0(q^2_{\rm max}),   \quad g^T_\perp(q^2_{\rm max}) =g_0^T(q^2_{\rm max}),
\end{eqnarray}
for $q^2_{\rm max}=(M-m)^2$. These relations are method independent and therefore must be satisfied regardless of the computational approach. In the following, they will be used as consistency checks of our numerical calculations. For a detailed derivation of the end point relations in baryonic decays, the reader is referred to Ref.~\cite{Hiller:2021zth}.

\subsection{Angular distribution and observables } \label{sec:obe}
We consider the quasi-four-body decay chain $\mathcal{B}_i \to \mathcal{B}_f(\to \mathcal{B}_h P)\ell^+\ell^-$, in which the intermediate baryon $\mathcal{B}_f$ subsequently undergoes a weak decay into a lighter baryon $\mathcal{B}_h$ and a meson $P$. For an unpolarized initial baryon $\mathcal{B}_i$, the corresponding fourfold differential branching fraction can be decomposed into a set of trigonometric angular functions~\cite{Gutsche:2013pp, Boer:2014kda, Detmold:2016pkz, Blake:2017une}
\begin{eqnarray}\label{eq:dB4}
  \frac{d^4\mathcal{B}}{dq^2d\cos\theta_\ell d\cos\theta_hd \phi}&=&\frac{3}{8\pi}\{ K_1 \sin^2 \theta_\ell +K_2 \cos^2 \theta_\ell+K_3 \cos \theta_\ell \nonumber\\&+&(K_4 \sin^2 \theta_\ell +K_5 \cos^2 \theta_\ell+K_6 \cos \theta_\ell)\cos\theta_h  \nonumber\\ &+& (K_7\sin\theta_\ell\cos\theta_\ell+K_8\sin\theta_\ell)\sin\theta_h\cos \phi \nonumber\\&+&(K_9\sin\theta_\ell\cos\theta_\ell+K_{10}\sin\theta_\ell)\sin\theta_h\sin \phi\},
\end{eqnarray}
where $\theta_\ell$ and $\theta_h$ denote the polar angles of the negatively charged lepton and the daughter baryon, respectively, while $\phi$ is the azimuthal angle between the decay planes of the hadronic and leptonic systems. The angular variables are defined in the ranges $0<\theta_{\ell,h}<\pi$ and $0<\phi<2\pi$. The angular coefficients $K_i$ encode the underlying short-distance dynamics as well as the hadronic transition form factors. Their explicit expressions in terms of transversity amplitudes are adopted from Ref.~\cite{Blake:2017une}:
\begin{eqnarray}\label{eq:kk}
K_{1}&=& \frac{1}{4}(|A_{\parallel 1}^L|^2+|A_{\perp 1}^L|^2+|A_{\parallel 1}^R|^2+|A_{\perp 1}^R|^2)
+\frac{1}{4}(1+\beta_\ell^2)(|A_{\parallel 0}^L|^2+|A_{\perp 0}^L|^2+|A_{\parallel 0}^R|^2+|A_{\perp 0}^R|^2)\nonumber\\ &+&
\frac{1}{2}(1-\beta_\ell^2)\Re(A_{\parallel 1}^R A_{\parallel 1}^{L*}+A_{\perp 1}^R A_{\perp 1}^{L*}+A_{\parallel 0}^R A_{\parallel 0}^{L*}+A_{\perp 0}^R A_{\perp 0}^{L*})
+\frac{1}{2}(1-\beta_\ell^2)(|A_{\parallel t}|^2+|A_{\perp t}|^2)),\nonumber\\
K_{2}&=& \frac{1}{4}(1+\beta_\ell^2)(|A_{\parallel 1}^L|^2+|A_{\perp 1}^L|^2+|A_{\parallel 1}^R|^2+|A_{\perp 1}^R|^2)\nonumber\\
&+&\frac{1}{4}(1-\beta_\ell^2)(|A_{\parallel 0}^L|^2+|A_{\perp 0}^L|^2+|A_{\parallel 0}^R|^2+|A_{\perp 0}^R|^2)\nonumber\\ &+&
\frac{1}{2}(1-\beta_\ell^2)\Re(A_{\parallel 1}^R A_{\parallel 1}^{L*}+A_{\perp 1}^R A_{\perp 1}^{L*}+A_{\parallel 0}^R A_{\parallel 0}^{L*}+A_{\perp 0}^R A_{\perp 0}^{L*})
+\frac{1}{2}(1-\beta_\ell^2)(|A_{\parallel t}|^2+|A_{\perp t}|^2)),\nonumber\\
K_{3}&=&-\beta_\ell \Re(A_{\perp 1}^R A_{\parallel 1}^{R*}-A_{\perp 1}^L A_{\parallel 1}^{L*}),\nonumber\\
K_{4}&=&\frac{1}{2}\alpha_h\Re(A_{\perp 1}^R A_{\parallel 1}^{R*}+A_{\perp 1}^L A_{\parallel 1}^{L*})+\frac{1}{2}\alpha_h(1+\beta_\ell^2)\Re(A_{\perp 0}^R A_{\parallel 0}^{R*}+A_{\perp 0}^L A_{\parallel 0}^{L*})\nonumber\\&+&\frac{1}{2}\alpha_h(1-\beta_\ell^2)\Re(A_{\perp 1}^R A_{\parallel 1}^{L*}+A_{\parallel 1}^R A_{\perp 1}^{L*}+A_{\perp 0}^R A_{\parallel 0}^{L*}+A_{\parallel 0}^R A_{\perp 0}^{L*})
+\alpha_h(1-\beta_\ell^2)\Re(A_{\perp t} A_{\parallel t}^{*}),\nonumber\\
K_{5}&=&\frac{1}{2}\alpha_h(1+\beta_\ell^2)\Re(A_{\perp 1}^R A_{\parallel 1}^{R*}+A_{\perp 1}^L A_{\parallel 1}^{L*})+\frac{1}{2}\alpha_h(1-\beta_\ell^2)\Re(A_{\perp 0}^R A_{\parallel 0}^{R*}+A_{\perp 0}^L A_{\parallel 0}^{L*})\nonumber\\&+&\frac{1}{2}\alpha_h(1-\beta_\ell^2)\Re(A_{\perp 1}^R A_{\parallel 1}^{L*}+A_{\parallel 1}^R A_{\perp 1}^{L*}+A_{\perp 0}^R A_{\parallel 0}^{L*}+A_{\parallel 0}^R A_{\perp 0}^{L*})
+\alpha_h(1-\beta_\ell^2)\Re(A_{\perp t} A_{\parallel t}^{*}),\nonumber\\
K_{6}&=&-\frac{1}{2}\alpha_h\beta_\ell(|A_{\parallel 1}^R|^2+|A_{\perp 1}^R|^2-|A_{\parallel 1}^L|^2-|A_{\perp 1}^L|^2),\nonumber\\
K_{7}&=& \frac{1}{\sqrt{2}}\alpha_h\beta_\ell^2\Re(A_{\perp 1}^R A_{\parallel 0}^{R*}-A_{\parallel 1}^R A_{\perp 0}^{R*}+A_{\perp 1}^L A_{\parallel 0}^{L*}-A_{\parallel 1}^L A_{\perp 0}^{L*}),\nonumber\\
K_{8}&=& \frac{1}{\sqrt{2}}\alpha_h\beta_\ell^2\Re(A_{\perp 1}^R A_{\perp 0}^{R*}-A_{\parallel 1}^R A_{\parallel 0}^{R*}-A_{\perp 1}^L A_{\perp 0}^{L*}+A_{\parallel 1}^L A_{\parallel 0}^{L*}),\nonumber\\
K_{9}&=& \frac{1}{\sqrt{2}}\alpha_h\beta_\ell^2\Im(A_{\perp 1}^R A_{\parallel 0}^{R*}-A_{\parallel 1}^R A_{\perp 0}^{R*}+A_{\perp 1}^L A_{\parallel 0}^{L*}-A_{\parallel 1}^L A_{\perp 0}^{L*}),\nonumber\\
K_{10}&=& \frac{1}{\sqrt{2}}\alpha_h\beta_\ell^2\Im(A_{\perp 1}^R A_{\perp 0}^{R*}-A_{\parallel 1}^R A_{\parallel 0}^{R*}-A_{\perp 1}^L A_{\perp 0}^{L*}+A_{\parallel 1}^L A_{\parallel 0}^{L*}),
\end{eqnarray}
where $\beta_\ell=\sqrt{1-\frac{4m_{\ell}^2}{q^2}}$ denotes the relativistic velocity of the lepton in the dilepton rest frame, and $m_{\ell}$ is the corresponding lepton mass. The indices $L$ and $R$ refer to the left- and right-handed chiralities of the dilepton system, respectively. The parameter $\alpha_h$ is an angular asymmetry associated with the secondary decay $\mathcal{B}_f\to \mathcal{B}_h P$, arising from the hadronic matrix element of the parity-violating weak decay of the $\mathcal{B}_f$ baryon. In this work, we adopt the experimental values $\alpha_\Xi=-0.367^{+0.005}_{-0.006}$ for $\Xi^-\to \Lambda \pi^-$, $\alpha_\Sigma=-0.068\pm0.008$ for $\Sigma^-\to n \pi^-$, and $\alpha_\Lambda=0.746\pm0.008$ for $\Lambda\to p \pi^-$~\cite{ParticleDataGroup:2024cfk}. In terms of the helicity form factors, the ten transversity amplitudes take particularly simple forms~~\cite{Das:2018iap,Meinel:2016grj}.
\begin{eqnarray}\label{eq:amp}
A_{\perp 1}^{L(R)}&=&-2 N\sqrt{s_-}[f_\perp(C_9^{\text{eff}}\mp C^{\text{eff}}_{10})+\frac{2m_b}{q^2}f_\perp^T(M+m)C_7^{\text{eff}}],\nonumber\\
A_{\parallel 1}^{L(R)}&=&2 N\sqrt{s_+}[g_\perp(C_9^{\text{eff}}\mp C^{\text{eff}}_{10})+\frac{2m_b}{q^2}g_\perp^T(M-m)C_7^{\text{eff}}],\nonumber\\
A_{\perp 0}^{L(R)}&=& N\sqrt{\frac{2s_-}{q^2}}[f_0(M+m)(C_9^{\text{eff}}\mp C^{\text{eff}}_{10})+2m_bf_0^TC_7^{\text{eff}}],\nonumber\\
A_{\parallel 0}^{L(R)}&=&- N\sqrt{\frac{2s_+}{q^2}}[g_0(M-m)(C_9^{\text{eff}}\mp C^{\text{eff}}_{10})+2m_bg_0^TC_7^{\text{eff}}],\nonumber\\
A_{\perp t}&=& -2 N\sqrt{\frac{2s_+}{q^2}}f_t(M-m)C^{\text{eff}}_{10},\nonumber\\
A_{\parallel t}&=& 2 N\sqrt{\frac{2s_-}{q^2}}g_t(M+m)C^{\text{eff}}_{10}.
\end{eqnarray}
The $q^2$-dependent normalization factor $N$ is defined as
\begin{equation}
N(q^2)=G_F\alpha_eV_{tQ}^*V_{tb}\sqrt{\frac{\tau_{\mathcal{B}_i} q^2\beta_\ell\sqrt{\lambda_q}}{3\times2^{11}M^3\pi^5}},
\end{equation}
where $\lambda_q=M^4+m^4+q^4-2(M^2m^2+M^2q^2+m^2q^2)$ is the usual phase-space function. For brevity, the explicit $q^2$ dependence of the form factors has been omitted in the above expressions.

Integrating the fourfold differential branching fraction in Eq.~(\ref{eq:dB4}) over the angular variables $\theta_\ell$, $\theta_h$, and $\phi$ yields the $q^2$-differential branching fraction,
\begin{equation}\label{eq:dBB}
 \frac{d\mathcal{B}}{dq^2}=2 K_{1} + K_{2},
\end{equation}
which serves as the basis for defining a set of normalized angular observables. We further introduce the following ratios, which are sensitive to LFU violation,
\begin{equation}\label{eq:RR}
\mathcal{R}_{\mu(\tau)}= \frac{d\mathcal{B}_{\mu(\tau)}}{dq^2}/\frac{d\mathcal{B}_e}{dq^2},
\end{equation}
where the subscripts $e$, $\mu$, and $\tau$ indicate that the differential branching ratio is evaluated for the corresponding lepton species. A variety of physical observables can be constructed through weighted angular integrals of the distribution in Eq.~(\ref{eq:dB4}). The longitudinal polarization fraction of the dilepton system is defined as
\begin{equation}\label{eq:o1}
F^\ell_L(q^2)=\int\frac{d^4\mathcal{B}}{dq^2d\cos\theta_\ell d\cos\theta_hd \phi}(2-5\cos^2\theta_\ell) d\cos\theta_\ell d\cos\theta_hd \phi=
\frac{2 K_{1} - K_{2}}{2 K_{1} + K_{2}}. 
\end{equation}
Analogously, the hadronic side longitudinal polarization fraction reads
\begin{equation}\label{eq:o11}
F^h_L(q^2)=\int\frac{d^4\mathcal{B}}{dq^2d\cos\theta_\ell d\cos\theta_hd \phi}(2-5\cos^2\theta_h) d\cos\theta_\ell d\cos\theta_hd \phi=\frac{1}{3}.
\end{equation}
The forward-backward asymmetry with respect to the leptonic scattering angle is defined as
\begin{equation}\label{eq:o2}
A_{\text{FB}}^\ell(q^2)=\frac{1}{2 K_{1} + K_{2}}\left[\int_{0}^{1}-\int_{-1}^{0}\right]\frac{d^2\mathcal{B}}{dq^2d\cos \theta_\ell}d\cos \theta_\ell=\frac{3K_3}{2(2 K_{1} + K_{2})}.
\end{equation}
The corresponding asymmetry for the hadronic scattering angle reads
\begin{equation}\label{eq:o3}
A_{\text{FB}}^{h}(q^2)=\frac{1}{2 K_{1} + K_{2}}\left[\int_{0}^{1}-\int_{-1}^{0}\right]\frac{d^2\mathcal{B}}{dq^2d\cos \theta_h}d\cos \theta_h=\frac{2K_4+K_5}{2(2 K_{1} + K_{2})},
\end{equation}
which is the only angular observable that can be extracted from the angular distribution without requiring information on either the lepton helicity angle or the azimuthal angle between the decay planes~\cite{Boer:2019zmp}. Finally, a combined forward-backward asymmetry is obtained by integrating over both angular variables,
\begin{eqnarray}\label{eq:o4}
A_{\text{FB}}^{h\ell}(q^2)&=&\frac{1}{2 K_{1} + K_{2}}\left[\int_{0}^{1}-\int_{-1}^{0}\right]\left[\int_{0}^{1}-\int_{-1}^{0}\right]\frac{d^3\mathcal{B}}{dq^2d\cos \theta_hd\cos \theta_\ell}d\cos \theta_hd\cos \theta_\ell\nonumber\\&=&\frac{3K_6}{4(2 K_{1} + K_{2})}.
\end{eqnarray}
The $q^2$-integrated values of these observables, denoted by $\langle O\rangle$ for a generic quantity $O(q^2)$, are obtained by integrating the numerators and denominators separately over $q^2$. For $\langle\mathcal{R}_{\mu(\tau)}\rangle$, we adopt the same lower integration limit $4m_{\mu(\tau)}^2$ in both the numerator and denominator in order to remove phase-space effects in the ratio, since below this threshold only the electron mode is kinematically allowed.

The angular distribution and the definitions of the observables introduced for the dilepton mode can be directly applied to the dineutrino channel as well~\cite{Das:2023kch}. The angular coefficients $K_i$ for the dineutrino decay can be obtained from the dilepton expressions in the limit $m_\ell=0$. By comparing the decay amplitudes for the dilepton and dineutrino modes (see Eqs.~(\ref{eq:amplitude}) and~(\ref{eq:amplitude2})), the transversity amplitudes for the dineutrino channel can be derived from Eq.~(\ref{eq:amp}) through the replacements $C_9^{\text{eff}}\to C_L$, $C_{10}^{\text{eff}}\to -C_L$, and $C_7^{\text{eff}}\to 0$. This correspondence arises because neutrinos and left-handed charged leptons are related by $SU(2)_L$ symmetry~\cite{Buras:2014fpa}. In this scenario, the right-handed chirality amplitudes vanish, and the left-handed ones simplify to \cite{Das:2023kch}
\begin{eqnarray}\label{eq:ampv}
A_{\perp 1}^{L}&=&-4N'\sqrt{s_-}C_Lf_\perp\nonumber\\
A_{\parallel 1}^{L}&=&4N'\sqrt{s_+}C_L g_\perp\nonumber\\
A_{\perp 0}^{L}&=&2N'\sqrt{\frac{2s_-}{q^2}}C_Lf_0(M+m)\nonumber\\
A_{\parallel 0}^{L}&=&-2N'\sqrt{\frac{2s_+}{q^2}}C_Lg_0(M-m)\nonumber\\
A_{\perp t}&=&2N'\sqrt{\frac{2s_+}{q^2}}C_Lf_t(M-m)\nonumber\\
A_{\parallel t}&=&-2N'\sqrt{\frac{2s_-}{q^2}}C_Lg_t(M+m),
\end{eqnarray}
with $N^\prime=N|_{m_\ell\to 0}$. The numerical evaluation of these observables will be presented in the following section.

\section{Numerical Results}\label{sec:results}
In this section, we present the numerical results of the form factors and various observables, including the decay branching fractions, LFU ratios, and angular asymmetries, for the rare decays under consideration. The main input parameters used in the numerical calculations are the following masses $M_{\Xi_b}=5.797$ GeV, $m_{\Sigma}=1.189$ GeV, $m_{\Xi}=1.315$ GeV,  $m_{\Lambda}=1.116$ GeV, and $m_\tau=1.777$ GeV~\cite{ParticleDataGroup:2024cfk}.    For the mass of the $b$-quark, we adopt the $\overline{\text{MS}}$ scheme value $m_b(\bar{m}_b)=4.18$ GeV~\cite{ParticleDataGroup:2024cfk}.
 The lifetimes of the $\Xi_b$ baryons are taken as $\tau_{\Xi_b^-}=1.572$ ps and $\tau_{\Xi_b^0}=1.48$ ps~\cite{ParticleDataGroup:2024cfk}. The CKM matrix elements entering the effective Hamiltonian are taken from the PDG: $|V_{tb}|=0.999$, $|V_{ts}|=0.041$, and $|V_{td}|=0.00858$~\cite{ParticleDataGroup:2024cfk}. Additional input parameters relevant for the baryonic LCDAs, along with their associated uncertainties, are summarized in Table~\ref{tab:para1}.
\begin{table}[htbp]
    \caption{Central values of input parameters in the baryonic LCDAs and their variations~\cite{Liu:2009uc, Ali:2012pn, Liu:2008yg}.}
    \label{tab:para1}
    \centering
    \begin{tabular}{l S[table-format=-1.2] S[table-format=-1.1] c}
    \hline\hline
    Parameter & \multicolumn{1}{c}{Central Value} & \multicolumn{1}{c}{Variation} & Unit \\
    \hline
    $A$                               & 0.5    & \pm0.2 & $\cdots$ \\
    $f_\Xi$                           & 9.9    & \pm0.4 &$10^{-3}~\text{GeV}^{2}$ \\
    $\lambda_1(\Xi)$                  & -2.8   & \pm0.1 & $10^{-2}~\text{GeV}^{2}$\\
    $\lambda_2(\Xi)$                  & 5.2    & \pm0.2 &$10^{-2}~\text{GeV}^{2}$ \\
    $\lambda_3(\Xi)$                  & 1.7    & \pm0.1 &$10^{-2}~\text{GeV}^{2}$ \\
    \midrule
    $f_\Sigma$                        & 9.4    & \pm0.4 & $10^{-3}~\text{GeV}^{2}$ \\
    $\lambda_1(\Sigma)$               & -2.5   & \pm0.1 &$10^{-2}~\text{GeV}^{2}$ \\
    $\lambda_2(\Sigma)$               & 4.4    & \pm0.1 &$10^{-2}~\text{GeV}^{2}$ \\
    $\lambda_3(\Sigma)$               & 2.0    & \pm0.1 & $10^{-2}~\text{GeV}^{2}$\\
    \midrule
    $f_\Lambda$                       & 6.0    & \pm0.3 & $10^{-3}~\text{GeV}^{2}$ \\
    $\lambda_1(\Lambda)$              & 1.0    & \pm0.3 &$10^{-2}~\text{GeV}^{2}$ \\
    $\lambda_2(\Lambda)$~\footnotemark[1]     & 0.83   & \pm0.05 &$10^{-2}~\text{GeV}^{2}$ \\
    $\lambda_3(\Lambda)$~\footnotemark[1]     & 0.83   & \pm0.05 &$10^{-2}~\text{GeV}^{2}$ \\
    \hline\hline
    \end{tabular}
    \vspace{4pt}
    \footnotetext[1]{ Only the absolute values are given in~\cite{Liu:2008yg}. We adopt the positive values, which yield satisfactory results for the $\Lambda_b \to \Lambda$ form factors~\cite{Yang:2025yaw}.}
\end{table}
\subsection{ Form factor shapes} \label{sec:FFsv}

There are four available models for the $\Xi_b$ baryon LCDAs, with their explicit expressions collected in~\cite{Rui:2025iwa}.
These models differ significantly in their theoretical motivations, mathematical formulations, and physical assumptions, particularly regarding SU(3) flavor symmetry.
We first examine the impact of these different models on the form factors. The fitted central values of $a_0$ and $a_1$ together with their standard errors  for the Weinberg form factors~\footnote{Here, we choose to fit the Weinberg form factors, from which the helicity form factors can be obtained via Eq.~(\ref{eq:HW}). This scheme avoids introducing kinematic singularities in the axial-vector and pseudotensor form factors at zero recoil.} of $\Xi_b\to \Xi,\Sigma,\Lambda$ transitions with the four different LCDA models are compared in Table~\ref{tab:para}.
 One can see that the variations of these fitted parameters across models are typically within the 5$\sigma$ error bands. For instance, the parameter $a_0$ of $f_1$ for the $\Xi_b \to \Xi$ transition ranges from $0.551$ to $0.696$, with uncertainties that overlap.
In addition, all models exhibit consistency in the signs (positive or negative) of $a_0$ and $a_1$ for each form factor. This agreement directly implies that the $q^2$ dependence of the form factors follows a similar trend across all models, thereby ensuring that physical observables, such as branching fractions and forward-backward asymmetries, are less sensitive to the choice of $\Xi_b$ baryon LCDA model.

\begin{figure}[htbp]
\begin{center}
\setlength{\abovecaptionskip}{0pt}
\centerline{
 \subfigure{\epsfxsize=8cm \epsffile{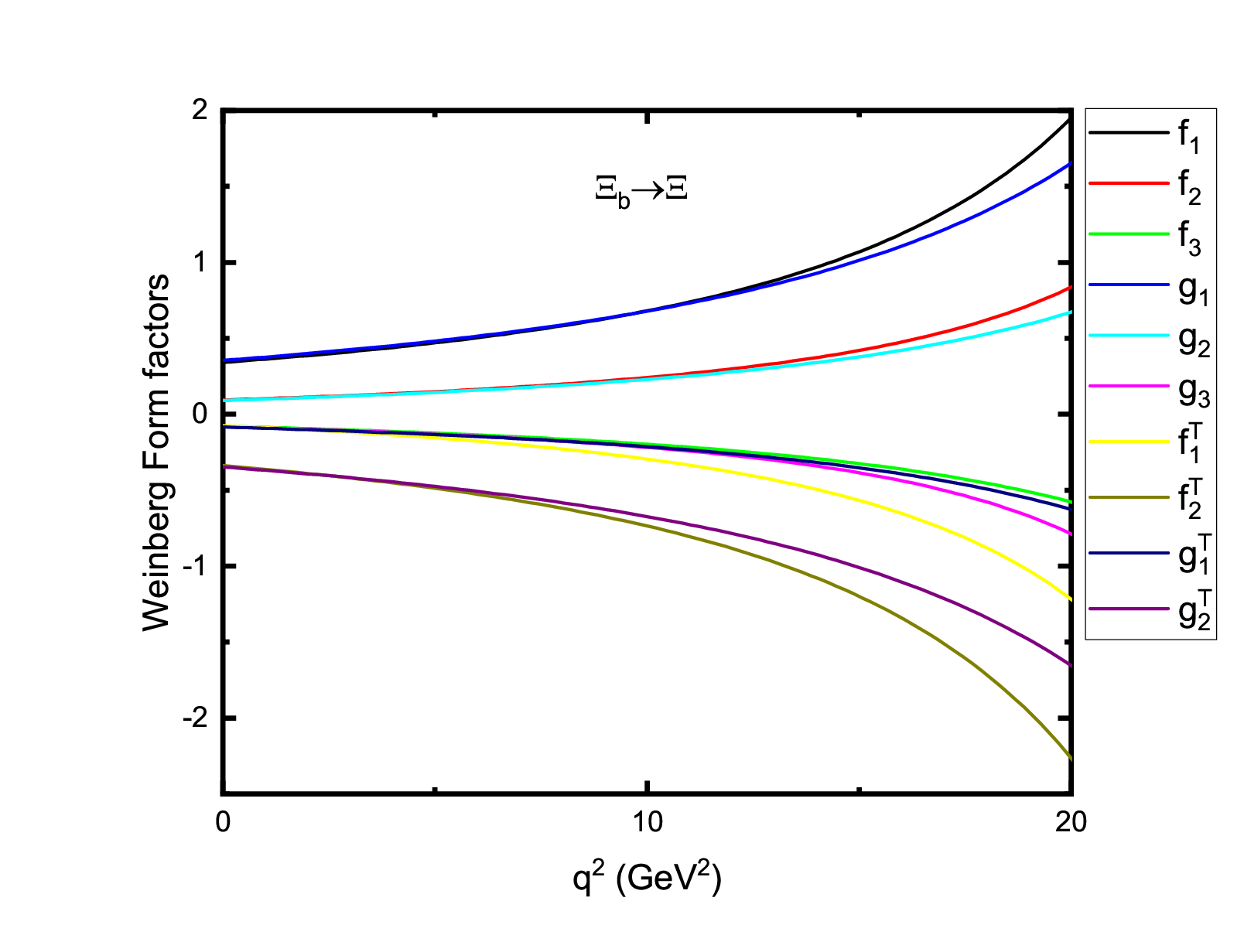} }
\subfigure{ \epsfxsize=8cm \epsffile{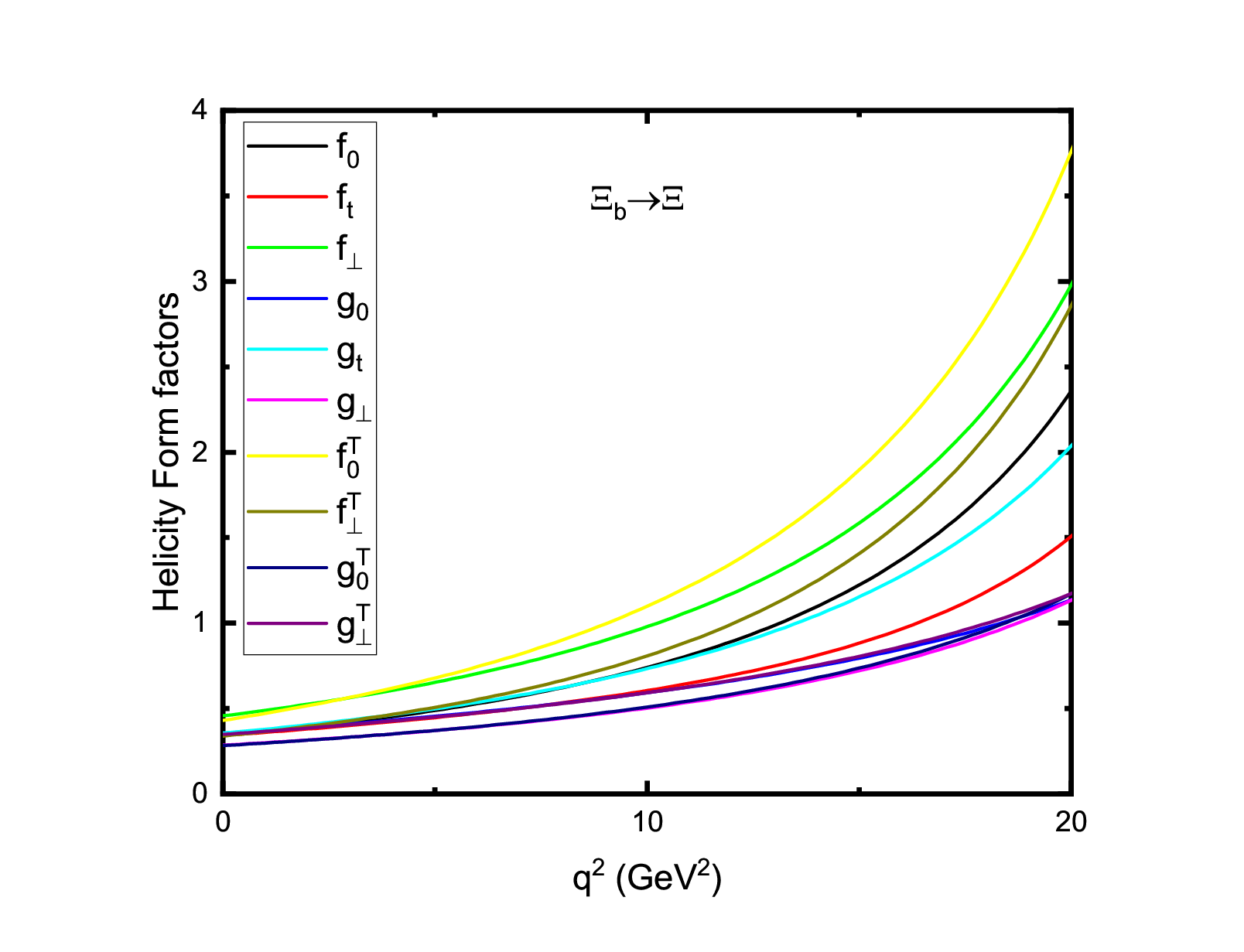}}}
\vspace{0.5cm}
\centerline{
\subfigure{\epsfxsize=8cm \epsffile{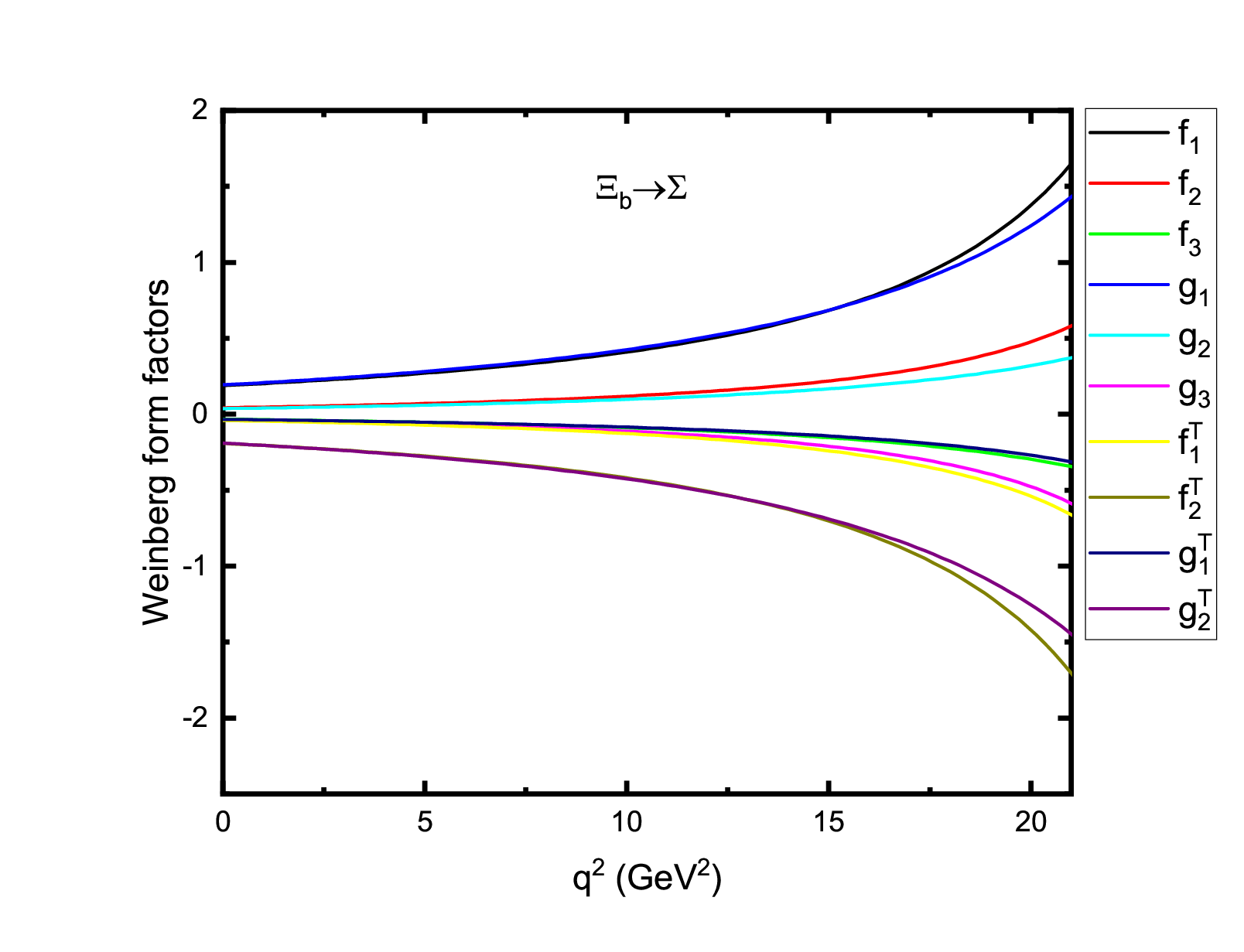} }
\subfigure{ \epsfxsize=8cm \epsffile{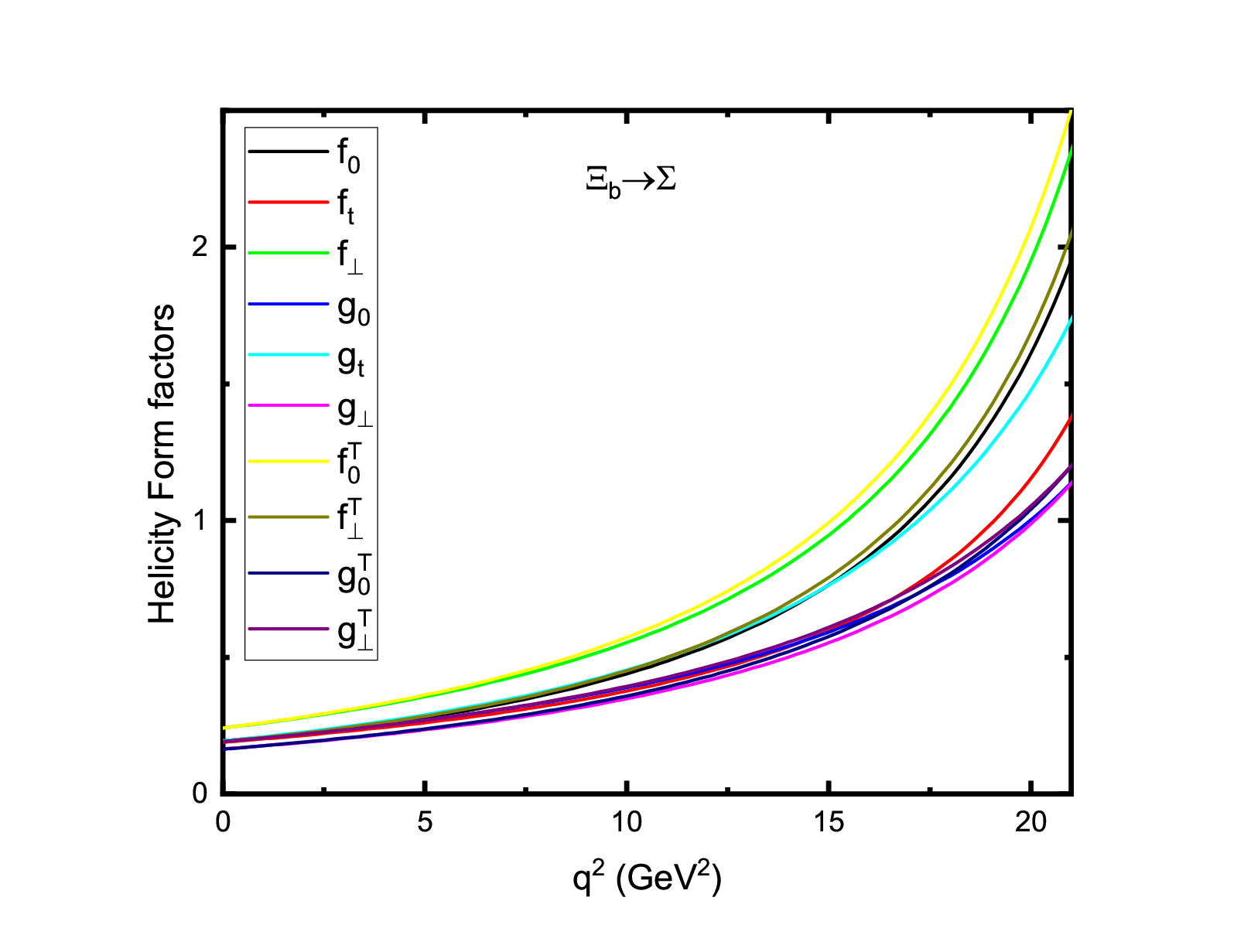}}}
\vspace{0.5cm}
\centerline{
\subfigure{\epsfxsize=8cm \epsffile{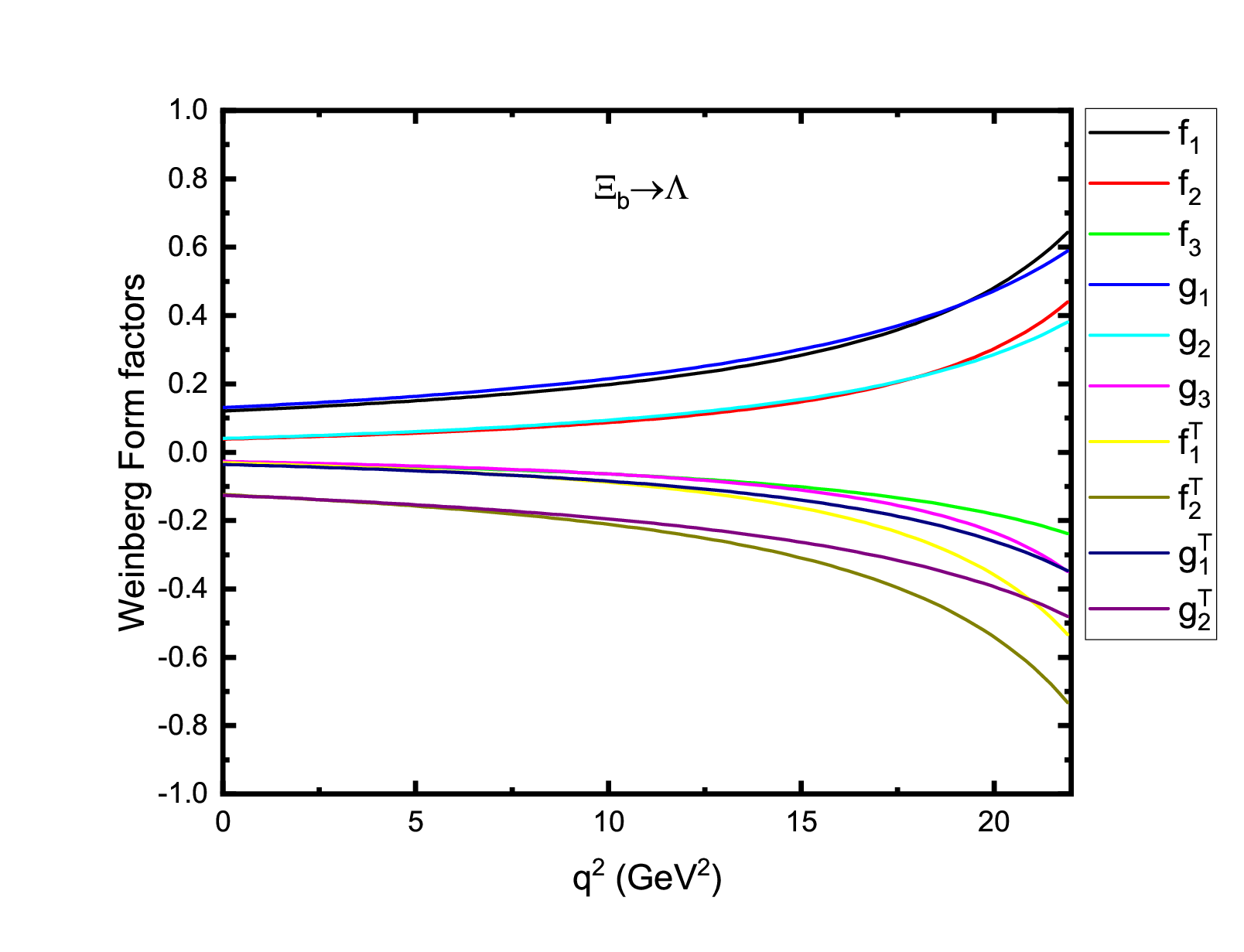} }
\subfigure{ \epsfxsize=8cm \epsffile{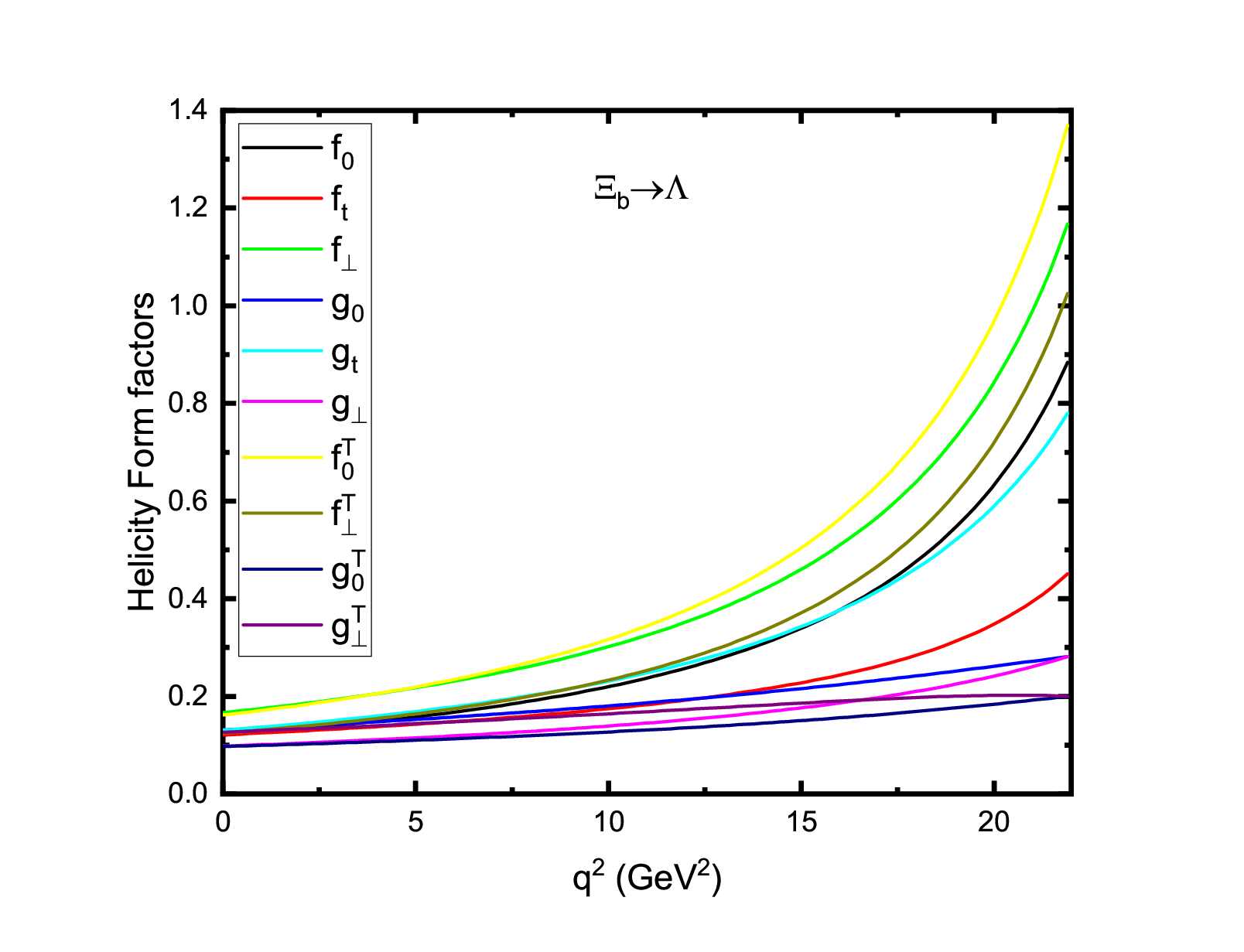}}}
\vspace{0.5cm}
\caption{The $q^2$ distributions of  the Weinberg form factors (left) and helicity form factors (right) with Gegenbauer model of baryonic LCDAs. Upper, middle, and lower panels correspond to $\Xi_b$ to $\Xi$, $\Sigma$, and $\Lambda$ transitions, respectively. Here, the uncertainties in these curves are not shown for clarity.}
 \label{fig:FFs}
\end{center}
\end{figure}

 As noted in~\cite{Ali:2012pn}, the SU(3)-breaking effects due to the strange-quark mass are incorporated in the $\Xi_b$ baryonic LCDAs within the Gegenbauer model, providing detailed information on the quark distributions inside the baryon. Moreover, the Gegenbauer model demonstrates good stability with respect to its parameters~\cite{Ali:2012pn}, and the resulting uncertainties are minimal~\cite{Rui:2023fiz,Rui:2025iwa}. For these reasons, we adopt the Gegenbauer model in the subsequent calculations. Using the fitted results in Table~\ref{tab:para}, we obtain the $q^2$-dependence of the form factors for the $\Xi_b\to \Xi,\Sigma,\Lambda$ transitions, as illustrated in Fig.~\ref{fig:FFs}. For clarity, uncertainties are not displayed in these curves but will be discussed later. The left and right panels correspond to the Weinberg and helicity form factors, respectively. For the Weinberg form factors, $f_{1,2},g_{1,2}$ are positive, while $f_3, g_3,f_{1,2}^T,g_{1,2}^T$ are negative. All helicity form factors remain positive throughout the allowed $q^2$ region. Curves such as $f_1$ and $g_1$ exhibit nearly identical shapes at low $q^2$, splitting only at higher $q^2$. Overall, Fig.~\ref{fig:FFs} shows that the magnitudes of all form factors steadily increase with $q^2$, consistent with expectations from weak decays.

\begin{sidewaystable}[htbp]
\caption{Comparison of the fitted $z$-expansion parameters describing the Weinberg form factors for the concerned transitions with different models of baryonic LCDAs.}\label{tab:para}
    \centering
    \footnotesize 
    \setlength{\tabcolsep}{3pt} 
	\begin{tabular}[t]{lcccccccccccc}
	\hline\hline
Transition &Model   & Parameter     &$f_1$ &$f_2$ &$f_3$ &$g_1$ &$g_2$ &$g_3$ &$f_1^T$ &$f_2^T$&$g_1^T$ &$g_2^T$\\ \hline
$\Xi_b\to \Xi$ &Exponential &$a_0 $     &0.551(21)  &0.253(12)  &-0.216(13)  &0.627(6)   &0.264(4)  &-0.225(4)&-0.235(9)&-0.560(6)&-0.254(6)&-0.619 (5) \\
            &&$a_1 $     &-1.057(99) &-0.716(57) &0.640(59)   &-1.328 (26) &-0.784(17) &0.660(17) &0.667(44) &1.068(26) &0.780 (29)&1.333(23)\\
&Gegenbauer  &$a_0 $     &$0.622(12)$  &0.267(5)  &-0.239(5)  &0.682(15)   &0.278(3)  &-0.242(4)&-0.389(9)&-0.723(12)&-0.259(13)&-0.683(12)  \\
            &&$a_1 $     &$-1.240(58)$ &-0.767(25) &0.713(22)   &-1.447(69)  &-0.826(15) &0.711(19) &1.390(40) &1.698(56) &0.773(60) &1.482(58)\\
&QCDSR       &$a_0 $     &0.582 (19) &0.269(4)  &-0.232(3)  &0.677 (11)  &0.277(4)  &-0.230(3)&-0.259(10)&-0.590(5)&-0.247(11)&-0.649(12)  \\
            &&$a_1 $     &-1.173(90) &-0.773(18) &0.701 (12)  &-1.514 (52) &-0.829(16) &0.671(13) &0.760(45) &1.169(23) &0.739(50) &1.431(56)\\
&Free Parton &$a_0 $     &0.696 (16) &0.339(4)  &-0.290(5)  &0.816 (12)  &0.344(7) &-0.295(6)&-0.323(12)&-0.747(7)&-0.314(10)&-0.828(10)  \\
            &&$a_1 $     &-1.303(74) &-0.977(20) &0.875(25)   &-1.743(58)  &-1.024 (31)&0.871 (28)&0.948(57) &1.487(33) &0.943(46) &1.850(46)\\
$\Xi_b\to \Sigma$ &Exponential
&$a_0 $      &0.469(11)  &0.172(2)  &-0.135(4)  &0.489(9)   &0.165(4)  &-0.155(2) &-0.173(18) &-0.442(11) &-0.158(9) &-0.504(15)  \\
&&$a_1 $     &-0.868(37) &-0.391(7) &0.316(14)   &-0.899(29)  &-0.386(14) &0.363(8)  &0.407(59)  &0.765(37)  &0.384 (32) &0.965(50)\\
&Gegenbauer
&$a_0 $      &0.432(22) &0.152(5)  &-0.127(5)  &0.520 (14)  &0.135(4)  &-0.146(4) &-0.173(19) &-0.446(14) &-0.113(16) &-0.526(9)  \\
&&$a_1 $     &-0.785 (74) &-0.355 (15)&0.309(18)   &-1.056 (48) &-0.318 (13)&0.352(13)  &0.431(67)  &0.825 (46) &0.260(53)  &1.086(31)\\
&QCDSR
&$a_0 $      &0.444 (13) &0.175(2)  &-0.143(4)  &0.511 (10)  &0.172(4)  &-0.160(3) &-0.169(13) &-0.443 (9)&-0.150(9) &-0.500 (10) \\
&&$a_1 $     &-0.759(44)  &-0.396 (7)&0.339(12)   &-0.949(35)  &-0.404(12) &0.375(10)  &0.389 (42) &0.745(30)  &0.354(30)  &0.928(34)\\
&Free Parton
&$a_0 $      &0.528(14)  &0.225(6)  &-0.179(6)  &0.652(23)   &0.216(3)  &-0.210(6) &-0.208(12) &-0.566 (15)&-0.196(13))&-0.635(12)  \\
&&$a_1 $     &-0.856(48)  &-0.515 (19)&0.423(19)   &-1.230(77)  &-0.508(10) &0.503(20)  &0.471(42)  &0.965(50)  &0.468(43)  &1.191(41)\\
$\Xi_b\to \Lambda$ &Exponential
&$a_0 $      &0.113(9)  &0.081(3)  &-0.057(3)  &0.152(7)   &0.093(2)  &-0.051(3) &-0.082(10) &-0.126 (7)&-0.075(7) &-0.128(5)  \\
&&$a_1 $     &-0.109(28) &-0.162(9) &0.116 (10)  &-0.199(23)  &-0.194(7) &0.099(8)  &0.180(30)  &0.135(21)  &0.155(23)  &0.140(16)\\
&Gegenbauer
&$a_0 $      &0.108(22)  &0.087(3)  &-0.067(4)  &0.165(13)   &0.114 (3) &-0.051(4) &-0.097(14) &-0.111(13)&-0.100 (8)&-0.116 (9) \\
&&$a_1 $     &-0.004(67) &-0.155 (10)&0.120 (12)  &-0.146(42)  &-0.231 (10)&0.079(12)  &0.206(43)  &0.003 (40) &0.203 (24) &0.009(29)\\
&QCDSR
&$a_0 $      &0.114(11)  &0.086 (2) &-0.065(2)  &0.167(6)   &0.100 (2) &-0.051(2)&-0.058(20) &-0.113(14) &-0.083(3) &-0.145(7)  \\
&&$a_1 $     &-0.109 (34)&-0.174 (6)&0.135 (6)  &-0.239(19)  &-0.214(6) &0.099(7)  &0.104(63)  &0.090 (43) &0.177(11)  &0.189(22)\\
&Free Parton
&$a_0 $      &0.127(10)  &0.107 (5) &-0.082(6)  &0.196(9)   &0.123(2)  &-0.064(2) &-0.096(18) &-0.158(12) &-0.104(5) &-0.173(5)  \\
&&$a_1 $     &-0.084(30) &-0.216(17) &0.174(19)   &-0.256(29)  &-0.257(8) &0.123(7)  &0.203(55)  &0.161(38)  &0.220(16)  &0.207(15)\\
		\hline\hline
	\end{tabular}
\end{sidewaystable}

\begin{figure}[htbp]
\begin{center}
\setlength{\abovecaptionskip}{0pt}
\centerline{
 \subfigure{\epsfxsize=6.5cm \epsffile{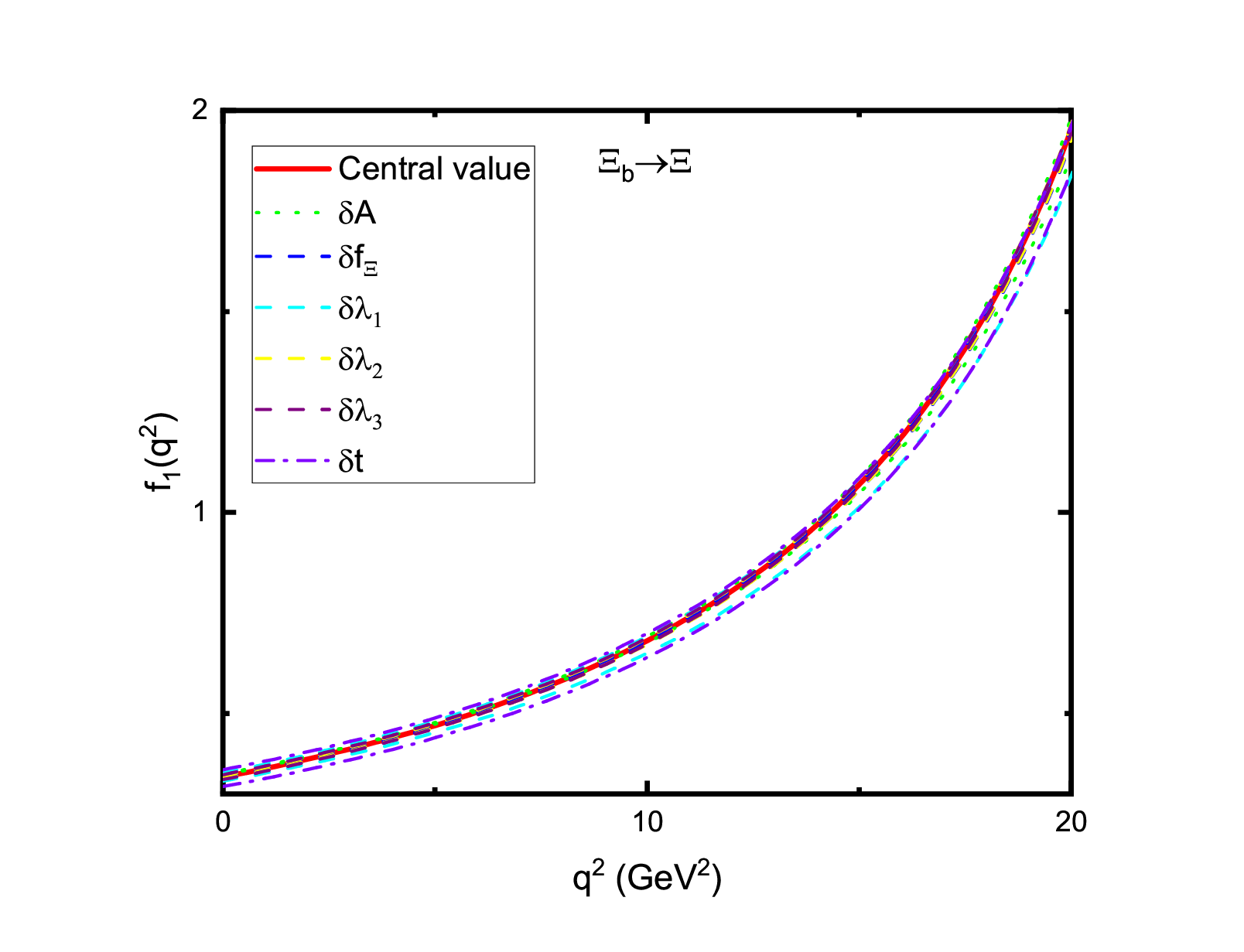} }
\hspace{-1.5cm}\subfigure{\epsfxsize=6.5cm \epsffile{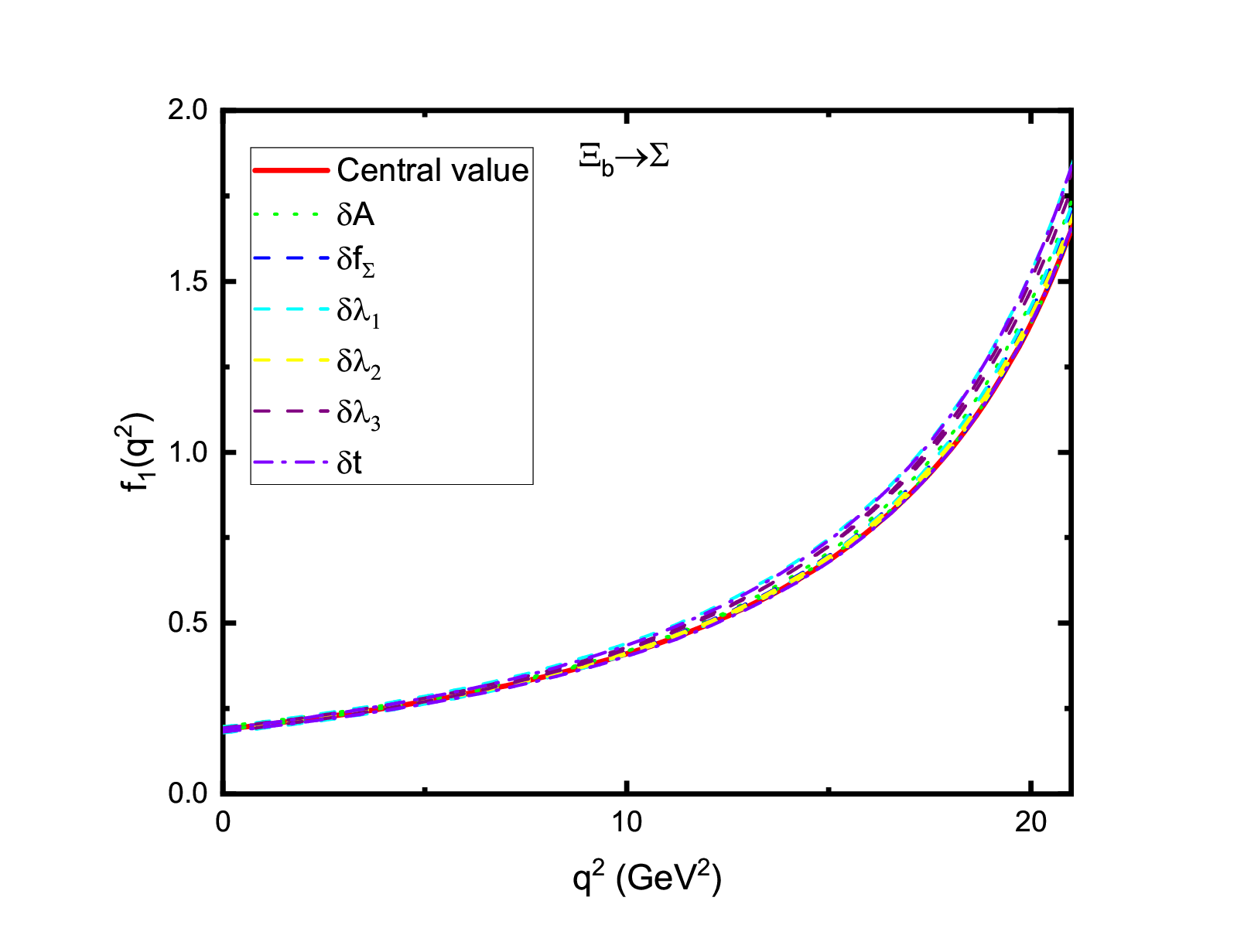} }
\hspace{-1.7cm}\subfigure{ \epsfxsize=6.5cm \epsffile{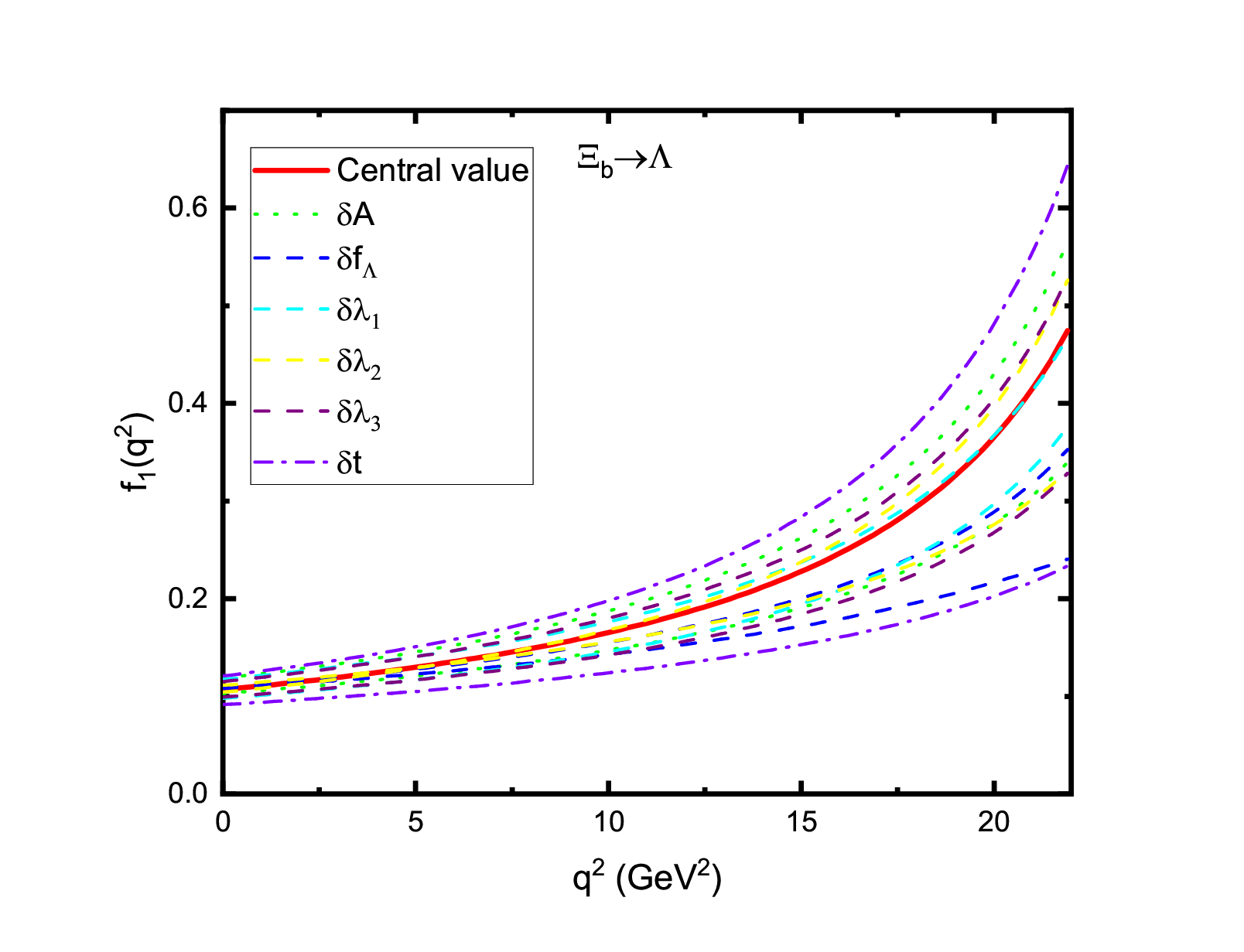}}}
\vspace{0.5cm}
\caption{
The theoretical uncertainties in the PQCD predictions for the form factor $f_1(q^2)$ of $\Xi_b\to\Xi$ (left), $\Xi_b\to\Sigma$ (middle), and $\Xi_b\to\Lambda$ (right) transitions.   For a detailed description of these curves, see the text.}
 \label{fig:f1}
\end{center}
\end{figure}

\begin{table}[htbp]
	\caption{Form factor predictions at the large ($q^2=0$) and zero ($q^2=(M-m)^2$) recoils for the $\Xi_b\to \Xi, \Sigma, \Lambda$ transitions. The errors are obtained by varying all parameters in a given range as shown in Table~\ref{tab:para1}. The errors from the model parameters in the final baryon LCDAs have been added in quadratures.}
	\label{tab:form1}
     \centering
	\begin{tabular}[t]{lccc}
	\hline\hline
Form factor        &  $\Xi_b\to \Xi$    & $\Xi_b^-\to \Sigma^-$~\footnotemark[1]   & $\Xi_b\to \Lambda$   \\ \hline
$f_1(0) $        & $0.342^{+0.008}_{-0.000}\pm0.013^{+0.018}_{-0.024} $ 
                 & $0.189^{+0.008+0.008+0.002}_{-0.002-0.013-0.007}$
                 & $0.106^{+0.012+0.014+0.014}_{-0.004-0.012-0.015}$                    \\
$f_2(0) $        & $0.094^{+0.004}_{-0.001}\pm0.004^{+0.004}_{-0.006}$  
                 & $0.043\pm0.003\pm0.003^{+0.001}_{-0.002}$ 
                 & $0.036^{+0.002+0.004+0.002}_{-0.000-0.005-0.003}$                      \\
$f_3(0) $        & -$0.078^{+0.000}_{-0.001}\pm0.004^{+0.007}_{-0.005}$  
                 & -$0.032^{+0.001+0.002}_{-0.003-0.003}\pm0.002$  
                 & -$0.027^{+0.001+0.003+0.004}_{-0.002-0.002-0.003}$                  \\
$g_1(0) $        & $0.354^{+0.010+0.013+0.016}_{-0.000-0.016-0.020}$
                 & $0.193^{+0.010+0.010+0.004}_{-0.001-0.011-0.005}$
                 & $0.117^{+0.011+0.015+0.013}_{-0.003-0.013-0.015}$                    \\
$g_2(0) $        & $0.091^{+0.004+0.004+0.005}_{-0.001-0.005-0.006}$
                 & $0.037^{+0.004+0.003+0.001}_{-0.003-0.004-0.002}$
                 & $0.038^{+0.002}_{-0.000}\pm0.004^{+0.002}_{-0.003}$ \\ 
$g_3(0) $        &-$0.081^{+0.000}_{-0.002}\pm0.003^{+0.007}_{-0.005}$   
                 &-$0.038^{+0.001}_{-0.002}\pm0.002^{+0.002}_{-0.001}$    
                 &-$0.025^{+0.001}_{-0.002}\pm0.003^{+0.004}_{-0.002}$                     \\ 
$f_1^T(0) $      & -$0.075^{+0.000+0.000+0.000}_{-0.015-0.021-0.015}$
                 & -$0.041^{+0.001+0.002+0.001}_{-0.005-0.005-0.003}$
                 & -$0.029^{+0.000+0.003+0.003}_{-0.002-0.005-0.002}$                     \\
$f_2^T(0) $      & -$0.338^{+0.000+0.004+0.014}_{-0.015-0.025-0.025}$
                 & -$0.191^{+0.002+0.012+0.005}_{-0.008-0.010-0.004}$
                 & -$0.112^{+0.004+0.013+0.016}_{-0.012-0.014-0.012}$                      \\
$g_1^T(0) $      & -$0.084^{+0.000+0.005+0.006}_{-0.002-0.004-0.005}$
                 & -$0.034\pm0.001^{+0.005+0.002}_{-0.002-0.000}$   
                 & -$0.033^{+0.001+0.003+0.004}_{-0.003-0.004-0.003}$                     \\
$g_2^T(0) $      & -$0.348^{+0.000+0.014+0.023}_{-0.007-0.012-0.018}$
                 & -$0.190^{+0.003}_{-0.009}\pm0.011^{+0.005}_{-0.004}$   
                 & -$0.113^{+0.006}_{-0.011}\pm0.014^{+0.016}_{-0.013}$                    \\  
$f_1(q^2_{\text{max}})$    & $1.978^{+0.028+0.125+0.007}_{-0.067-0.019-0.105}$
                           & $1.711^{+0.085+0.265+0.192}_{-0.003-0.000-0.000}$
                           & $0.475^{+0.090+0.074+0.169}_{-0.135-0.327-0.241}$                    \\
$f_2(q^2_{\text{max}}) $   & $0.849^{+0.060+0.033+0.079}_{-0.000-0.054-0.064}$
                           & $0.604^{+0.067+0.052+0.000}_{-0.037-0.036-0.009}$
                           & $0.384^{+0.041+0.119+0.056}_{-0.000-0.059-0.020}$                    \\
$f_3(q^2_{\text{max}}) $   &-$0.584^{+0.000+0.017+0.055}_{-0.030-0.040-0.071}$
                           &-$0.356^{+0.027+0.042+0.038}_{-0.011-0.029-0.005}$
                           &-$0.199^{+0.000+0.040+0.031}_{-0.021-0.087-0.040}$                    \\
$g_1(q^2_{\text{max}})$     & $1.670^{+0.024+0.058+0.192}_{-0.081-0.000-0.331}$
                           & $1.471^{+0.009+0.093+0.000}_{-0.067-0.122-0.047}$
                           & $0.498^{+0.036+0.079+0.091}_{-0.037-0.139-0.168}$                    \\
$g_2(q^2_{\text{max}}) $    & $0.681^{+0.035+0.035+0.033}_{-0.017-0.046-0.044}$
                           & $0.383^{+0.057+0.068+0.014}_{-0.017-0.015-0.033}$
                           & $0.347^{+0.017+0.065+0.055}_{-0.001-0.055-0.032}$                    \\
$g_3(q^2_{\text{max}}) $    &-$0.801^{+0.003+0.038+0.081}_{-0.030-0.031-0.044}$
                           &-$0.611^{+0.004+0.022+0.033}_{-0.034-0.066-0.030}$
                           &-$0.345^{+0.020+0.063+0.037}_{-0.011-0.061-0.038}$                     \\
$f_1^T(q^2_{\text{max}}) $  &-$1.238^{+0.364+0.571+0.401}_{-0.000-0.000-0.000}$
                           &-$0.688^{+0.127+0.271+0.135}_{-0.000-0.000-0.000}$
                           &-$0.429^{+0.000+0.095+0.080}_{-0.065-0.163-0.105}$                      \\
$f_2^T(q^2_{\text{max}}) $  &-$2.297^{+0.262+0.374+0.388}_{-0.000-0.000-0.000}$
                           &-$1.768^{+0.000+0.048+0.000}_{-0.098-0.166-0.009}$
                           &-$0.488^{+0.029+0.112+0.169}_{-0.067-0.131-0.245}$                       \\
$g_1^T(q^2_{\text{max}}) $  &-$0.635^{+0.015+0.036+0.077}_{-0.024-0.058-0.030}$
                           &-$0.322^{+0.067+0.038+0.031}_{-0.093-0.113-0.090}$
                           &-$0.302^{+0.000+0.065+0.068}_{-0.004-0.008-0.046}$                      \\
$g_2^T(q^2_{\text{max}}) $  &-$1.672^{+0.000+0.056+0.119}_{-0.023-0.031-0.106}$
                           &-$1.488^{+0.000+0.153+0.083}_{-0.064-0.006-0.013}$
                           &-$0.349^{+0.000+0.119+0.109}_{-0.050-0.079-0.131}$                      \\
$f_0(0) $        & $0.342^{+0.008}_{-0.000}\pm0.013^{+0.018}_{-0.024}$   
                 & $0.189^{+0.008+0.008+0.002}_{-0.002-0.013-0.007}$
                 & $0.106^{+0.012+0.014+0.014}_{-0.004-0.012-0.015}$                     \\
$f_\perp(0) $    & $0.457^{+0.010+0.018+0.023}_{-0.000-0.017-0.031}$
                 & $0.241^{+0.012+0.012+0.003}_{-0.005-0.016-0.009}$
                 & $0.149^{+0.014+0.019+0.017}_{-0.004-0.018-0.019}$                     \\
$f_t(0) $        & $0.342^{+0.008}_{-0.000}\pm0.013^{+0.018}_{-0.024}$   
                 & $0.189^{+0.008+0.008+0.002}_{-0.002-0.013-0.007}$
                 & $0.106^{+0.012+0.014+0.014}_{-0.004-0.012-0.015}$                     \\
$g_0(0) $        & $0.354^{+0.010+0.013+0.015}_{-0.000-0.016-0.020}$
                 & $0.193^{+0.010}_{-0.001}\pm0.011^{+0.003}_{-0.005}$   
                 & $0.117^{+0.011+0.015+0.013}_{-0.003-0.014-0.016}$                     \\
$g_\perp(0) $    & $0.284^{+0.005+0.012+0.012}_{-0.000-0.010-0.015}$
                 & $0.164^{+0.007}_{-0.000}\pm0.008^{+0.002}_{-0.004}$  
                 & $0.086^{+0.010+0.012+0.012}_{-0.003-0.010-0.013}$                    \\
$g_t(0) $        & $0.354^{+0.010+0.013+0.015}_{-0.000-0.016-0.020}$
                 & $0.193^{+0.010}_{-0.001}\pm0.011^{+0.003}_{-0.005}$   
                 & $0.117^{+0.011+0.015+0.013}_{-0.003-0.014-0.016}$                    \\
$f_0^T(0) $      & $0.429^{+0.033+0.029+0.043}_{-0.000-0.000-0.009}$
                 & $0.240^{+0.014+0.015+0.007}_{-0.004-0.013-0.006}$
                 & $0.147^{+0.014+0.020+0.014}_{-0.003-0.017-0.019}$                     \\
$g_0^T(0) $      & $0.282^{+0.006+0.010+0.014}_{-0.000-0.009-0.018}$
                 & $0.164^{+0.008+0.009+0.004}_{-0.002-0.007-0.003}$
                 & $0.086^{+0.010+0.012+0.012}_{-0.003-0.010-0.013}$                     \\
$f_\perp^T(0) $  & $0.338^{+0.015+0.009+0.024}_{-0.000-0.000-0.015}$
                 & $0.191^{+0.008+0.009+0.004}_{-0.002-0.011-0.005}$
                 & $0.112^{+0.012+0.014+0.012}_{-0.004-0.013-0.016}$                      \\
$g_\perp^T(0) $  & $0.348^{+0.007+0.014+0.018}_{-0.000-0.012-0.023}$
                 & $0.190^{+0.009}_{-0.003}\pm0.011^{+0.003}_{-0.004}$   
                 & $0.113^{+0.011}_{-0.006}\pm0.014^{+0.013}_{-0.016}$                      \\ 
$f_0(q^2_{\text{max}}) $        & $2.391^{+0.058+0.137+0.045}_{-0.054-0.039-0.137}$
                                & $2.027^{+0.066+0.283+0.190}_{-0.000-0.000-0.000}$
                                & $0.685^{+0.098+0.098+0.199}_{-0.112-0.305-0.252}$        \\
$f_\perp(q^2_{\text{max}}) $    & $3.019^{+0.102+0.076+0.104}_{-0.034-0.156-0.184}$
                 & $2.440^{+0.078+0.308+0.186}_{-0.000-0.000-0.005}$
                 & $0.932^{+0.108+0.159+0.236}_{-0.086-0.281-0.265}$                      \\
$f_t(q^2_{\text{max}}) $         & $1.525^{+0.005+0.010+0.000}_{-0.087-0.115-0.063}$
                 & $1.428^{+0.107+0.264+0.188}_{-0.011-0.010-0.000}$
                 & $0.314^{+0.079+0.086+0.137}_{-0.152-0.349-0.216}$                      \\
$g_0(q^2_{\text{max}}) $  & $1.144^{+0.036+0.134+0.165}_{-0.108-0.000-0.296}$
                 & $1.166^{+0.000+0.048+0.000}_{-0.053-0.137-0.024}$
                 & $0.220^{+0.019+0.044+0.061}_{-0.028-0.094-0.137}$                     \\
$g_\perp(q^2_{\text{max}}) $     & $1.144^{+0.036+0.134+0.166}_{-0.108-0.000-0.296}$
                 & $1.166^{+0.000+0.048+0.000}_{-0.053-0.137-0.024}$
                 & $0.220^{+0.019+0.044+0.061}_{-0.028-0.094-0.137}$                    \\
$g_t(q^2_{\text{max}}) $         & $2.060^{+0.022+0.140+0.213}_{-0.066-0.000-0.370}$
                 & $1.791^{+0.027+0.122+0.004}_{-0.070-0.129-0.064}$
                 & $0.628^{+0.055+0.123+0.151}_{-0.027-0.158-0.207}$                       \\
$f_0^T(q^2_{\text{max}}) $      & $3.815^{+0.000+0.000+0.000}_{-0.709-1.391-0.881}$
                 & $2.598^{+0.000+0.105+0.000}_{-0.054-0.291-0.153}$
                 & $0.999^{+0.093+0.267+0.370}_{-0.000-0.220-0.265}$                      \\
$g_0^T(q^2_{\text{max}}) $        & $1.181^{+0.035+0.078+0.082}_{-0.002-0.062-0.060}$
                 & $1.232^{+0.118+0.008+0.000}_{-0.073-0.190-0.058}$
                 & $0.105^{+0.047+0.079+0.094}_{-0.000-0.074-0.054}$                     \\
$f_\perp^T(q^2_{\text{max}}) $   & $2.899^{+0.000+0.000+0.000}_{-0.439-0.832-0.584}$
                 & $2.128^{+0.032+0.113+0.000}_{-0.000-0.139-0.063}$
                 & $0.722^{+0.079+0.180+0.302}_{-0.000-0.160-0.213}$                     \\
$g_\perp^T(q^2_{\text{max}}) $   & $1.181^{+0.035+0.078+0.082}_{-0.002-0.062-0.060}$
                 & $1.232^{+0.118+0.008+0.000}_{-0.073-0.190-0.058}$
                 &  $0.105^{+0.047+0.079+0.094}_{-0.000-0.074-0.054}$                    \\
		\hline\hline
	\end{tabular}
\footnotetext[1]{The $\Xi_b^0\to \Sigma^0$ transition form factors receive a factor $1/\sqrt{2}$ relative to the $\Xi_b^-\to \Sigma^-$ ones due to isospin. }
\end{table}

In Fig.~\ref{fig:f1}, we illustrate the various sources of uncertainty in the form factors, taking $f_1$  as an example, by varying all input parameters within the ranges listed in Table~\ref{tab:para1}. The solid red curves correspond to the central values. The dotted green curves show the impact of varying the Gegenbauer model parameter $A=0.5\pm0.2$ for the $\Xi_b$ baryon. Variations in the daughter baryon LCDA parameters $f_{\mathcal{B}_f}$, $\lambda_1$, $\lambda_2$, and $\lambda_3$ are displayed by the dashed blue, cyan, yellow, and purple curves, respectively. The dot-dashed violet curves reflect the effect of varying the hard scale $t$ between$0.8t$ and $1.2t$. It is evident that the dominant source of uncertainty arises from the scale $t$, followed by the parameters of the daughter baryon LCDAs. The form factor $f_1(q^2)$ is largely insensitive to variations of the $\Xi_b$ baryon Gegenbauer model parameter for all three modes, consistent with our earlier observations.
The observed uncertainty pattern in Fig.~\ref{fig:f1} for $f_1$ holds for all other form factors.
These uncertainties show a clear $q^2$ dependence, steadily increasing with $q^2$, which reflects the fact that PQCD calculations are most reliable at low $q^2$. In the high $q^2$ region, results are obtained via extrapolation, which tends to overestimate parametric uncertainties. Future improvements can be achieved by combining lattice QCD results for high $q^2$ to reduce the form factor uncertainties. The parametrization also may be improved by  imposing additional unitarity constraints on the $z$-series parameters~\cite{Blake:2022vfl}, which is worth exploring when the extrapolation from large to low recoil is available in the future.
The extracted values of the form factors at zero and maximum recoil are presented in Table~\ref{tab:form1}, where the three uncertainties correspond to the LCDA parameters of the initial and final baryons and the hard scale $t$, respectively. Our predictions satisfy the end point relations among the form factors given in Eqs.~(\ref{eq:FFC1}) and (\ref{eq:FFC2}). It is also instructive to consider the low- and large-recoil limits. In the low-recoil region, where heavy quark effective theory (HQET) is applicable~\cite{Mannel:1990vg}, spin symmetry reduces the number of independent form factors to two~\cite{Mannel:1997xy}. In this limit, the ten helicity form factors are related to the two leading Isgur-Wise functions, $\xi_1$ and $\xi_2$, as follows~\cite{Feldmann:2011xf,Descotes-Genon:2019dbw}:
\begin{eqnarray}
 f_0 &\approx& f_\perp\approx g_t\approx f_0^T\approx f_\perp^T=\xi_1-\xi_2, \nonumber\\
 g_0 &\approx& g_\perp\approx f_t\approx g_0^T\approx g_\perp^T=\xi_1+\xi_2.
\end{eqnarray}
The right panel of Fig.~\ref{fig:FFs} shows these form factors clustering into two distinct groups at high $q^2$, consistent with HQET expectations, which is further supported by the numbers at zero recoil in Table~\ref{tab:form1}. In the large-recoil region, where soft-collinear effective theory (SCET) applies,  $\xi_1$ becomes the leading universal soft form factor, while $\xi_2$ is subleading and can be neglected~\cite{Charles:1998dr,Beneke:2000wa,Bauer:2000yr,Beneke:2002ph}. Consequently, all helicity form factors are approximately equal at leading order:
\begin{eqnarray}\label{eq:app}
 f_0(0)\approx    f_\perp(0)\approx g_0(0) \approx f_t(0)\approx g_t(0) \approx g_\perp(0) \approx f_0^T(0) \approx g_0^T(0)\approx f_\perp^T(0)\approx g_\perp^T(0).
\end{eqnarray}
Our large-recoil predictions in Fig.~\ref{fig:FFs} follow this approximation closely. The small differences between, for example, $f_\perp(0)$ and $g_\perp(0)$ in Table~\ref{tab:form1} may indicate the presence of sizable subleading corrections, such as nonfactorizable contributions~\cite{Boer:2014kda}.

It is of considerable interesting to compare our form factors with the corresponding results of QCD sum rules presented in~\cite{Azizi:2011mw}. Note that the authors in~\cite{Azizi:2011mw} adopt a parametrization for the form factors that differs from ours. After converting their results into our conventions according to the definitions in Eqs.~(\ref{eq:Weinberg}) and~(\ref{eq:helicity0}), we obtain the following central values:
\begin{eqnarray}\label{eq:form11}
f_1(0)&=&0.142, \quad f_2(0)=0.116, \quad f_3(0)=-0.058, \nonumber\\
g_1(0)&=&0.160, \quad g_2(0)=0.052, \quad g_3(0)=-0.017, \nonumber\\
f_0^T(0)&=&0.229, \quad f_\perp^T(0)=0.157, \quad g_0^T(0)=0.110, \quad g_\perp^T(0)=0.155,
\end{eqnarray}
for the $\Xi_b\to \Xi$ transition, and
\begin{eqnarray}\label{eq:form22}
f_1(0)&=&0.046, \quad f_2(0)=0.191, \quad f_3(0)=-0.139, \nonumber\\
g_1(0)&=&0.067, \quad g_2(0)=0.139, \quad g_3(0)=-0.157, \nonumber\\
f_0^T(0)&=&0.084, \quad f_\perp^T(0)=0.049, \quad g_0^T(0)=-0.002, \quad g_\perp^T(0)=0.013,
\end{eqnarray}
for the $\Xi_b\to \Sigma$ transition.  In general, our form factors have the same signs as those in Ref.~\cite{Azizi:2011mw}, except for $g_0^T(0)$ in the $\Xi_b \to \Sigma$ transition. For the $\Xi_b \to \Xi$ channel, the magnitudes of their form factors are approximately a factor of two smaller than our predictions listed in Table~\ref{tab:form1}. Moreover, the discrepancies between our results for the $\Xi_b \to \Sigma$ transition and those given in Eq.~(\ref{eq:form22}) are particularly pronounced. As expected from the heavy-quark limit~\cite{Mannel:1990vg}, the form factors $f_{2,3}$ and $g_{2,3}$ should be suppressed relative to the dominant contributions $f_1$ and $g_1$. However, as evident from Eq.~(\ref{eq:form22}), the magnitudes of $f_{2,3}$ and $g_{2,3}$ exceed those of $f_1$ and $g_1$ by factors of about $2$--$3$. In addition, the four tensor form factors exhibit substantial variations; in particular, $g_0^T(0)$ is negative, which is inconsistent with the end point relations in Eqs.~(\ref{eq:FFC1}) and~(\ref{eq:app}). These unexpected features call for further investigation.
We also anticipate that additional theoretical studies of these form factors will help to improve our understanding of these weak baryonic decays. With the form factors at hand, we are now in a position to investigate the decay branching fractions and angular observables, which will be discussed in the following subsections.

\subsection{Decay branching fractions and LFU Ratios}\label{sec:ratio}

\begin{figure}[htbp]
\begin{center}
\setlength{\abovecaptionskip}{0pt}
\centerline{
\subfigure{\epsfxsize=6.5cm \epsffile{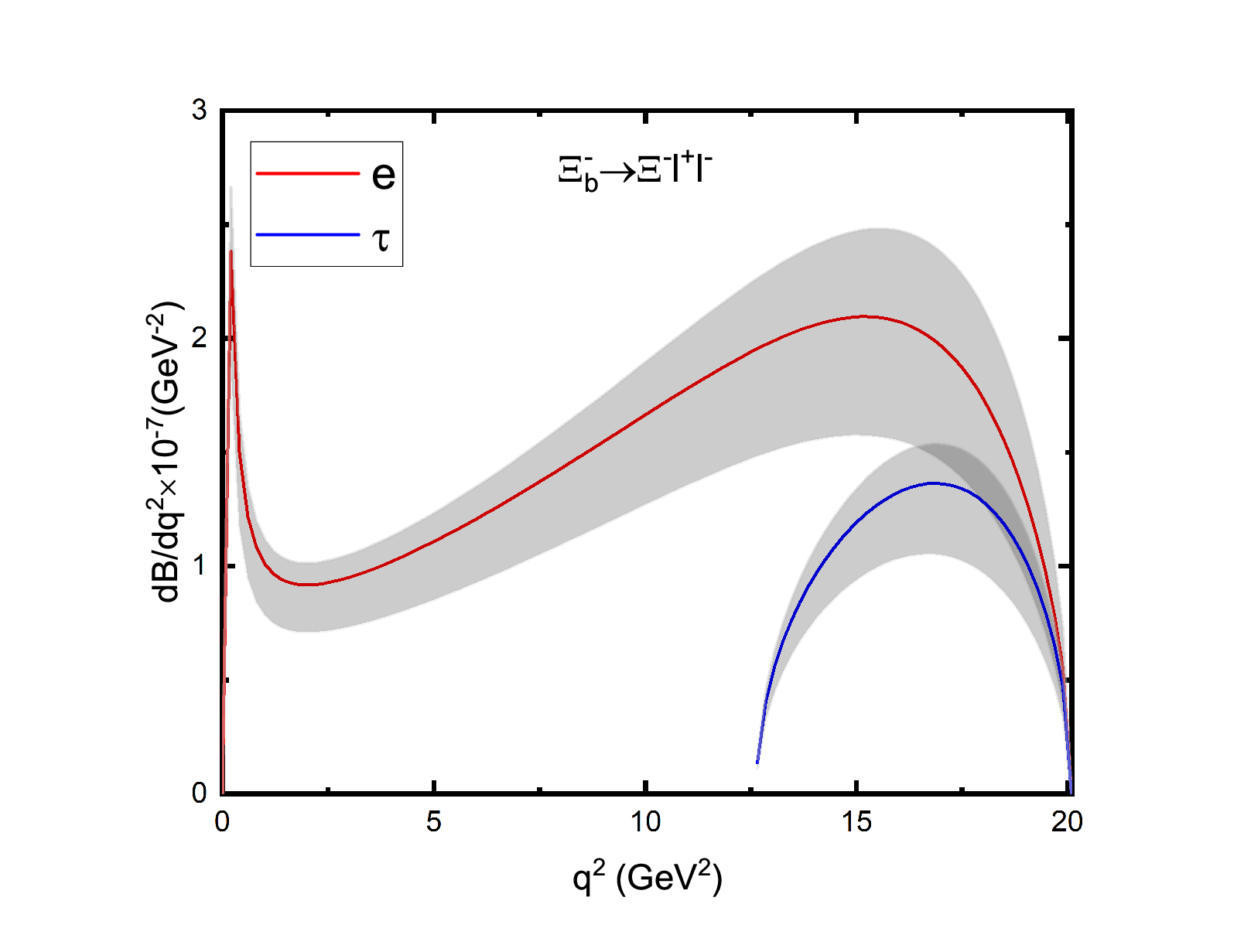} }
\hspace{-1.5cm}\subfigure{\epsfxsize=6.5cm \epsffile{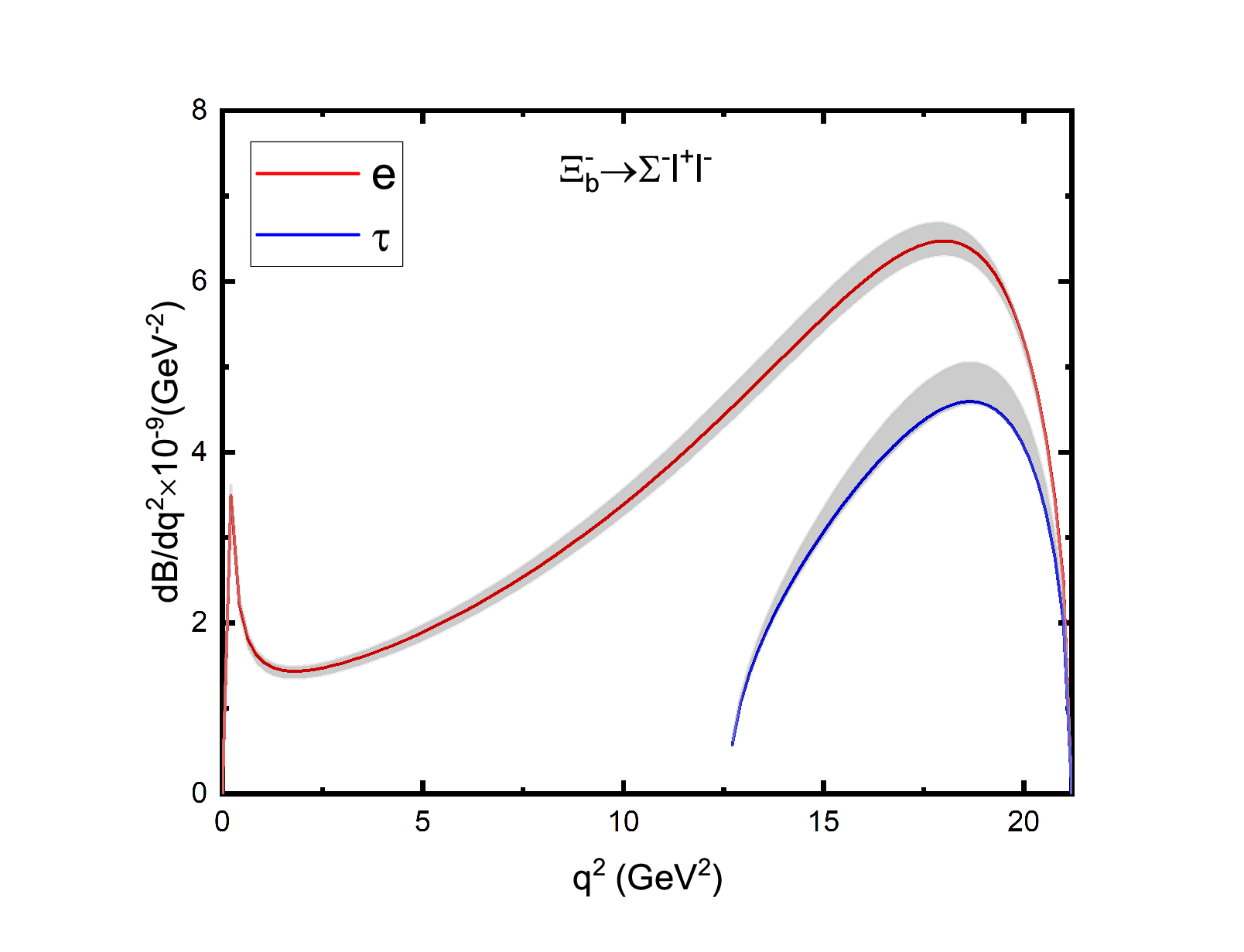} }
\hspace{-1.7cm}\subfigure{ \epsfxsize=6.5cm \epsffile{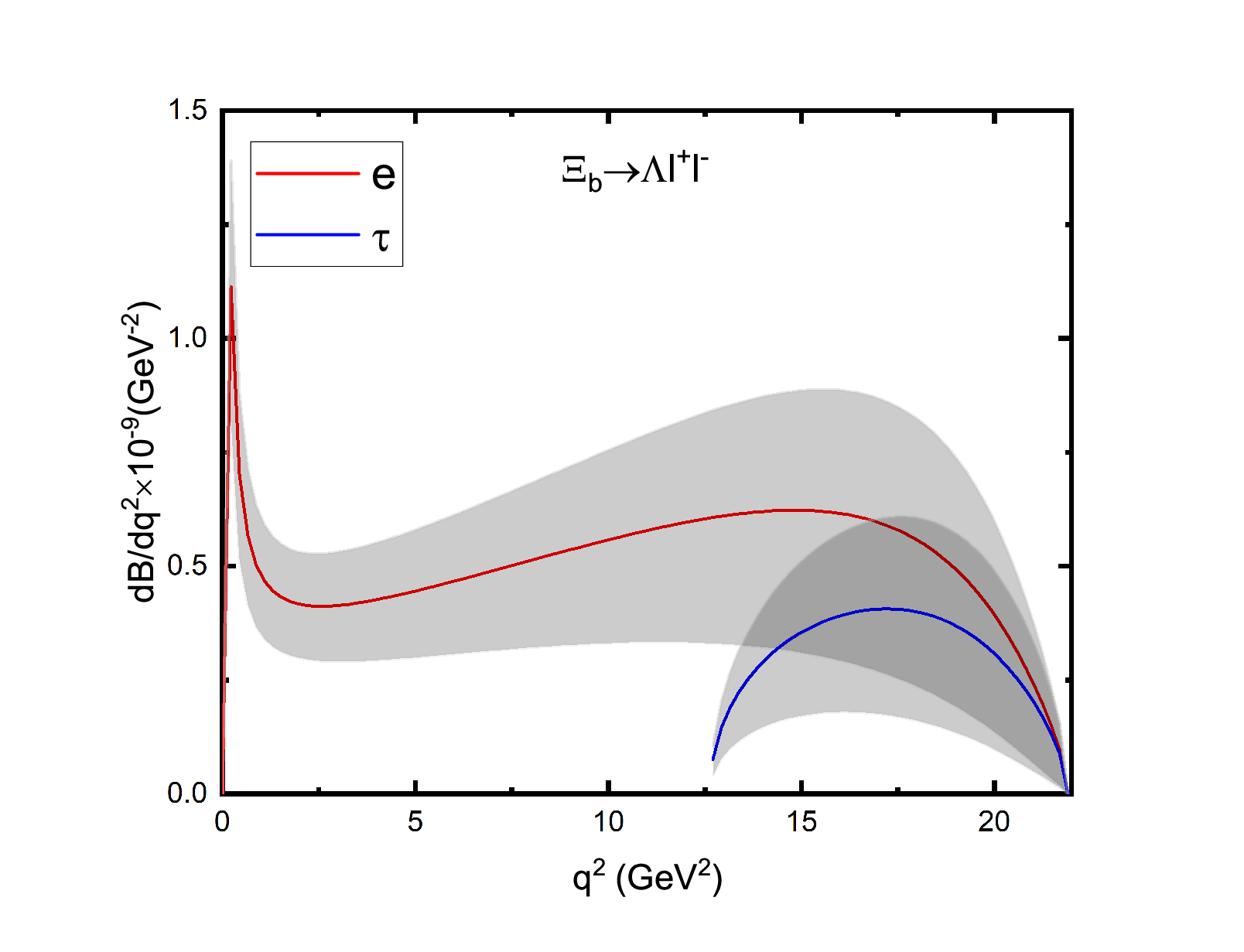}}}
\vspace{0.5cm}
\caption{Differential branching ratios of the semileptonic $\Xi^-_b\to\Xi^- \ell^+\ell^-$ (left), $\Xi^-_b\to\Sigma^- \ell^+\ell^-$ (middle), and $\Xi^0_b\to\Lambda \ell^+\ell^-$ (right) decays in the full $q^2$ kinematic region. The gray bands account for uncertainties induced by the scale $t$. The solid lines in the center of the band refer to the PQCD results with default parameters.}
 \label{fig:dB}
\end{center}
\end{figure}

Figure.~\ref{fig:dB} displays the differential branching fractions $d\mathcal{B}/dq^2$ for the three decays under consideration, where the final states with $e$ and $\tau$ leptons are denoted by red and blue curves, respectively.\footnote{Since the electron and muon masses are negligible compared with the $\Xi_b$ baryon mass, the available phase space for these two modes is essentially identical to excellent accuracy, rendering their $q^2$ dependences indistinguishable within the SM. For this reason, we do not present them separately in the following discussion.} The shaded bands surrounding the curves represent the uncertainty associated with the choice of the hard scale $t$. We observe that, for the electronic modes, each curve exhibits a steplike behavior at the threshold $q^2=4m_e^2$. For both the $e$ and $\tau$ channels, the differential distributions are dominated by the large-$q^2$ region. At the zero-recoil point, however, all distributions approach zero due to the vanishing available phase space.

\begin{figure}[htbp]
\begin{center}
\setlength{\abovecaptionskip}{0pt}
\centerline{
 \subfigure{\epsfxsize=6.5cm \epsffile{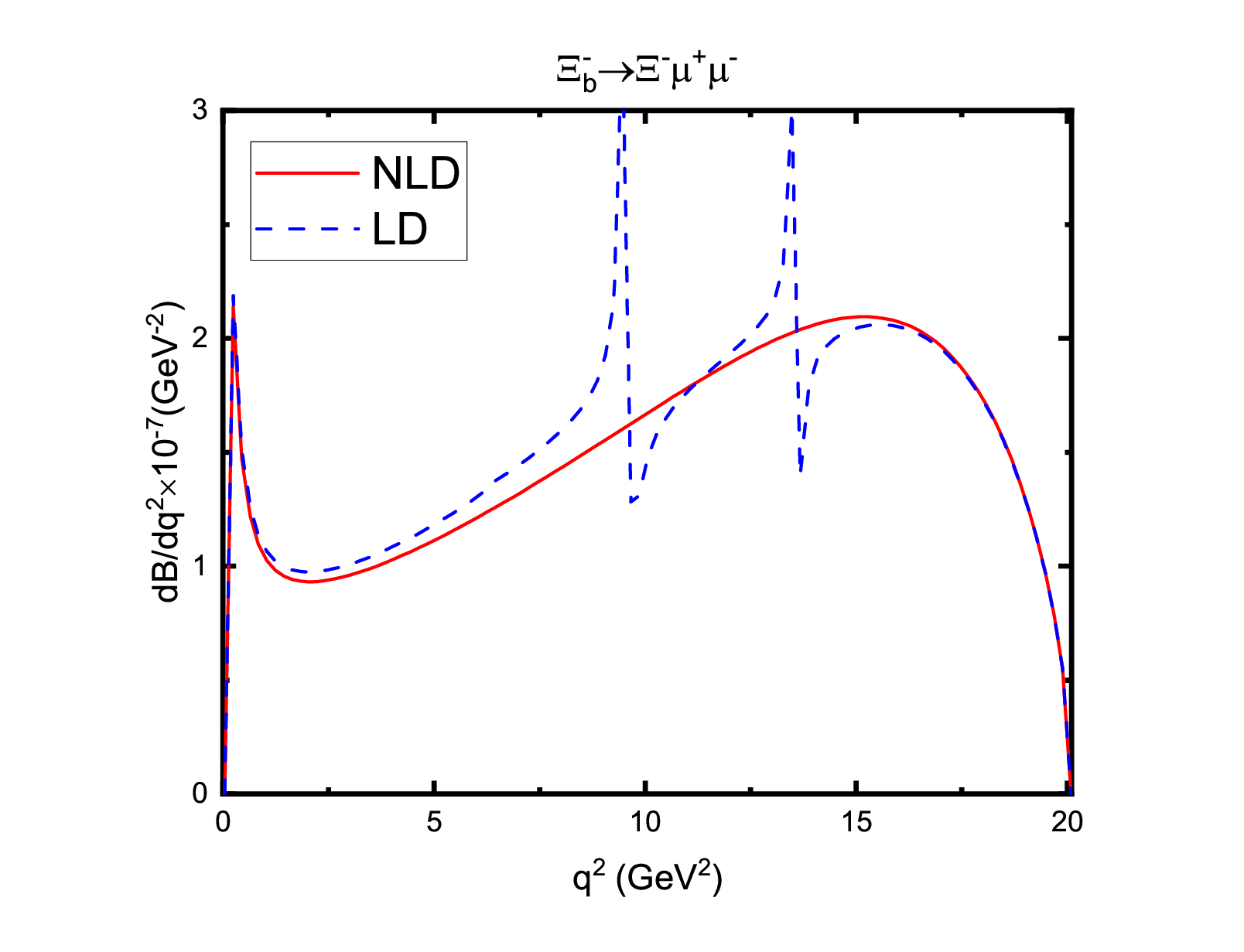} }
\hspace{-1.5cm}\subfigure{\epsfxsize=6.5cm \epsffile{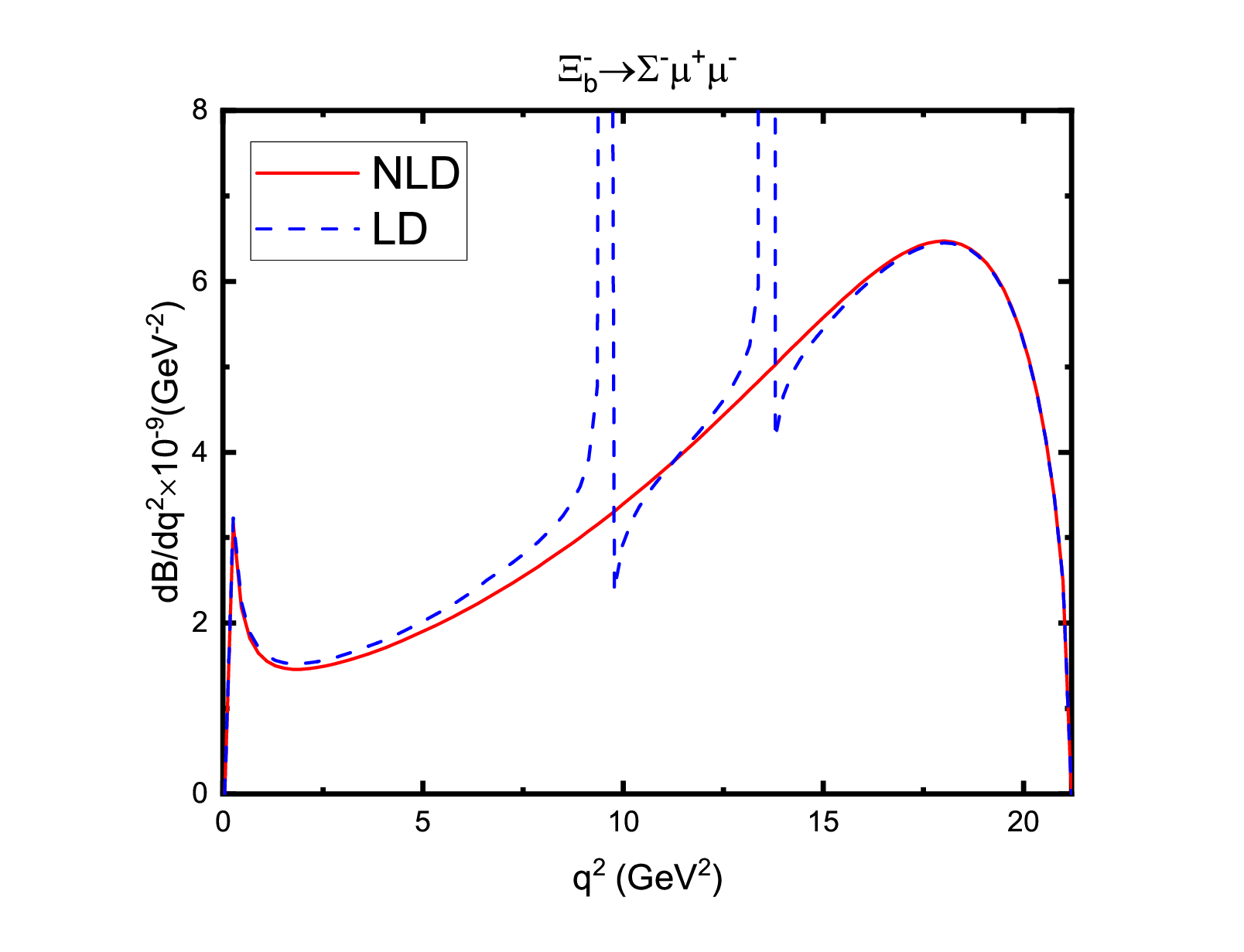} }
\hspace{-1.7cm}\subfigure{ \epsfxsize=6.5cm \epsffile{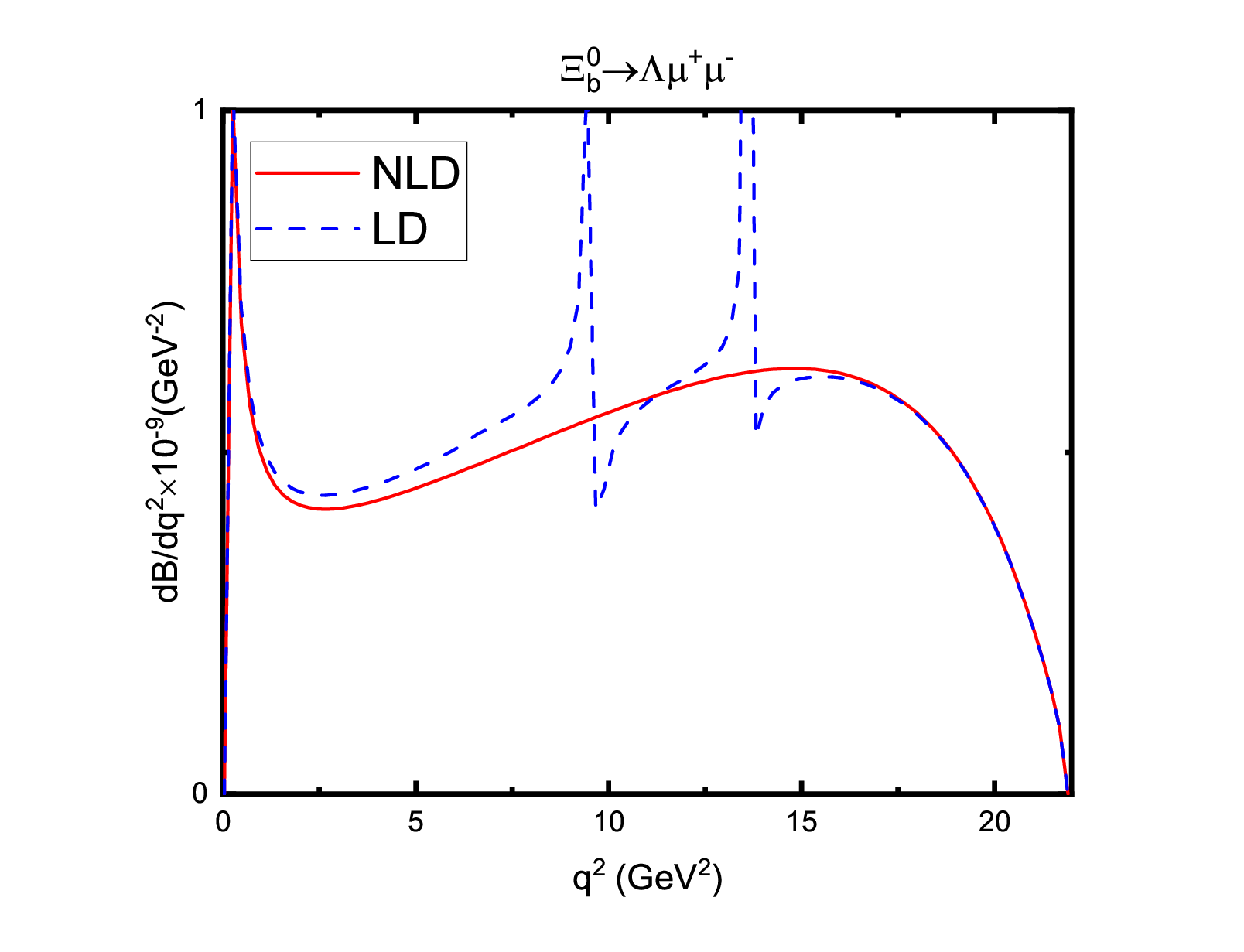}}}
\vspace{0.5cm}
\caption{ $q^2$ dependence of the differential branching ratios with and without including the long-distance contributions, labeled by ``LD" and ``NLD", respectively.}
 \label{fig:dBLD}
\end{center}
\end{figure}

We now turn to the $\ell=\mu$ case, which is experimentally the most accessible channel at hadron colliders, and examine the impact of long-distance effects on the differential branching fraction. To this end, we replace the Wilson coefficient $C_9^{\rm eff}$ by $C_9^{\rm eff}+Y_{SD}+Y_{LD}$, where $Y_{SD}$ ($Y_{LD}$) captures the short-distance (long-distance) contributions from four-quark operators far from (near) the charmonium resonance regions. In the present analysis, we include the dominant charmonium states $J/\psi$ and $\psi(2S)$, while higher radial excitations are neglected due to the strong suppression of their dileptonic decay rates. The explicit expressions for $Y_{SD}$ and $Y_{LD}$ are rather lengthy and can be found in Refs.~\cite{Mott:2011cx,Nayek:2020hna}; we do not reproduce them here. In Fig.~\ref{fig:dBLD}, the PQCD predictions obtained with (without) long-distance contributions are shown by solid red (dashed blue) curves. It is evident that the differential distributions are dominated by long-distance effects in the vicinity of the charmonium resonances, whereas away from these regions the impact of such effects is negligible.

\begin{figure}[htbp]
\begin{center}
\setlength{\abovecaptionskip}{0pt}
\centerline{
\subfigure{\epsfxsize=6.5cm \epsffile{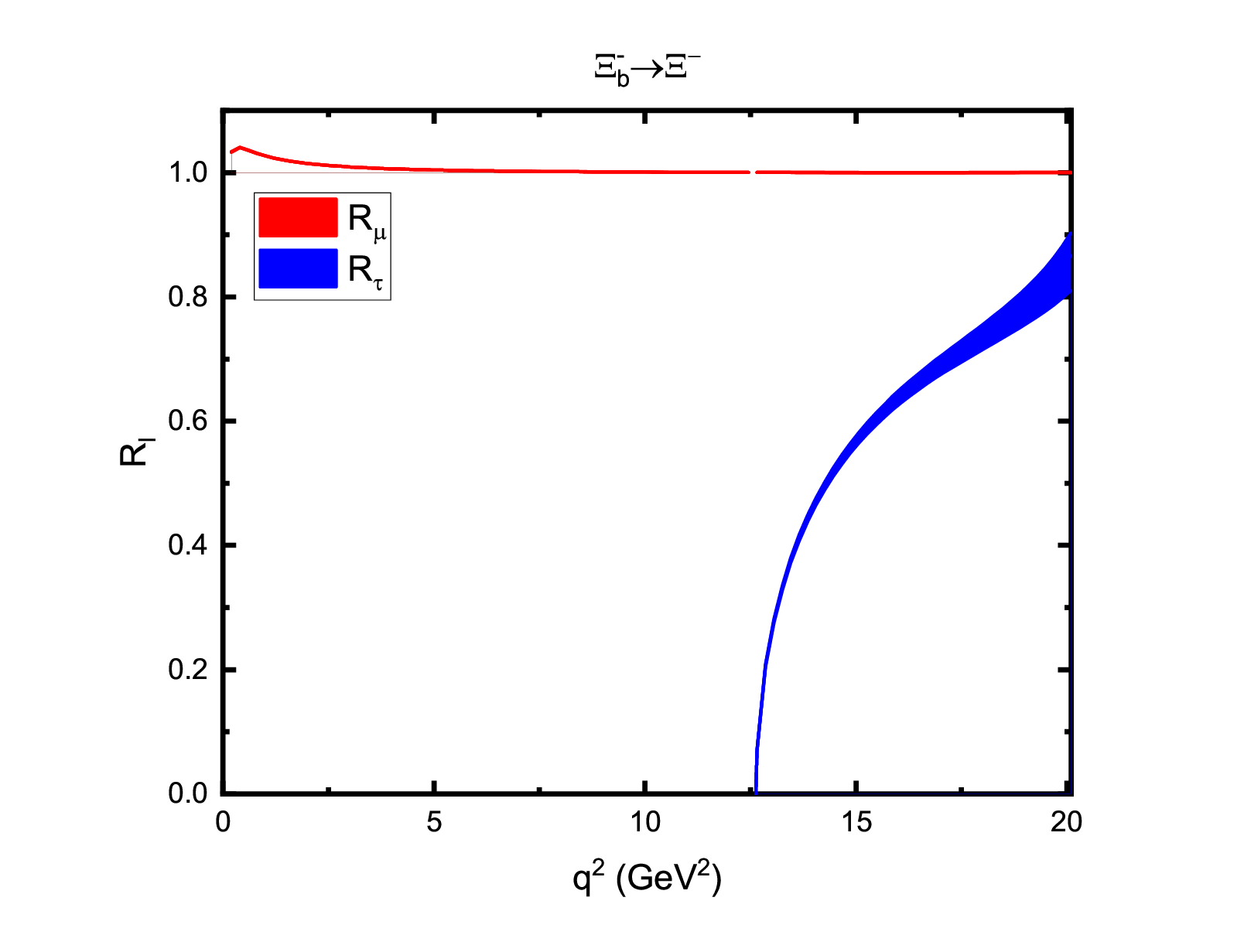} }
 \hspace{-1.5cm}\subfigure{\epsfxsize=6.5cm \epsffile{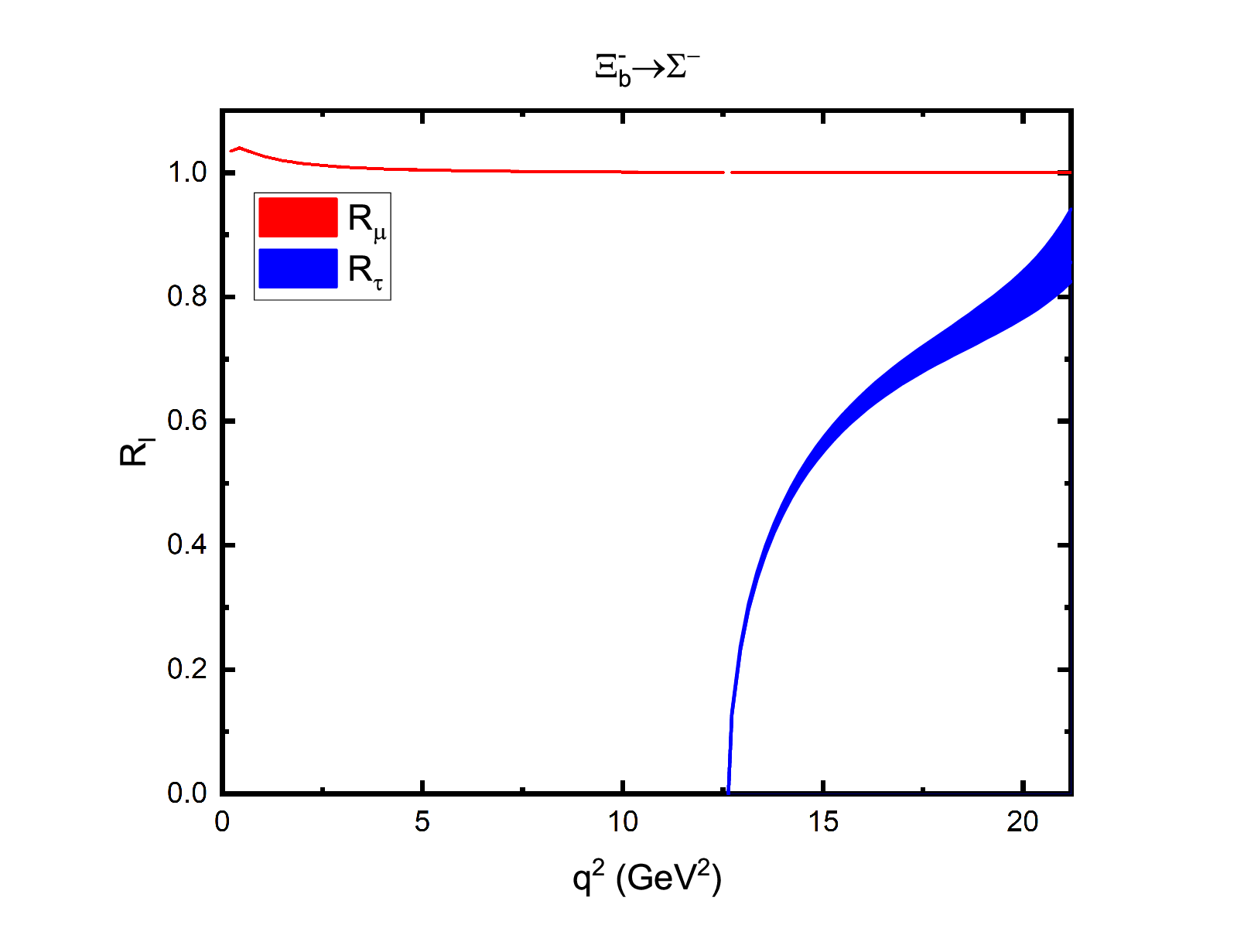} }
 \hspace{-1.7cm}\subfigure{ \epsfxsize=6.5cm \epsffile{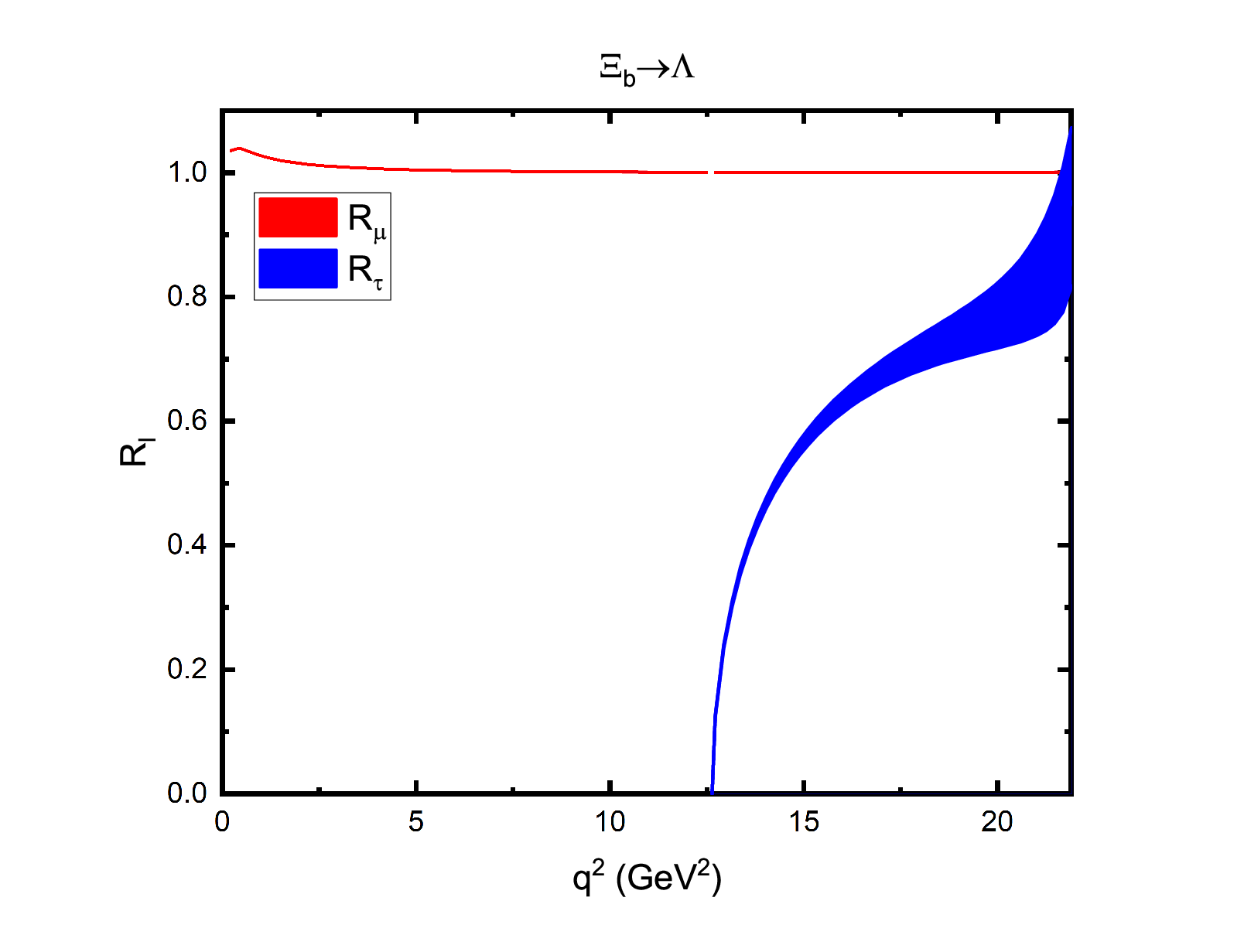}}}
\vspace{0.5cm}
\caption{ LFU ratios as a function of $q^2$, where the red and blue bands denote $\mathcal{R}_\mu$ and $\mathcal{R}_\tau$,
respectively.} 
 \label{fig:RR}
\end{center}
\end{figure}

Since most theoretical uncertainties largely cancel in the ratio defined in Eq.~(\ref{eq:RR}), one obtains a precise SM prediction for the LFU ratios, as illustrated in Fig.~\ref{fig:RR}. As expected, $\mathcal{R}_\mu$ remains very close to unity over the entire kinematic range, reflecting the near electron-muon flavor universality. This also indicates that nonperturbative form-factor effects cancel efficiently between the numerator and denominator in $\mathcal{R}_\mu$. In contrast, $\mathcal{R}_\tau$ exhibits a sizable deviation from unity throughout the allowed $q^2$ region due to the presence of lepton-mass terms in the decay amplitude, with the deviation becoming more pronounced at lower $q^2$. Moreover, $\mathcal{R}_\tau$ shows an overall positive slope across the full kinematic range and vanishes at the threshold $q^2=4m_\tau^2$.

\begin{table}[htbh]
	\caption{The predicted branching fractions and integrated LFU ratios for the semileptonic $\mathcal{B}_i\to \mathcal{B}_f \ell^+ \ell^-$ decays with $\ell$ denoting a charged lepton. All the considered theoretical uncertainties  have been combined in quadrature. The first number is the value integrated over the full $q^2$ range, while the one which follows in square brackets is the result when the ranges $m_{J/\psi}\pm 0.001\text{GeV}$ and $m_{\psi(2S)}\pm 0.001\text{GeV}$ are excluded. The predictions from QCDSR~\cite{Azizi:2011mw,Sarac:2013rpa} are also listed for a comparison.}
	\label{tab:branching}
	\begin{tabular}[t]{lccc}
	\hline\hline
          &  $\Xi^-_b\to \Xi^- \ell^+ \ell^-$   & $\Xi^-_b\to \Sigma^-\ell^+ \ell^-$     &  $\Xi_b^0\to \Lambda\ell^+ \ell^-$     \\ \hline
 $\mathcal{B}_e$          &  $3.4\pm0.7[3.8]\times10^{-6}$      &  $8.5^{+1.1}_{-0.9} [9.3]\times10^{-8}$        &  $1.3^{+0.6}_{-0.7}[1.5]\times10^{-8}$   \\
 QCDSR~\cite{Azizi:2011mw,Sarac:2013rpa}     &  $(2.25\pm0.78)\times10^{-6}$               &  $(14.4\pm5.0)\times10^{-8}$             &$\cdots$\\
  $\mathcal{B}_\mu$       &  $3.1^{+0.6}_{-0.7}[3.5]\times10^{-6}$      &  $7.9^{+1.0}_{-0.8}[8.7]\times10^{-8}$        &  $1.1^{+0.6}_{-0.7}[1.3]\times10^{-8}$   \\
 QCDSR~\cite{Azizi:2011mw,Sarac:2013rpa}     &  $(2.23\pm0.67)\times10^{-6}$               &  $(14.0\pm4.9)\times10^{-8}$            &  $\cdots$\\
   $\mathcal{B}_\tau$     &  $7.9^{+1.3}_{-1.8}[8.0]\times10^{-7}$      &  $2.9^{+0.6}_{-0.2}[2.9]\times10^{-8}$        &  $2.9^{+1.7}_{-2.4}[2.9]\times10^{-9}$   \\
 QCDSR~\cite{Azizi:2011mw,Sarac:2013rpa}     &  $(3.6\pm1.1)\times10^{-7}$               &  $(9.8\pm2.9)\times10^{-8}$            &  $\cdots$\\
   $\langle \mathcal{R}_\mu\rangle$  &  $1.001\pm0.000[1.0]$      &  $1.000\pm0.000[1.0]$ &  $1.000\pm0.000[1.0]$    \\
   $\langle \mathcal{R}_\tau\rangle$  &  $0.592\pm0.053[0.58]$      &  $0.615^{+0.072}_{-0.001}[0.61]$ & $0.614^{+0.075}_{-0.132}[0.59]$   \\
  \hline\hline
	\end{tabular}
\end{table}

\enlargethispage{-\baselineskip}
Integrating over the full kinematically accessible $q^2$ range, we obtain the decay branching fractions $\mathcal{B}_\ell$ and the integrated LFU ratios $\langle \mathcal{R}_{\mu/\tau}\rangle$, which are summarized in Table~\ref{tab:branching}. All theoretical uncertainties considered in this work have been combined in quadrature. For comparison, results including long-distance effects are also quoted in square brackets. As emphasized in Ref.~\cite{Mott:2011cx}, the regions around intermediate charmonium resonances are experimentally vetoed, since in these regions the signals are overwhelmingly dominated by nonleptonic decays of the type $\mathcal{B}_i \to \mathcal{B}_f \psi(\to \ell^+\ell^-)$. Accordingly, when long-distance effects are included in our calculations, we exclude narrow windows around the $J/\psi$ and $\psi(2S)$ masses, namely $[m_{J/\psi}-0.001\text{GeV},m_{J/\psi}+0.001\text{GeV}]$ and $[m_{\psi(2S)}-0.001\text{GeV},m_{\psi(2S)}+0.001\text{GeV}]$\footnote{ The 0.001 GeV window corresponds to approximately 5 to 10 times the natural width of the concerned charmonium states. This choice is sufficiently wide to avoid contamination from resonant contributions while remaining numerically stable during integration.}. Without such vetoes, the branching fractions including long-distance contributions would be dramatically enhanced by the sharp resonance peaks. We find that, after excluding these regions, the branching fractions obtained with long-distance effects are only slightly larger than those computed without them. By contrast, the LFU ratios are essentially insensitive to long-distance contributions, since these effects largely cancel between the numerator and denominator.

From Table~\ref{tab:branching}, one observes that the branching fractions for the rare semileptonic decays $\Xi_b \to \Xi \ell^+\ell^-$ with $\ell=e,\mu$ are of order $10^{-6}$, while the corresponding semitauonic mode is suppressed to the level of $10^{-7}$. The other two rare decay modes, $\Xi_b \to (\Sigma,\Lambda)\ell^+\ell^-$, are further suppressed relative to $\Xi_b \to \Xi \ell^+\ell^-$ by the CKM factor $|V_{td}/V_{ts}|^2$, leading to branching fraction of order $10^{-8}$. The slightly smaller value of $\mathcal{B}_\mu$ compared to $\mathcal{B}_e$ is mainly attributable to lepton-mass-induced phase-space effects. Similarly, the obtained branching fraction for the $\tau$ channel is strongly phase-space suppressed; in addition, spin-flip contributions play a significant role in shaping $\mathcal{B}_\tau$. We also compare our predictions for the branching fractions with those obtained from QCD sum rules (QCDSR)~\cite{Azizi:2011mw,Sarac:2013rpa}, as listed in Table~\ref{tab:branching}. Our central values for the $\Xi$ ($\Sigma$) mode are slightly larger (smaller) than the corresponding QCDSR results. These differences can be traced primarily to the distinct form-factor inputs adopted in the two approaches, and may be further influenced by different choices of effective Wilson coefficients. In Ref.~\cite{Nayek:2020hna}, the authors investigated the decays $\Xi_b \to \Xi (\mu^+\mu^-,\tau^+\tau^-)$ within both the SM and a $Z'$ scenario, dividing the full kinematic region into several $q^2$ bins. A straightforward summation of their binned results yields the total branching fractions
\begin{eqnarray}\label{eq:xxx}
 \mathcal{B}_\mu=(2.33\pm0.53) \times10^{-6}, \quad \mathcal{B}_\tau=(3.71\pm0.33) \times10^{-5}.
\end{eqnarray}
The muonic result is consistent with our prediction within uncertainties, whereas the quoted semitauonic branching fraction is larger by almost an order of magnitude. Given that the kinematic range available for $\tau$ leptons is much narrower than that for light leptons, as illustrated in Fig.~\ref{fig:dB}, the $\tau$-lepton mass is expected to play a crucial role in suppressing the decay rate relative to the electron mode. The surprisingly large value of $\mathcal{B}_\tau$ in Eq.~(\ref{eq:xxx}) therefore warrants further scrutiny. To the best of our knowledge, the $\Lambda$ channel has received comparatively little attention in the literature, and none of the rare $\Xi_b$ decay modes considered here have yet been observed experimentally. Our predictions thus provide well-defined targets for future experimental investigations.

The integrated LFU ratios $\langle \mathcal{R}_\mu\rangle$ for all three decay modes deviate from unity only at the level of $10^{-3}$, in agreement with the behavior shown in Fig.~\ref{fig:RR}. Consequently, any departure of $\mathcal{R}_\mu$ from the SM expectation at the level of a few percent or larger would constitute a clear signal of NP. The corresponding $\mu/e$ ratios in $B\to K^{(*)}$ decays, defined as $\mathcal{R}(K^{(*)})=\mathcal{B}(B\to K^{(*)}\mu^+\mu^-)/\mathcal{B}(B\to K^{(*)}e^+e^-)$, have been measured in restricted nonresonant regions by several independent experiments~\cite{CMS:2024syx,LHCb:2019hip,BaBar:2012mrf,BELLE:2019xld}. More recently, LHCb measured $\mathcal{R}(K)$ in the high-$q^2$ region ($q^2>14.3~\text{GeV}^2$) and obtained $\mathcal{R}(K)=1.08^{+0.11}_{-0.09}(\text{stat})^{+0.04}_{-0.04}(\text{syst})$~\cite{LHCb:2025ilq}, which is consistent with the SM expectation. We anticipate that future measurements by Belle~II~\cite{deMarino:2023zsx} and the LHCb Run~3 program~\cite{LHCb-TALK-2023-276} will substantially improve the precision of these FCNC observables. From Table.~\ref{tab:branching}, the PQCD predictions for the integrated ratios $\langle \mathcal{R}_\tau\rangle$ are typically around 0.6, a pattern that is compatible with analogous ratios observed in charged-current semileptonic decays~\cite{Rui:2025bsu}. Since LFU tests based on the $\tau/e$ ratio in $b\to s$ FCNC transitions have not yet been performed, our results provide a complementary benchmark and may serve as useful guidance for future experimental studies of LFU in these rare processes.

\subsection{Differential and integrated angular observables}\label{sec:ang}

\begin{figure}[htbp]
\begin{center}
\setlength{\abovecaptionskip}{0pt}
\centerline{
\subfigure{\epsfxsize=6.5cm \epsffile{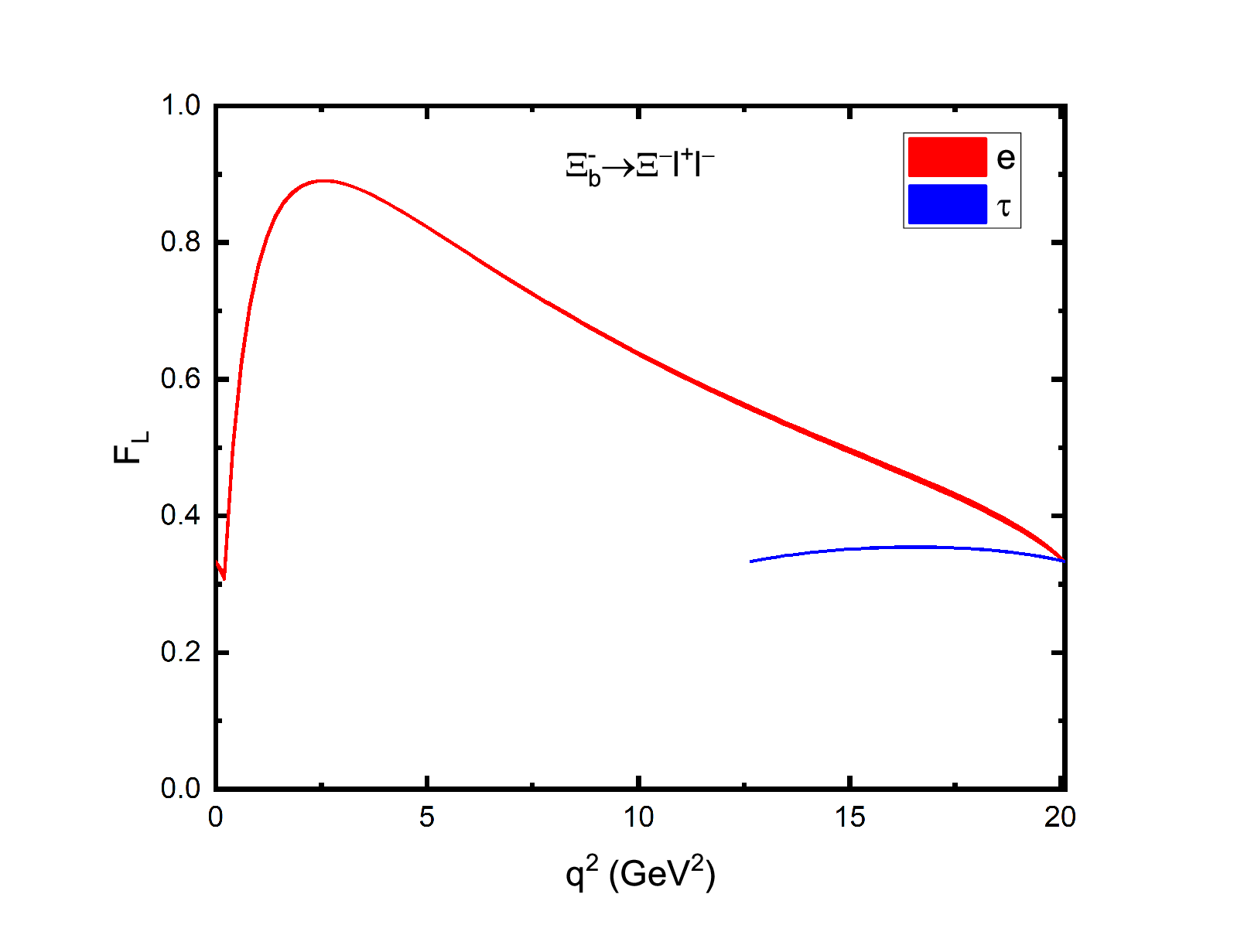} }
\hspace{-1.5cm}\subfigure{\epsfxsize=6.5cm \epsffile{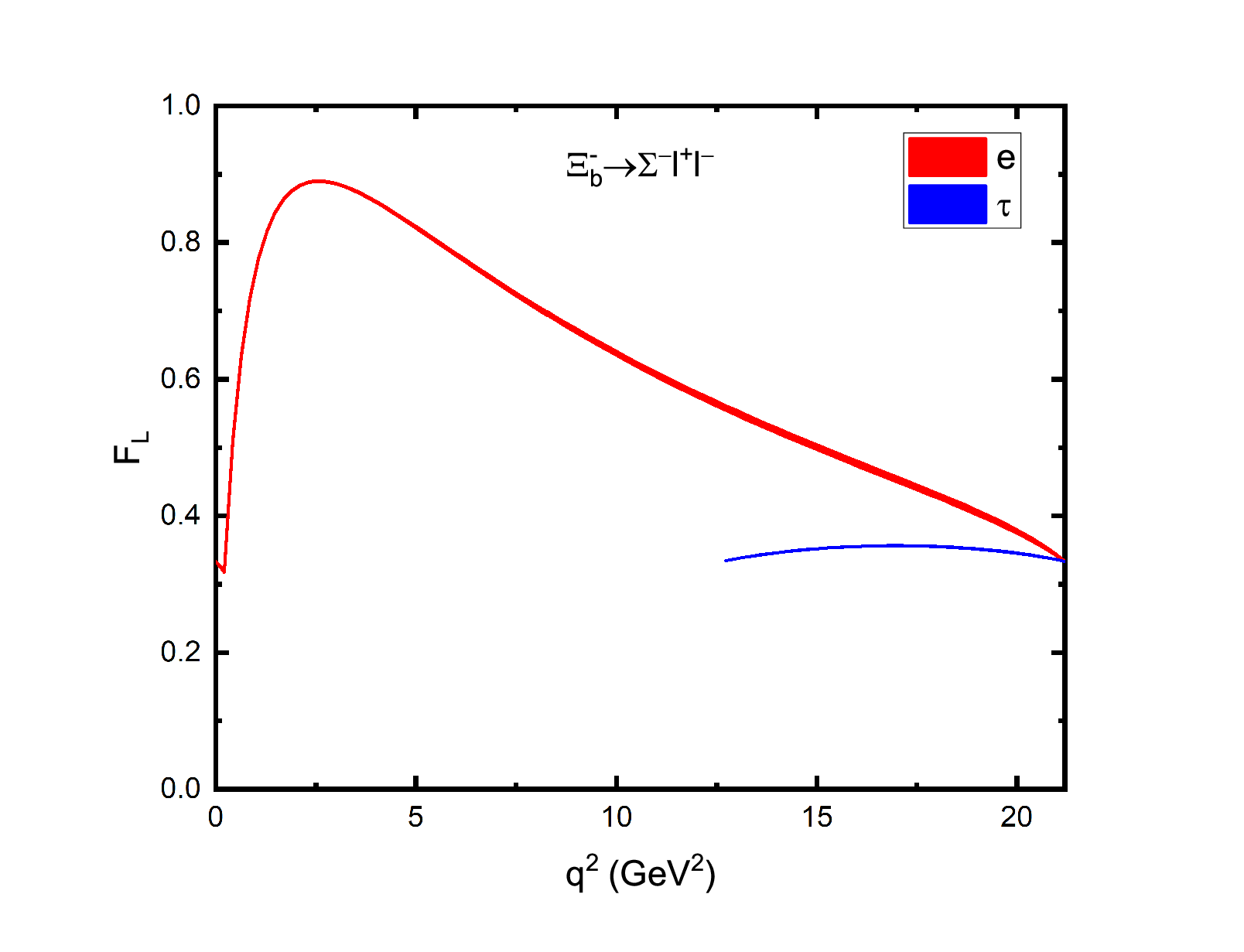} }
\hspace{-1.7cm}\subfigure{ \epsfxsize=6.5cm \epsffile{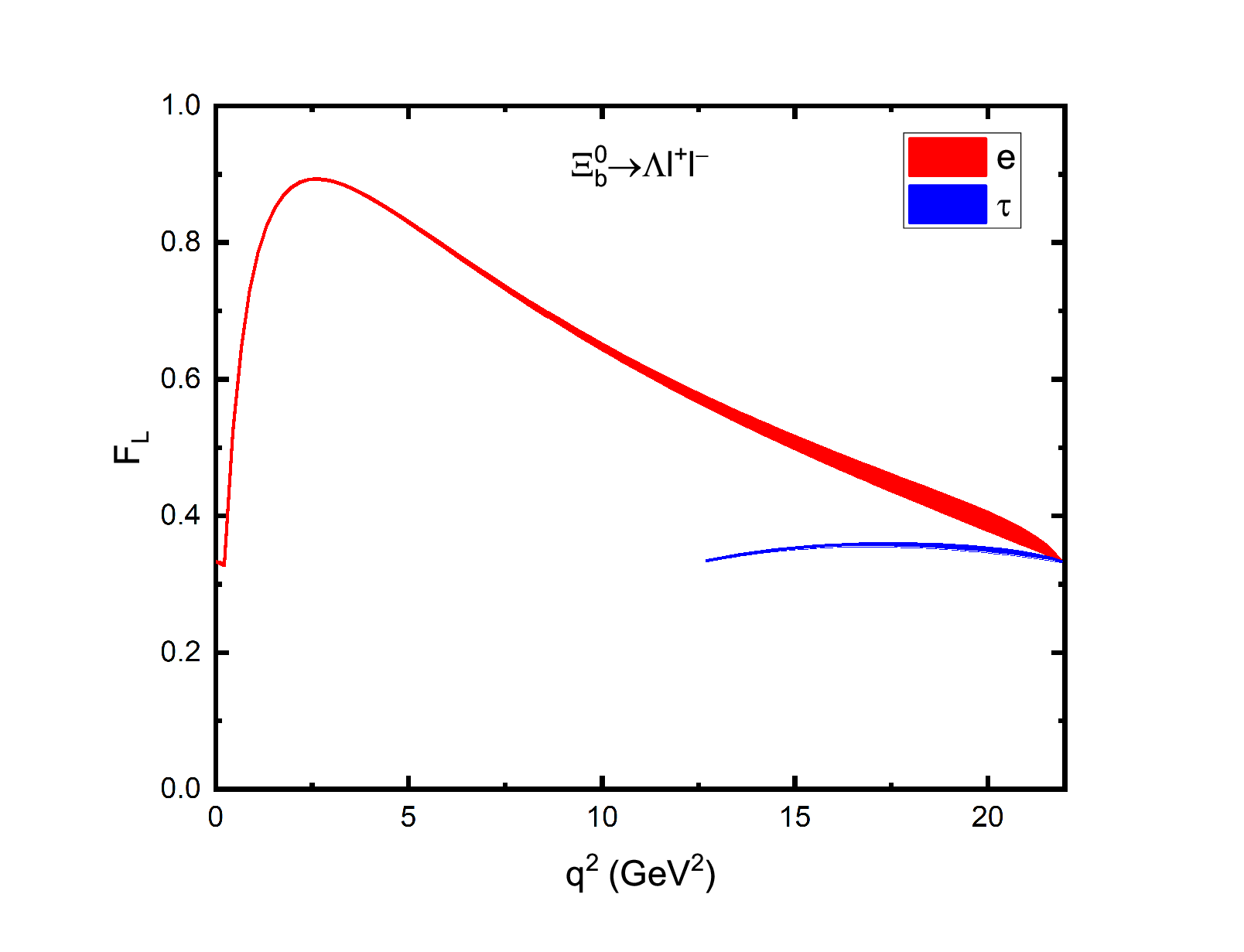}}}
\vspace{0.8cm}
\centerline{
 \subfigure{\epsfxsize=6.5cm \epsffile{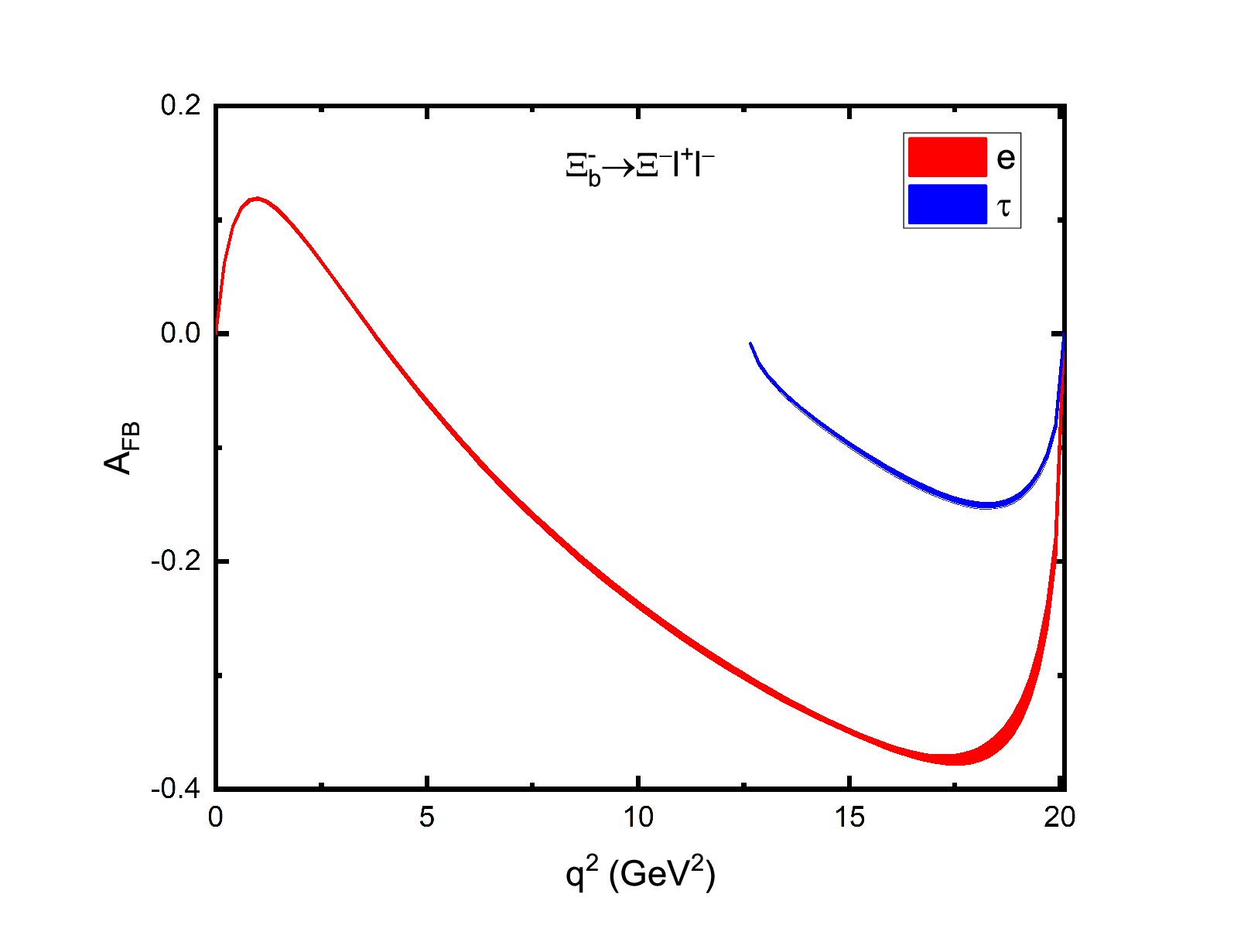} }
\hspace{-1.5cm}\subfigure{\epsfxsize=6.5cm \epsffile{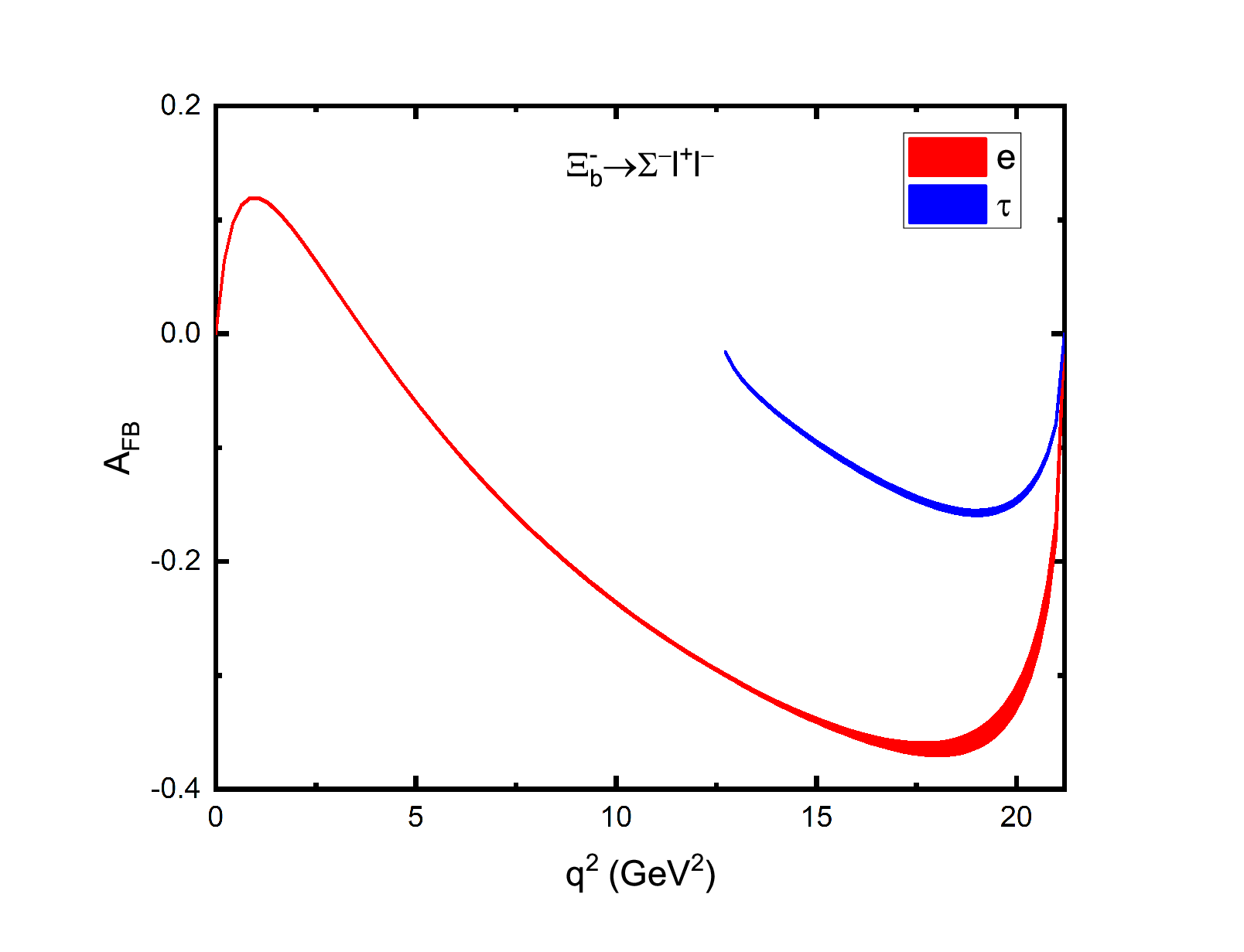} }
\hspace{-1.7cm}\subfigure{ \epsfxsize=6.5cm \epsffile{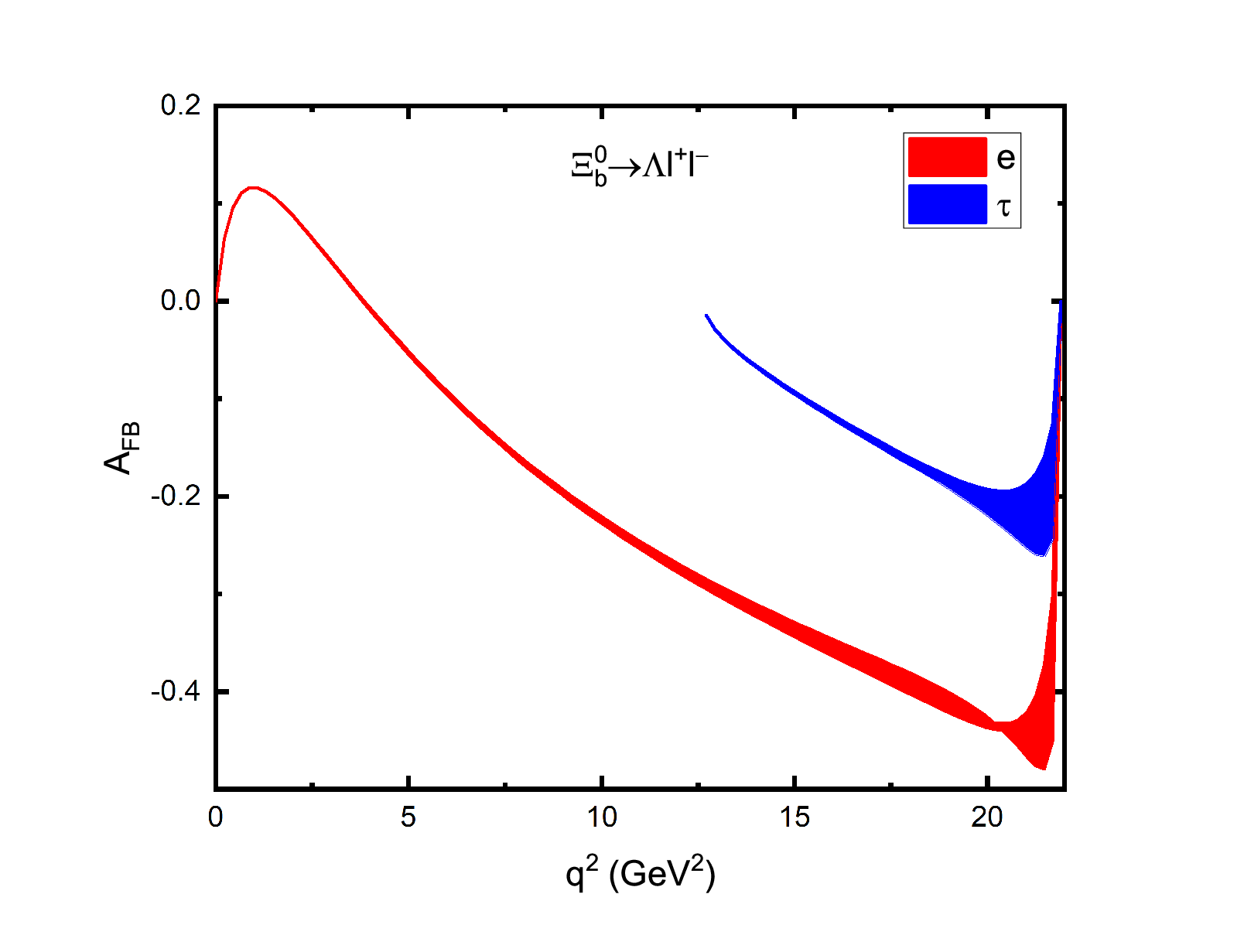}}}
\vspace{0.8cm}
\centerline{
 \subfigure{\epsfxsize=6.5cm \epsffile{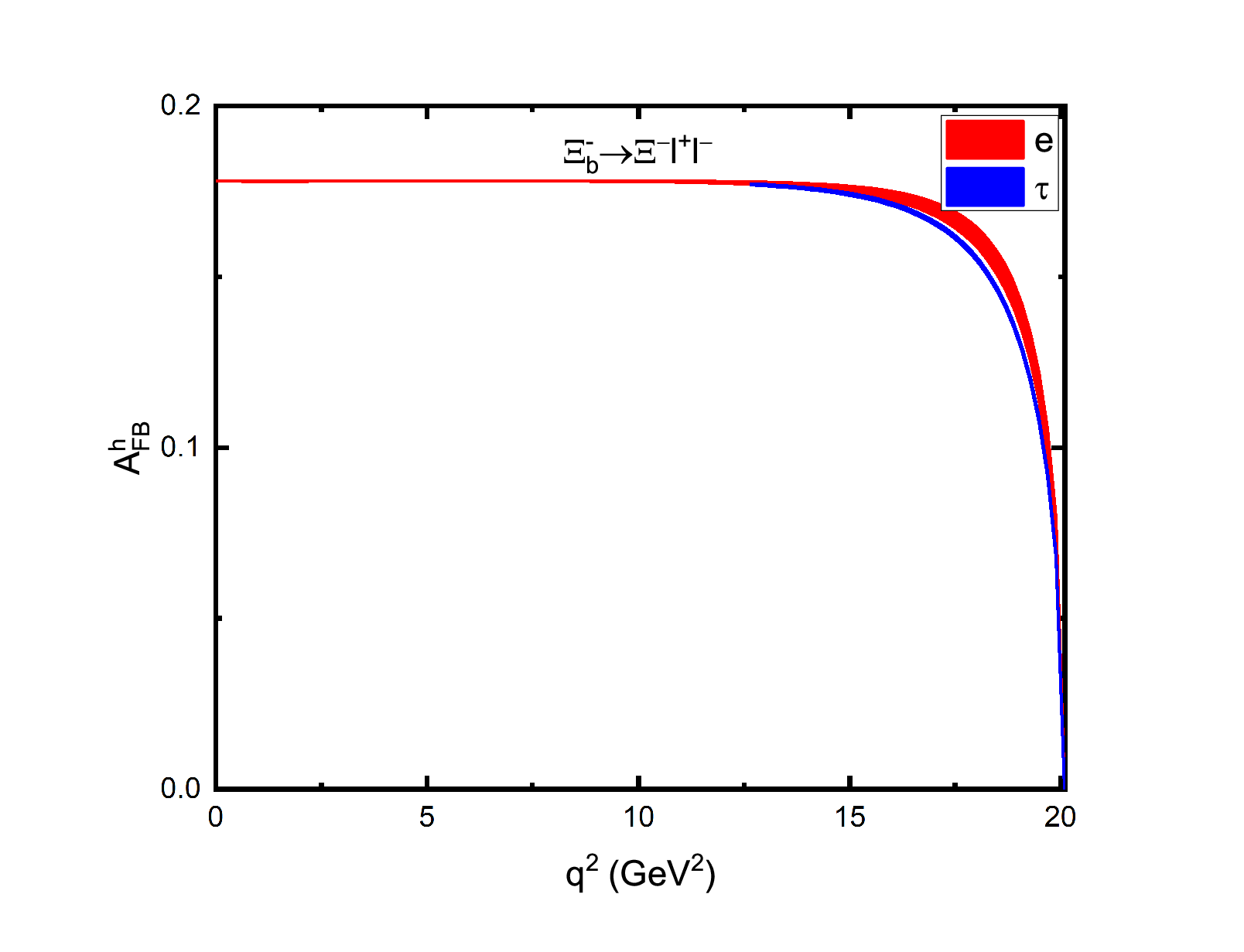} }
\hspace{-1.5cm}\subfigure{\epsfxsize=6.5cm \epsffile{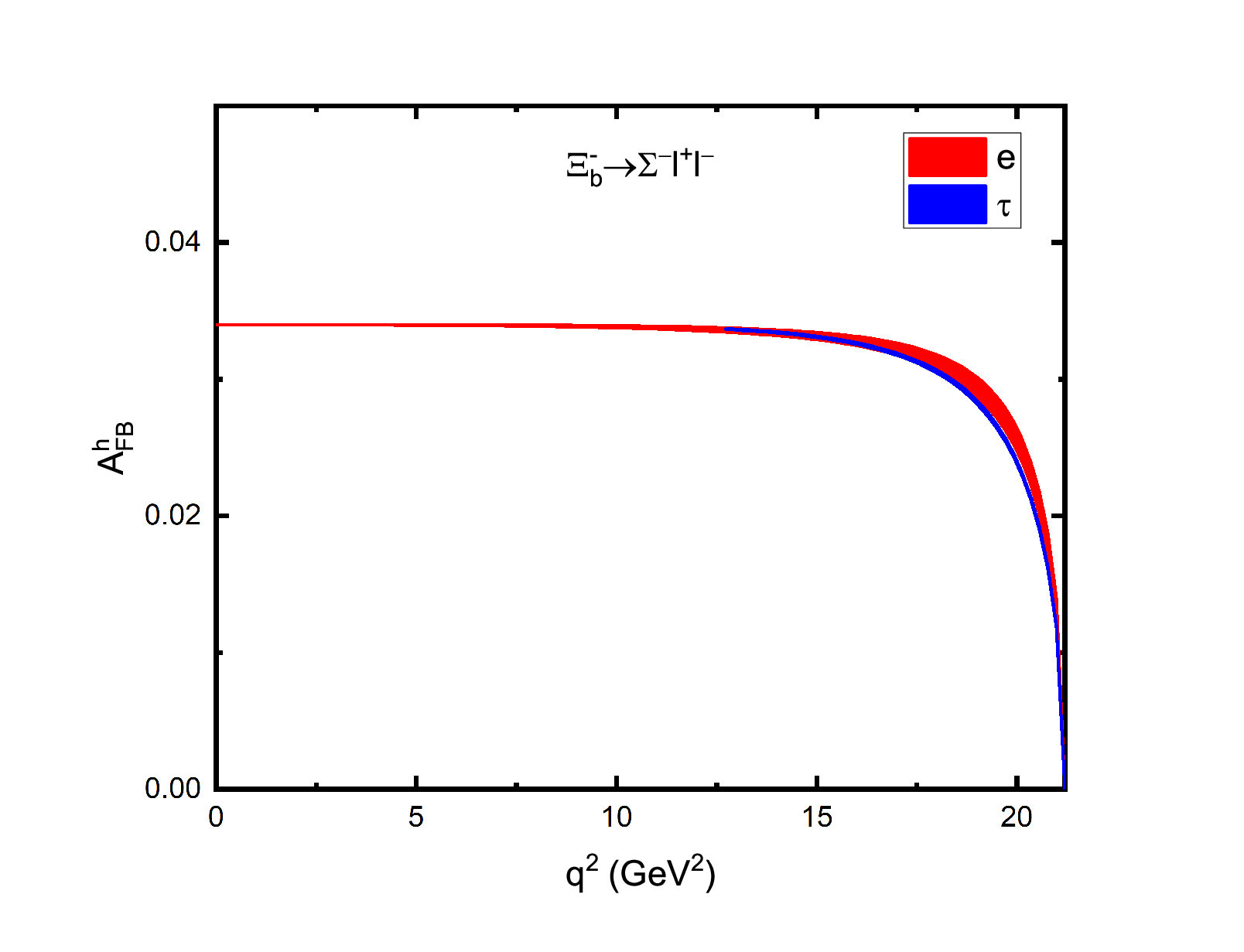} }
\hspace{-1.7cm}\subfigure{ \epsfxsize=6.5cm \epsffile{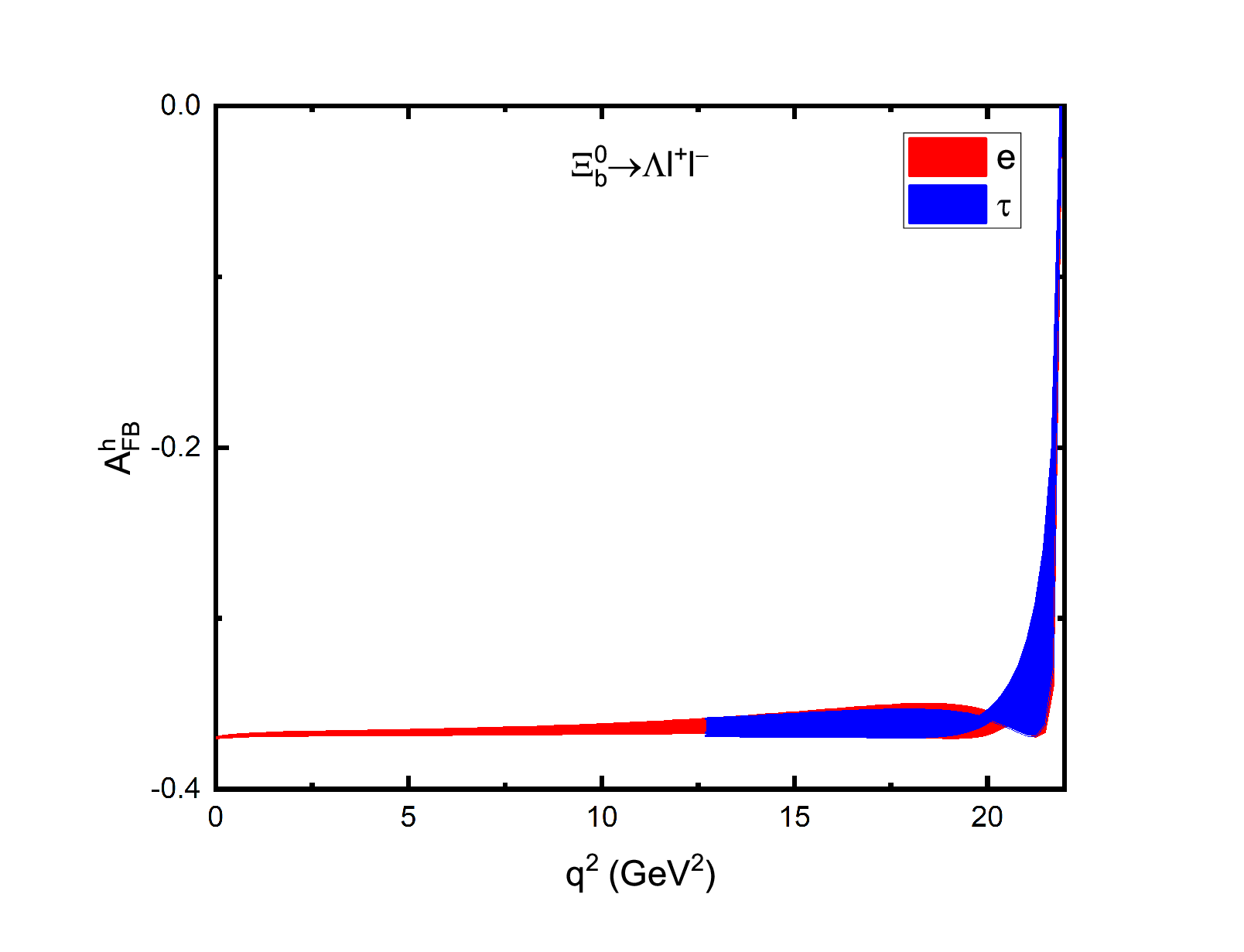}}}
\vspace{0.8cm}
\centerline{
 \subfigure{\epsfxsize=6.5cm \epsffile{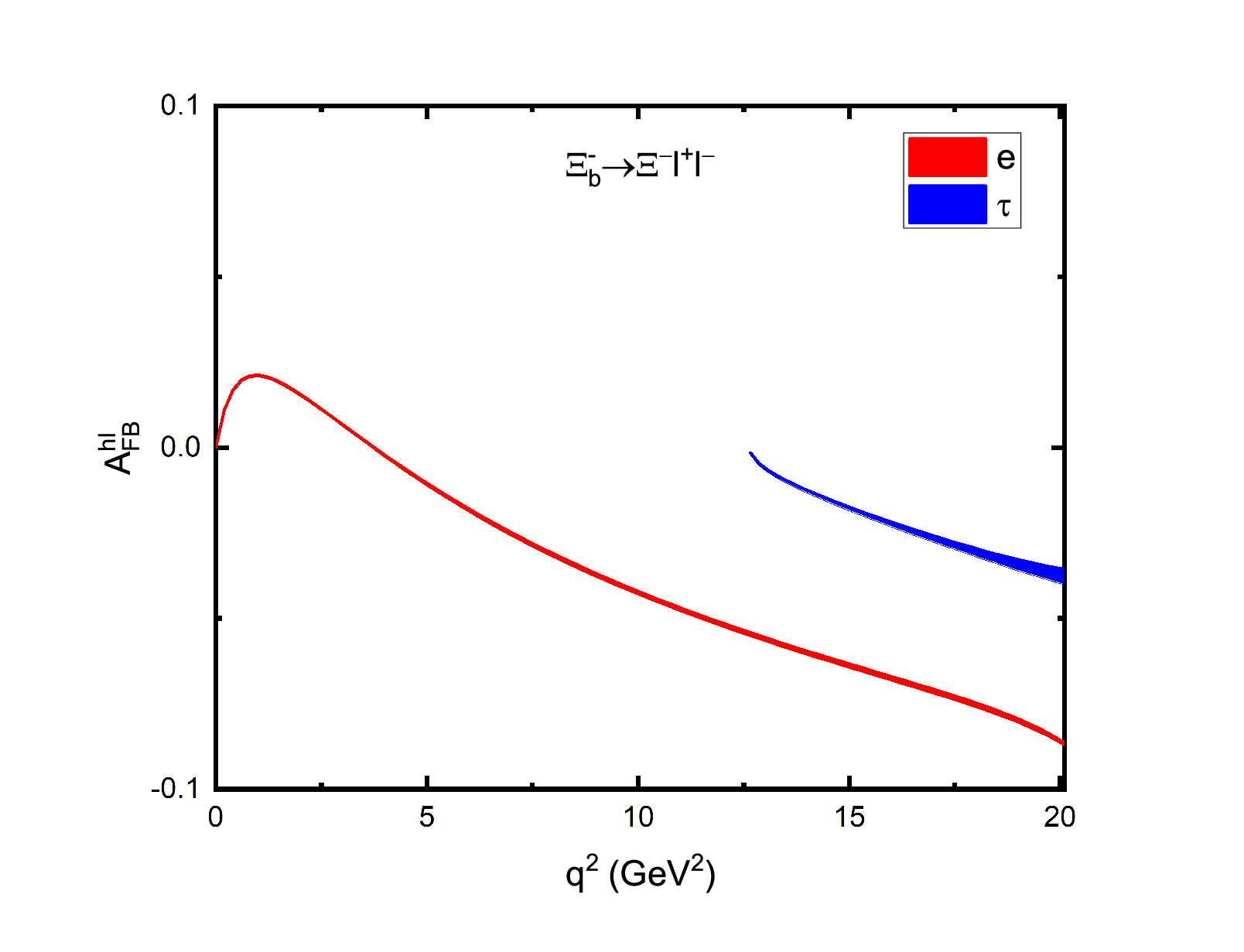} }
\hspace{-1.5cm}\subfigure{\epsfxsize=6.5cm \epsffile{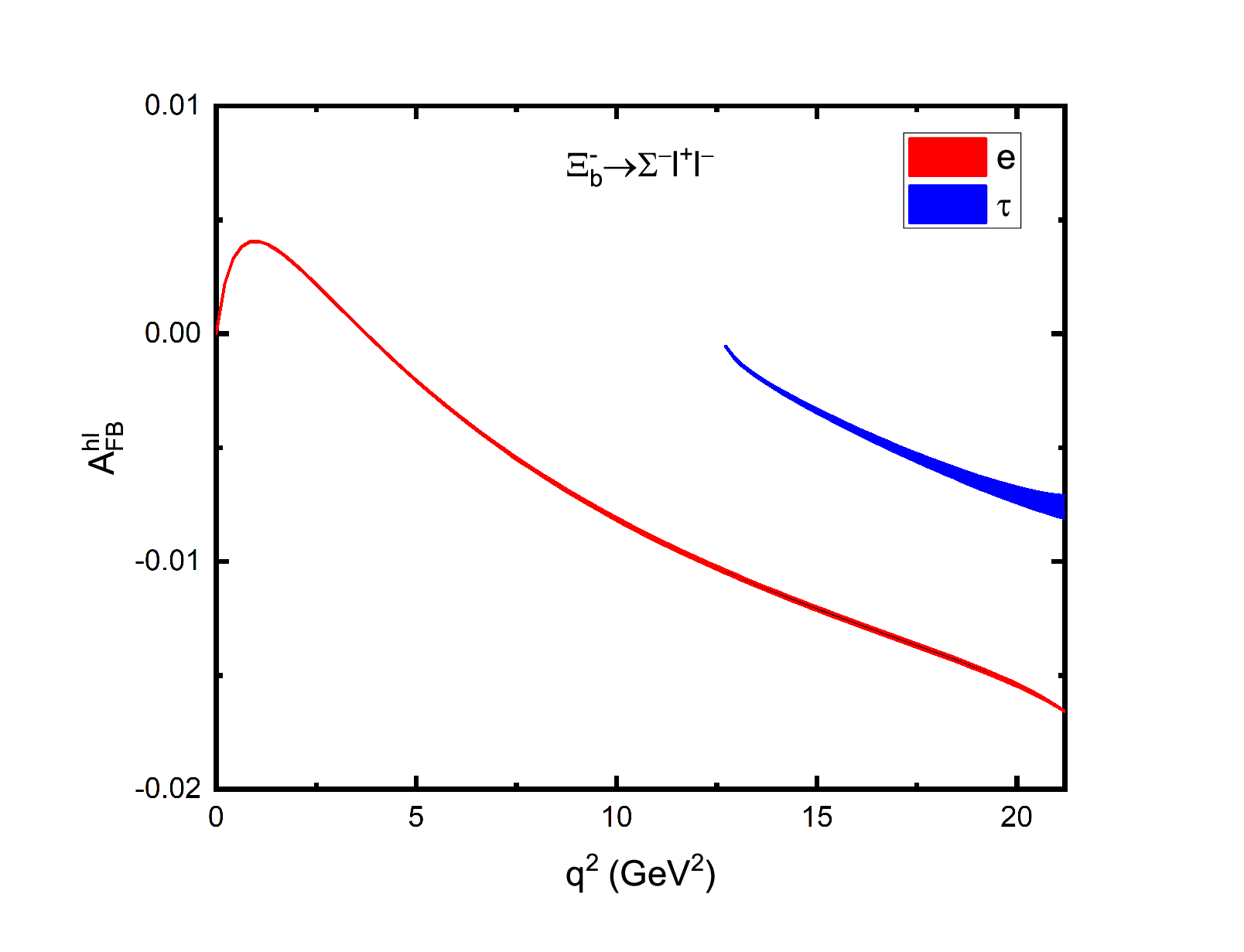} }
\hspace{-1.7cm}\subfigure{ \epsfxsize=6.5cm \epsffile{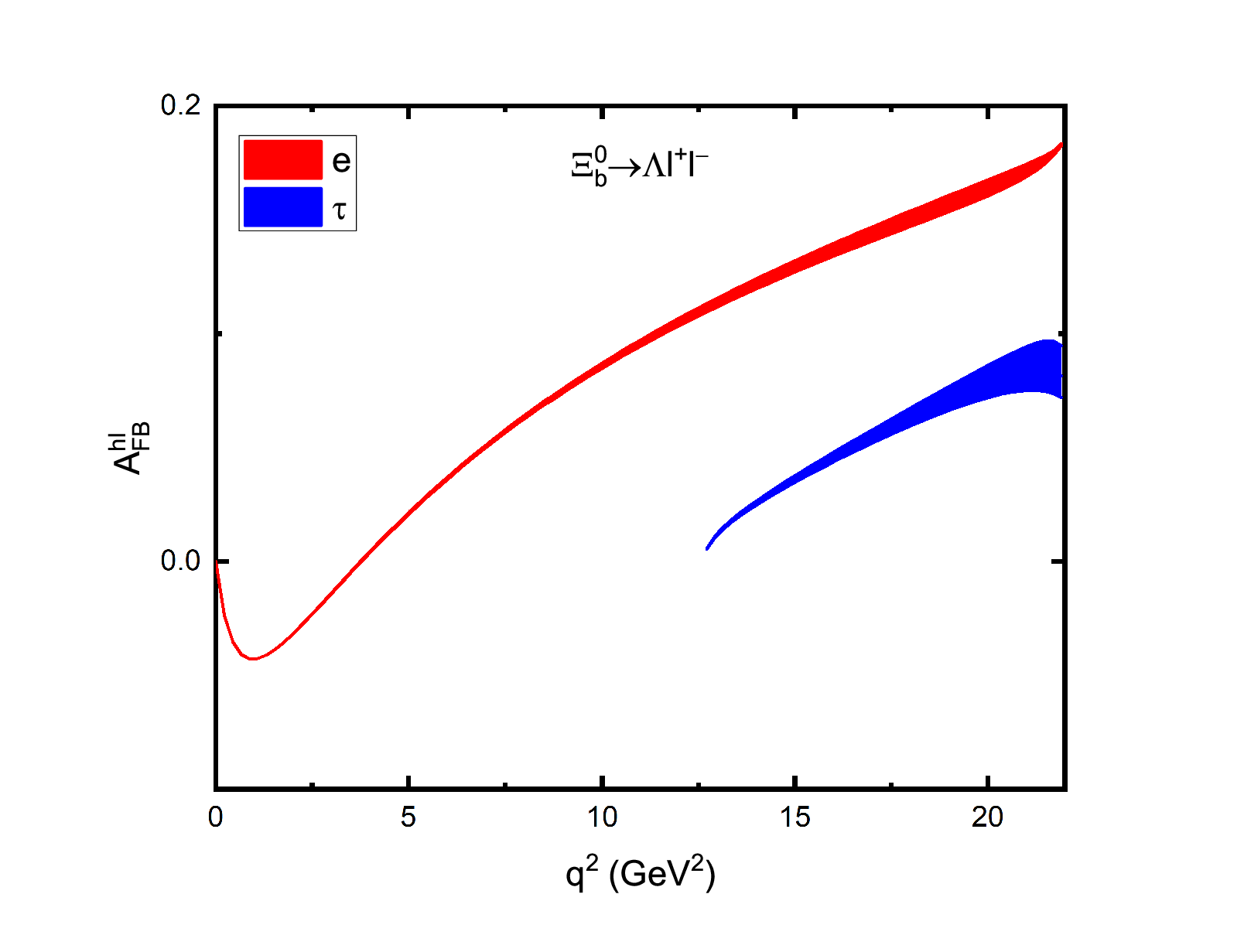}}}
\vspace{0.8cm}
\caption{$q^2$-distributions of various angular observables for the concerned decays. The bands correspond to the uncertainties sourced by the hard scale.}
 \label{fig:FL}
\end{center}
\end{figure}

 In this subsection, we evaluate a set of angular observables for the semileptonic transitions in the three charged-lepton channels. Figure~\ref{fig:FL} displays the longitudinal polarization fraction $F_L(q^2)$), the leptonic forward-backward asymmetry $A^{\ell}_{FB}$, the hadronic forward-backward asymmetry $A^h_{FB}$,  and a combined lepton-hadron forward-backward asymmetry $A^{h\ell}_{FB}$  as functions of $q^2$ in the kinematic range $4m_\ell^2<q^2<(M-m)^2$. The uncertainty associated with the choice of the hard scale is also indicated in these figures. It is evident that the theoretical uncertainties of these angular observables are significantly smaller than those of the branching franctions, which can be attributed to the partial cancellation of uncertainties in the ratios defined in Eqs.~(\ref{eq:o1}-\ref{eq:o4}). In the following, we present several remarks based on the $q^2$ distributions.

\begin{itemize}
  \item
  As shown in the upper panels of Fig.~\ref{fig:FL}, the longitudinal polarization fraction exhibits identical threshold behavior for all three decay modes, namely
  \begin{eqnarray}\label{eq:zero}
  F_L(4m^2_\ell)=F_L(q^2_{\text{max}})=\frac{1}{3}.
  \end{eqnarray}
  This feature can be readily understood from Eq.~(\ref{eq:o1}). At maximal recoil, where the dilepton invariant mass takes its minimal value $q^2=4m_\ell^2$ (equivalently $\beta_\ell=0$), Eq.~(\ref{eq:kk}) implies $K_1=K_2$, which directly leads to $F_L(4m_\ell^2)=1/3$. At the opposite kinematic end point, corresponding to zero recoil with $q^2_{\text{max}}=(M-m)^2$, the transversity amplitudes satisfy the relations $A_{\perp,1(0)}^{L(R)}=0$ and $A_{\parallel,1}^{L(R)}=\sqrt{2},A_{\parallel,0}^{L(R)}$, as follows from Eq.~(\ref{eq:amp}). These relations again enforce $K_1=K_2$, yielding $F_L(q^2_{\text{max}})=1/3$. The above threshold behaviors are therefore model independent and hold for all final-state lepton flavors. Away from the two kinematic endpoints, the $q^2$ dependence of $F_L$ differs markedly between the $e$ and $\tau$ channels. In the electron mode, $F_L$ rises rapidly in the low-$q^2$ region, reaching a maximum around $q^2 \simeq 2.5~\text{GeV}^2$, and then decreases gradually with increasing $q^2$ toward the zero-recoil point, where it approaches $1/3$. In contrast, the corresponding distribution for the $\tau$ channel is nearly flat, exhibiting a much weaker sensitivity to variations in $q^2$.

  \item
  The lepton-side forward-backward asymmetry constitutes a powerful probe for physics beyond the SM, with the position of its zero crossing being particularly sensitive to possible new-physics effects. For all lepton flavors, the leptonic forward-backward asymmetry vanishes at the two kinematic end points, $q^2=4m_\ell^2$ and $q^2=(M-m)^2$. In addition, we find that, in the large-recoil region, the asymmetry develops an extra zero-crossing point, whose position is determined by the implicit relation
  \begin{eqnarray}\label{eq:zero}
  q^2_0=-\frac{\Re[C_7^{\rm eff}C_{10}^{\rm eff*}]}{\Re[C_9^{\rm eff}C_{10}^{\rm eff*}] } m_b\left[(M+m)\frac{f^T_\perp}{f_\perp}+(M-m)\frac{g^T_\perp}{g_\perp}\right],
  \end{eqnarray}
  within the SM operator basis. It is evident that $q^2_0$ encodes information on both the Wilson coefficients and the ratios of form factors $f_\perp^T/f_\perp$ and $g_\perp^T/g_\perp$, rendering this observable especially informative. Since theoretical uncertainties associated with form factors largely cancel in these ratios, the predicted zero-crossing points for the three decay modes considered are very close to each other, as illustrated in the second panels of Fig.~\ref{fig:FL}. Consequently, the location of the zero is predominantly sensitive to the Wilson coefficients, which may receive sizable modifications from NP beyond the SM. In the SM, the opposite signs of $C_7^{\rm eff}$ and $C_9^{\rm eff}$ ensure that the zero occurs within the physical kinematic region. As noted in Ref.~\cite{Boer:2014kda}, to first approximation one has $f_\perp^T/f_\perp \approx g_\perp^T/g_\perp \approx 1$, such that Eq.~(\ref{eq:zero}) simplifies to a process-independent expression,
  \begin{eqnarray}
  q^2_0\approx-2m_bM\frac{\Re[C_7^{\rm eff}C_{10}^{\rm eff*}]}{\Re[C_9^{\rm eff}C_{10}^{\rm eff*}] },
  \end{eqnarray}
  which is well known from both exclusive and inclusive FCNC $B$-meson decays, such as $B\to K^{(*)}\ell^+\ell^-$~\cite{Antonelli:2009ws,Ali:1999mm}. Employing the SM Wilson coefficients together with the masses of the $\Xi_b$ baryon and the $b$ quark yields $q^2_0\simeq 4.0~\text{GeV}^2$, in remarkably good agreement with the value $q^2_0\simeq 3.8~\text{GeV}^2$ obtained from Eq.~(\ref{eq:zero}) using the form factors calculated in the previous section. This zero position is also consistent with that found in $\Lambda_b\to\Lambda\ell^+\ell^-$ decays~\cite{Mannel:2011xg,Wang:2008sm}. For a $\tau$ lepton in the final state, the sizable lepton mass restricts the accessible kinematic range to values far from the zero-crossing point. As a result, no additional zero appears, and the corresponding leptonic forward-backward asymmetry $A_{FB}^\tau$ remains negative definite throughout the physical region. Finally, we stress that the present analysis neglects long-distance contributions; therefore, one cannot conclude that there is only a single zero in the electron channel and none in the $\tau$ channel. In fact, long-distance effects can induce additional zero crossings in the resonance regions, as emphasized in Ref.~\cite{Mott:2011cx}.

  \item
  In contrast to the leptonic forward-backward asymmetry $A_{FB}^e$, the hadron-side forward-backward asymmetry $A_{FB}^h$ vanishes at zero recoil but remains nonzero in the low-$q^2$ region, as shown in the third panels of Fig.~\ref{fig:FL}. Its magnitude is process dependent, since it is proportional to the asymmetry parameter $\alpha_h$. If the final-state baryon decays through a strong, parity-conserving interaction, one has $\alpha_h=0$, and consequently $A_{FB}^h$ vanishes identically. In the decays considered here, however, the parity-nonconserving subprocess $\mathcal{B}_f \to \mathcal{B}_h$ induces a nonzero hadron-side forward-backward asymmetry. Because $\alpha_\Lambda$ carries the opposite sign compared to the other two cases, the corresponding curve exhibits a qualitatively different behavior. At maximal recoil, $A_{FB}^h$ takes a nonzero value and shows good stability against variations of $q^2$ in the region $q^2 \lesssim 15~\text{GeV}^2$; beyond this range, it decreases rapidly and approaches zero as the zero-recoil endpoint is reached. We further observe that, for each decay channel, the $q^2$ spectra of $A_{FB}^h$ for the $e$ and $\tau$ modes are very similar, differing mainly due to the available phase space. This feature motivates a test of lepton-flavor universality through the ratio
  \begin{eqnarray}
  \mathcal{R}(A^h_{FB})=\frac{A^h_{FB}(\ell=\tau)}{A^h_{FB}(\ell=e)},
  \end{eqnarray}
  in which the dependence on the parity-violating parameter $\alpha_h$ cancels completely. Since the predicted values of $\mathcal{R}(A_{FB}^h)$ are close to unity and exhibit only a weak $q^2$ dependence, we do not display their $q^2$ distributions. Any significant deviation from unity would constitute a clear signal of NP effects.

  \item
  The combined hadron-lepton forward-backward asymmetry $A^{h\ell}_{FB}$ approaches zero at maximum recoil in the SM, and exhibits an additional zero-crossing point at intermediate $q^2$. From Eq.~(\ref{eq:o4}), this zero is given by
  \begin{eqnarray}\label{eq:zero11}
  q^2_0=-\frac{\Re[C_7^{\rm eff}C_{10}^{\rm eff*}]}{\Re[C_9^{\rm eff}C_{10}^{\rm eff*}] }2m_b\left[(M+m)\frac{f^T_\perp }{f_\perp }\gamma+(M-m)\frac{g^T_\perp }{g_\perp}(1-\gamma)\right],
  \end{eqnarray}
  where \(\gamma=s_-f_\perp^2/(s_-f_\perp^2+s_+g_\perp^2)\). Numerically, $\gamma\simeq 0.5$ and exhibits only a weak dependence on $q^2$ over the entire kinematic range. As a result, the zero-crossing position of $A^{h\ell}_{FB}$ is nearly identical to that of $A^\ell_{FB}$ in Eq.~(\ref{eq:zero11}), and any significant shift in this position would again point to new-physics effects. As in the purely leptonic case, no zero crossing appears at high $q^2$ for the $\tau$ mode in the absence of long-distance contributions. At zero recoil, and in the limit of vanishing lepton mass, the endpoint relations in Eq.~(\ref{eq:FFC2}) lead to the simple expression
  \begin{eqnarray}\label{eq:zero2}
  A_{FB}^{h\ell}\approx-\frac{\alpha_h}{2}\frac{\Re[C_9^{\rm eff}C^{\rm eff*}_{10}]}{|C_9^{\rm eff}|^2+|C^{\rm eff}_{10}|^2},
  \end{eqnarray}
  where contributions from dipole operators have been neglected due to their smallness. A similar expression was obtained in Ref.~\cite{Hiller:2021zth} (see Eq.~(5.9)) upon setting the right-handed Wilson coefficients to zero. This observable therefore provides complementary information on the Wilson coefficients and offers additional sensitivity to NP scenarios that modify the effective weak Hamiltonian.
\end{itemize}

\begin{table}[htbh]
	\caption{Angular observables integrated over the entire kinematic region for the $\Xi_b\to (\Xi,\Sigma,\Lambda) \ell^+ \ell^-$ decays with different lepton flavors.}
	\label{tab:Observables}
    \centering
	\begin{tabular}[t]{lcccc}
	\hline\hline
 Observable        &$\ell$     &  $\Xi_b\to \Xi$   & $\Xi_b\to \Sigma$     &  $\Xi_b\to \Lambda$     \\ \hline
   $\langle F_L \rangle $ &e&  $0.529^{+0.006}_{-0.013}$      &  $0.509^{+0.013}_{-0.002}$ &  $0.518^{+0.009}_{-0.016}$    \\
  & $\mu$ &                    $0.592\pm0.018$      &  $0.543^{+0.016}_{-0.004}$ &  $0.604^{+0.021}_{-0.008}$    \\
  & $\tau$&                    $0.349\pm0.000$      &   $0.351^{+0.001}_{-0.000}$ &   $0.353^{+0.000}_{-0.003}$   \\
  $\langle A_{\text{FB}} \rangle$ &e& $-0.194^{+0.012}_{-0.013}$  & $-0.237^{+0.001}_{-0.014}$ & $-0.181^{+0.051}_{-0.008}$   \\
 & $\mu$&                             $-0.218^{+0.010}_{-0.013}$  & $-0.254^{+0.001}_{-0.013}$ & $-0.212^{+0.049}_{-0.007}$   \\
 & $\tau$                           & $-0.115^{+0.008}_{-0.011}$  & $-0.127^{+0.008}_{-0.002}$ & $-0.142^{+0.024}_{-0.030}$   \\
  $\langle A^h_{\text{FB}} \rangle$ &e& $0.174\pm0.003$  & $0.032^{+0.001}_{-0.000}$ & $-0.367^{+0.006}_{-0.000}$   \\
 & $\mu$&                               $0.173^{+0.004}_{-0.003}$  & $0.032^{+0.001}_{-0.000}$ & $-0.367^{+0.007}_{-0.000}$   \\
  & $\tau$&                             $0.160^{+0.003}_{-0.001}$  & $0.029^{+0.001}_{-0.000}$ & $-0.363^{+0.024}_{-0.002}$   \\
  $\langle A^{h\ell}_{\text{FB}} \rangle$ &e & $-0.037\pm0.004$  & $-9.2^{+0.5}_{-0.3}\times 10^{-3}$ & $0.068^{+0.003}_{-0.016}$   \\
   & $\mu$&                                    $-0.042\pm0.004$  & $-9.9^{+0.5}_{-0.2}\times 10^{-3}$ & $0.080^{+0.003}_{-0.015}$   \\
  & $\tau$&                                    $-0.023\pm0.004$  & $-5.4^{+0.8}_{-0.1}\times 10^{-3}$ & $0.054^{+0.012}_{-0.008}$   \\
  \hline\hline
	\end{tabular}
\end{table}

We also report integrated results obtained by integrating over the full $q^2$ range. The $q^2$-integrated angular observables for all lepton channels are collected in Table~\ref{tab:Observables}, quoted uncertainties have been combined in quadrature. Theoretical errors on these integrated quantities are substantially reduced relative to their differential counterparts. Lepton-side asymmetries display characteristic differences between electron, muon and tau modes owing to the sizable lepton-mass effects and the corresponding mass-dependent phase space. The particularly small values of  $\langle A^h_{\text{FB}} \rangle$ and $\langle A^{h\ell}_{\text{FB}} \rangle$ in the $\Sigma$ channel can be traced to the suppression induced by the asymmetry parameter $\alpha_\Sigma$ . These integrated predictions are amenable to experimental verification in future measurements.

\subsection{The dineutrino channels}\label{sec:dineu}
\begin{table}[htbh]
	\caption{ Branching fractions and  integrated angular observables for the $\Xi_b\to (\Xi,\Sigma,\Lambda) \nu\bar\nu$ decays.}
	\label{tab:Observables1}
    \centering
	\begin{tabular}[t]{lccc}
	\hline\hline
 Observable          &  $\Xi_b\to \Xi$   & $\Xi_b\to \Sigma$     &  $\Xi_b\to \Lambda$     \\ \hline
  $\mathcal{B}_\nu$      &$8.1^{+1.4}_{-1.9}\times10^{-6}$ &$2.1^{+0.3}_{-0.2}\times10^{-7}$ &$2.9^{+1.5}_{-1.7}\times10^{-8}$ \\
  $\langle R^{e/\nu}_{\mathcal{B}_f} \rangle $  &  $0.140^{+0.004}_{-0.000}$      &  $0.134^{+0.002}_{-0.001}$ &  $0.152^{+0.008}_{-0.004}$    \\
  $\langle F_L \rangle $  &  $0.577^{+0.018}_{-0.015}$      &  $0.527^{+0.015}_{-0.003}$ &  $0.600^{+0.026}_{-0.010}$    \\
  $\langle A_{\text{FB}} \rangle$ & $-0.303^{+0.004}_{-0.002}$  & $-0.318^{+0.000}_{-0.009}$ & $-0.294^{+0.031}_{-0.008}$   \\
  $\langle A^h_{\text{FB}} \rangle$ &  $0.174^{+0.003}_{-0.004}$  & $0.032^{+0.001}_{-0.000}$ & $-0.367^{+0.007}_{-0.000}$   \\
  $\langle A^{h\ell}_{\text{FB}} \rangle$   & $-0.056\pm0.002$  & $-0.012\pm0.000$ & $0.110^{+0.003}_{-0.007}$   \\
  \hline\hline
	\end{tabular}
\end{table}

\begin{figure}[htbh]
\centering
\setlength{\abovecaptionskip}{0pt}
\centerline{
 \subfigure{\epsfxsize=6.5cm \epsffile{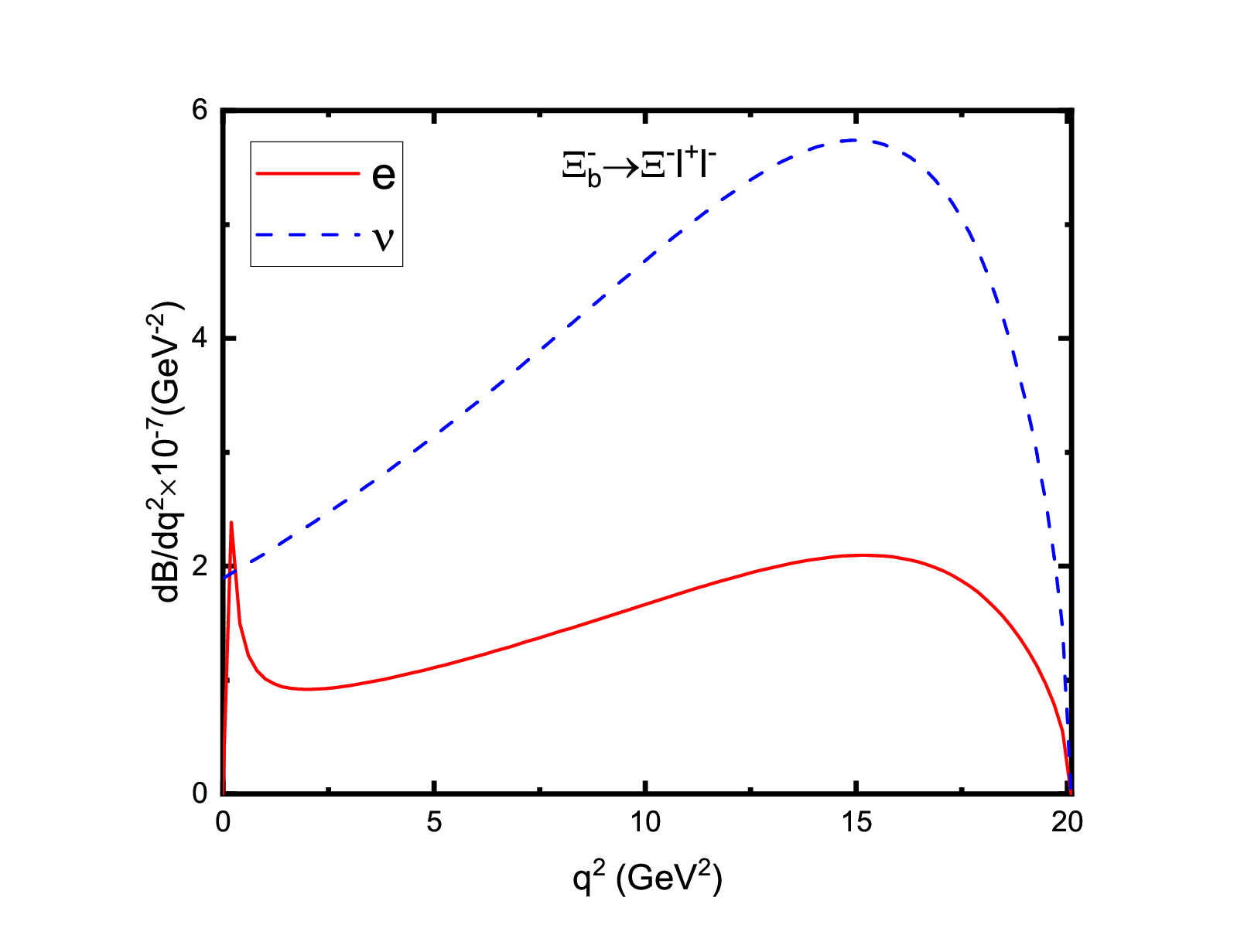} }
\hspace{-1.5cm}\subfigure{\epsfxsize=6.5cm \epsffile{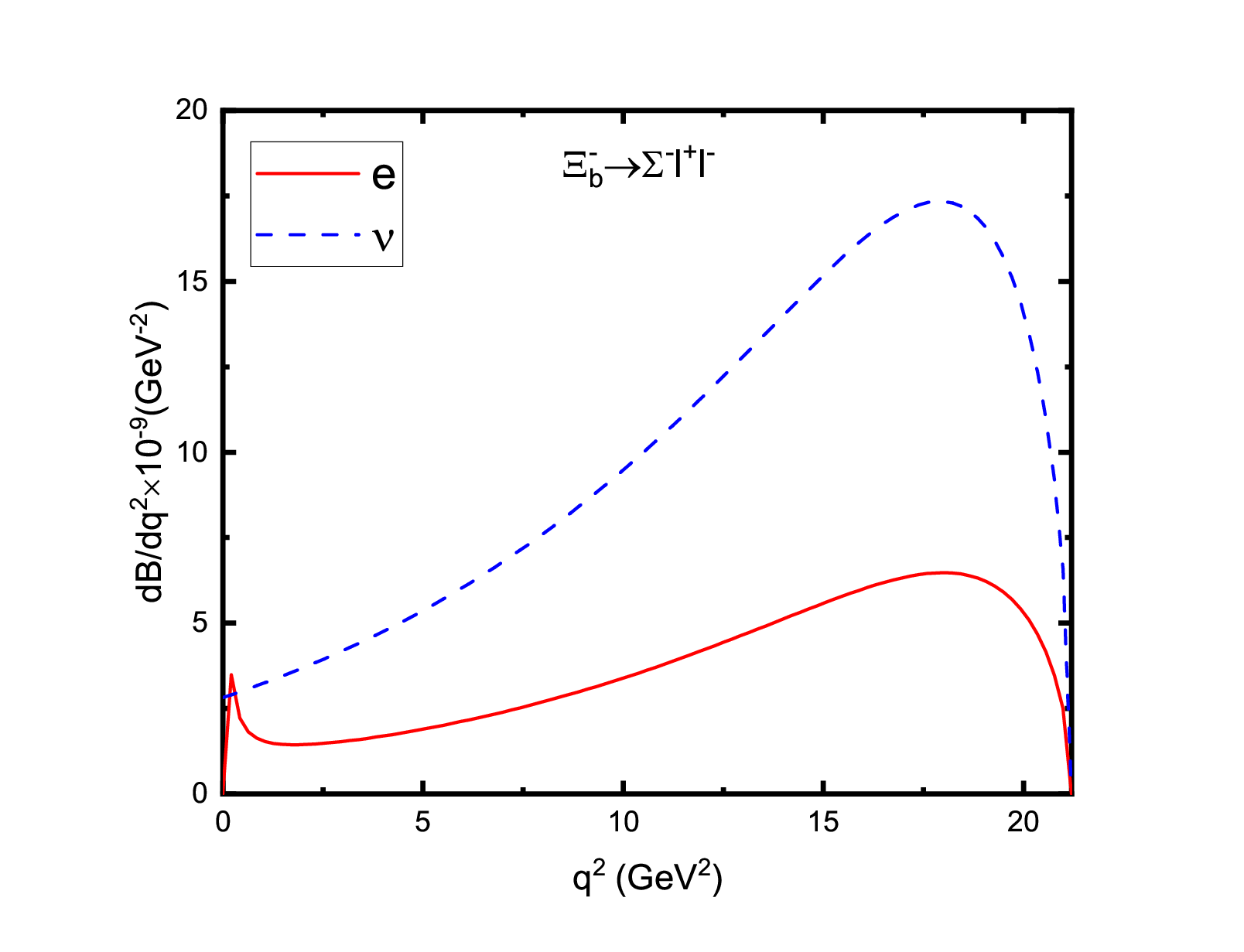} }
\hspace{-1.7cm}\subfigure{ \epsfxsize=6.5cm \epsffile{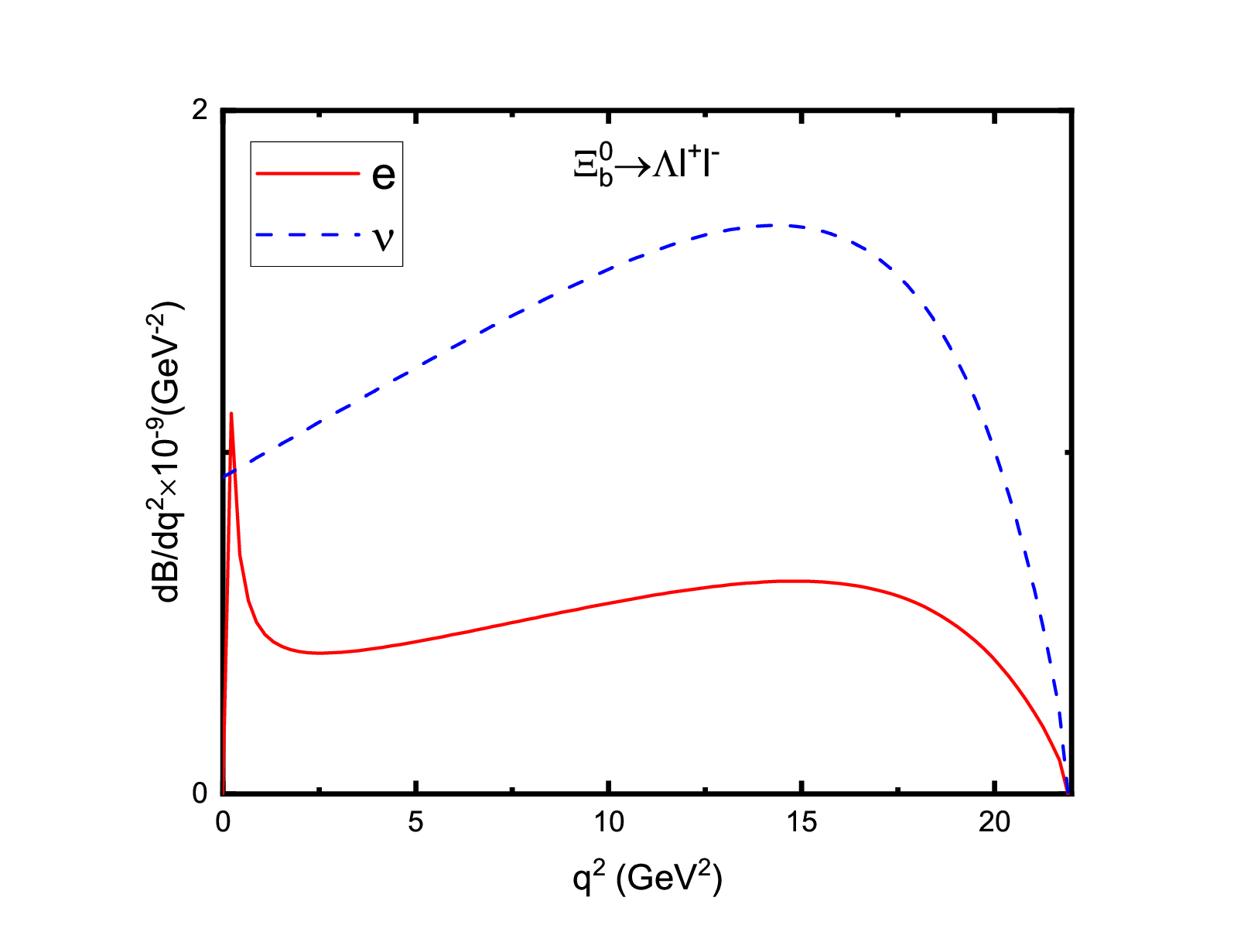}}}
\vspace{-0.8cm}
\centerline{
 \subfigure{\epsfxsize=6.5cm \epsffile{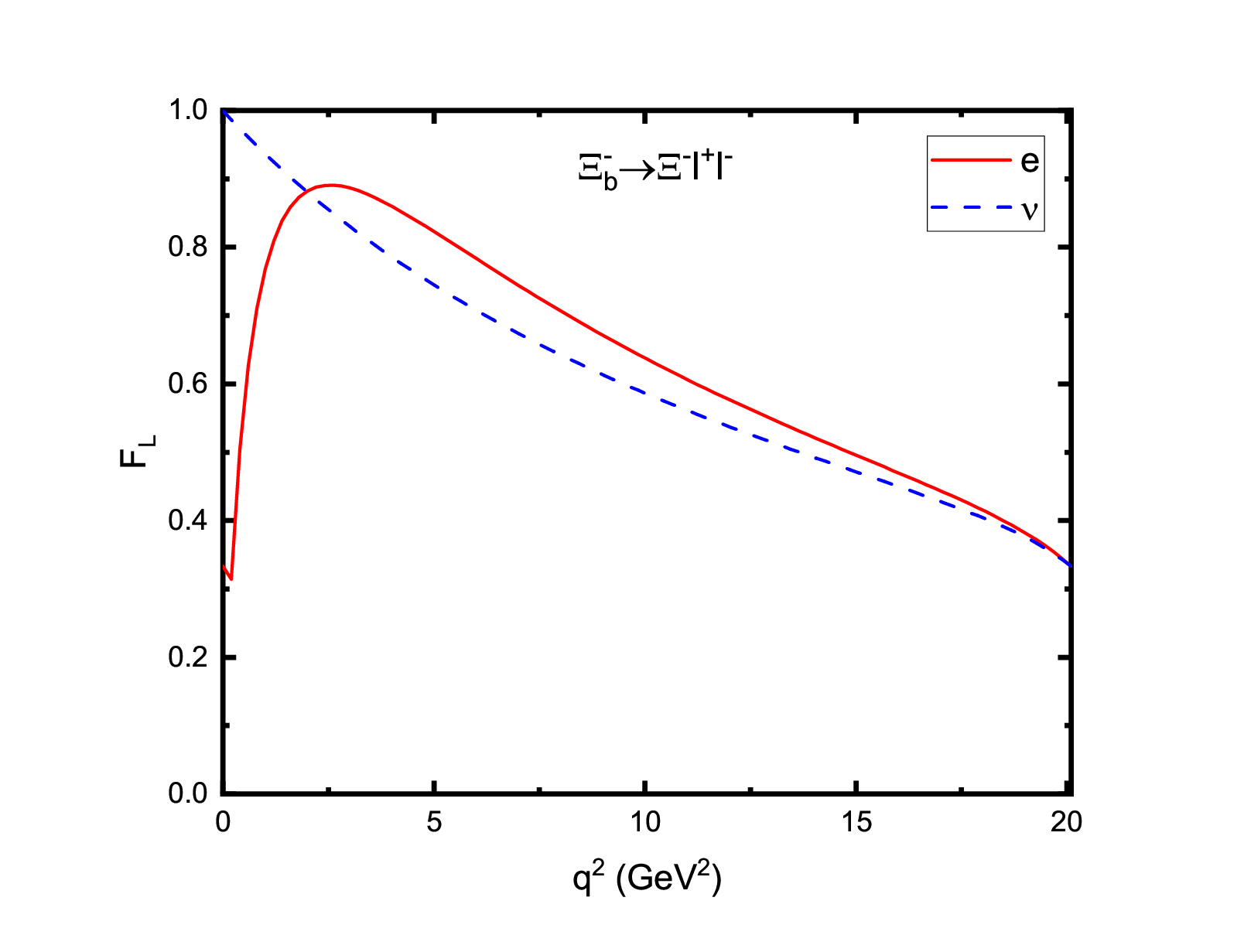} }
\hspace{-1.5cm}\subfigure{\epsfxsize=6.5cm \epsffile{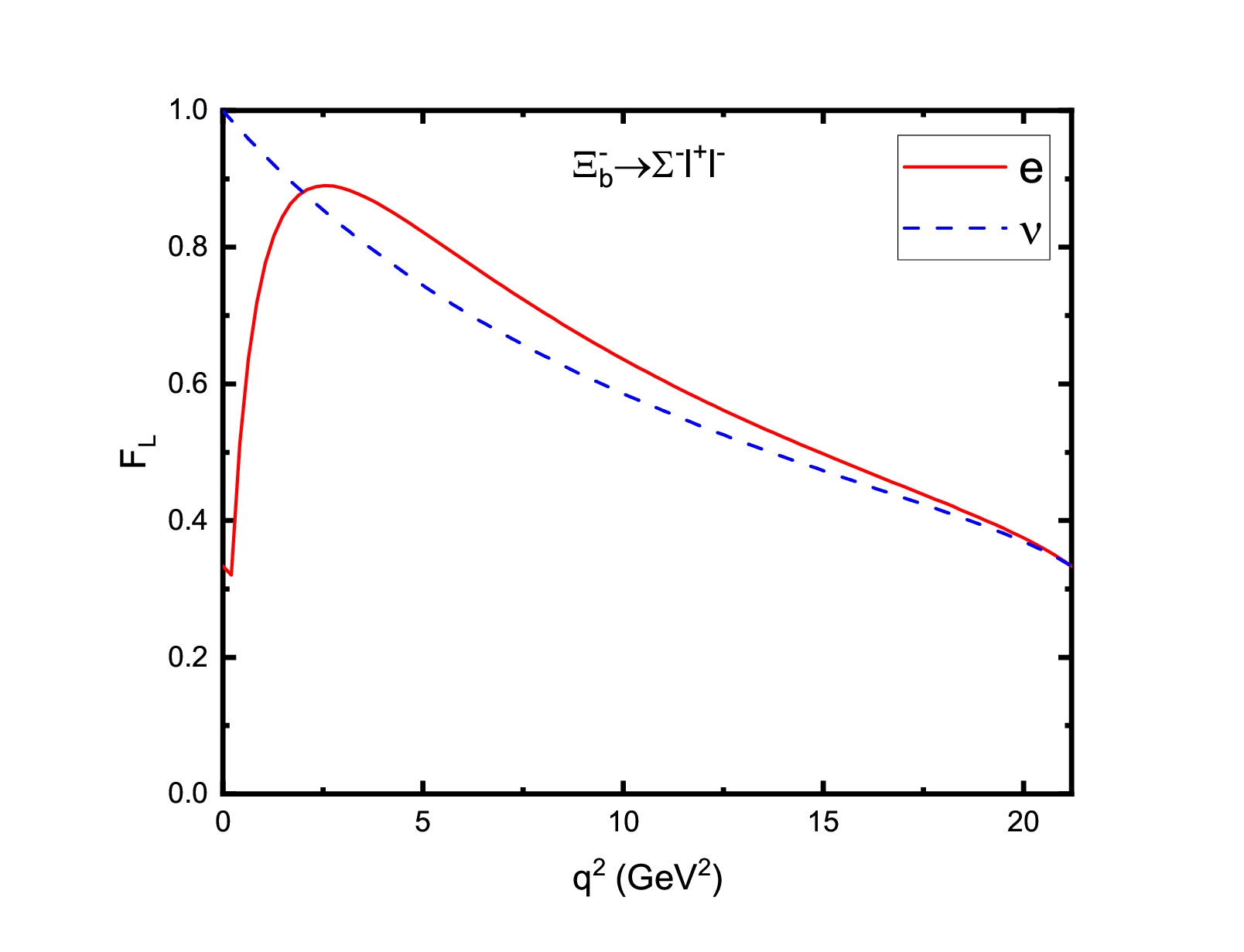} }
\hspace{-1.7cm}\subfigure{ \epsfxsize=6.5cm \epsffile{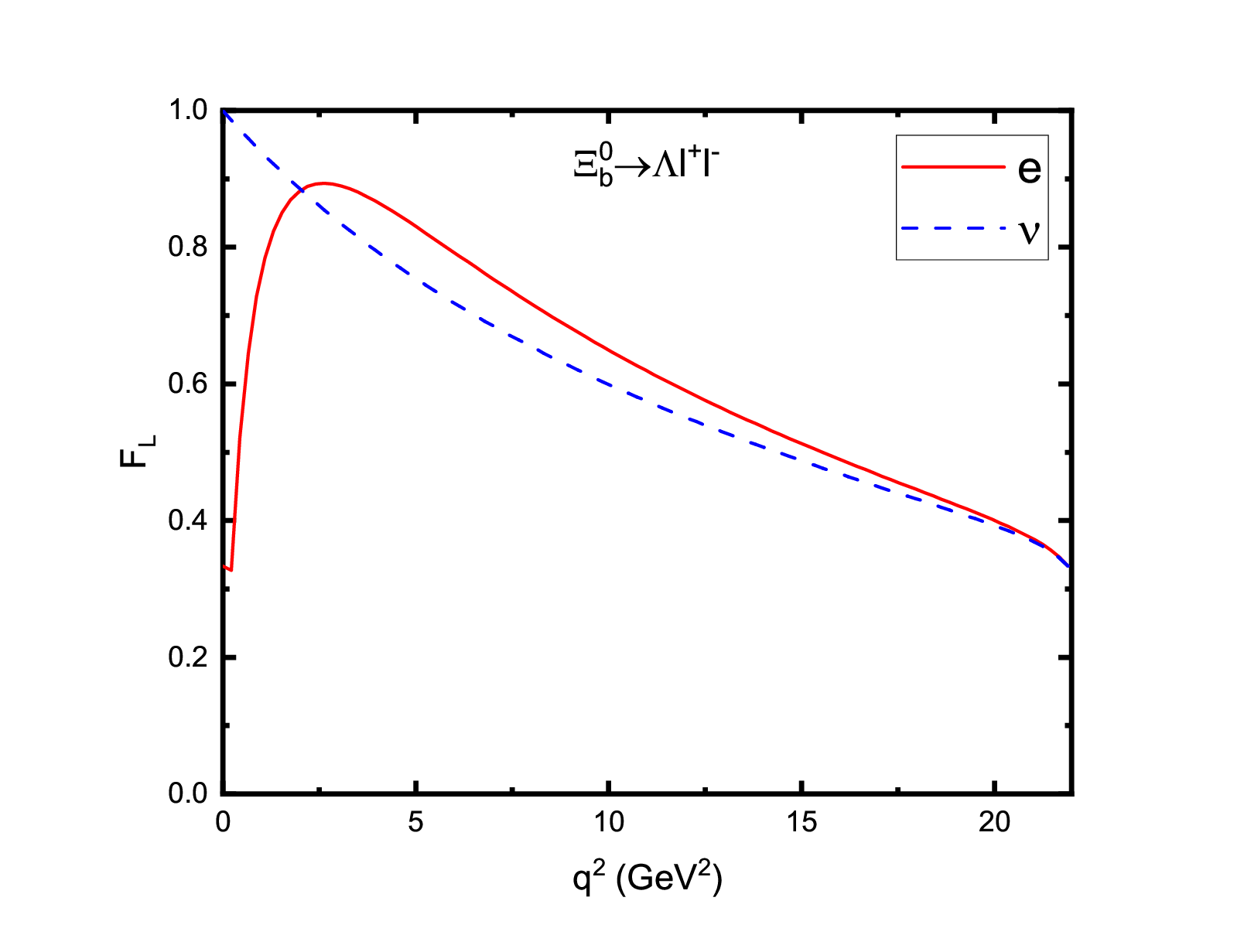}}}
\vspace{-0.8cm}
\centerline{
 \subfigure{\epsfxsize=6.5cm \epsffile{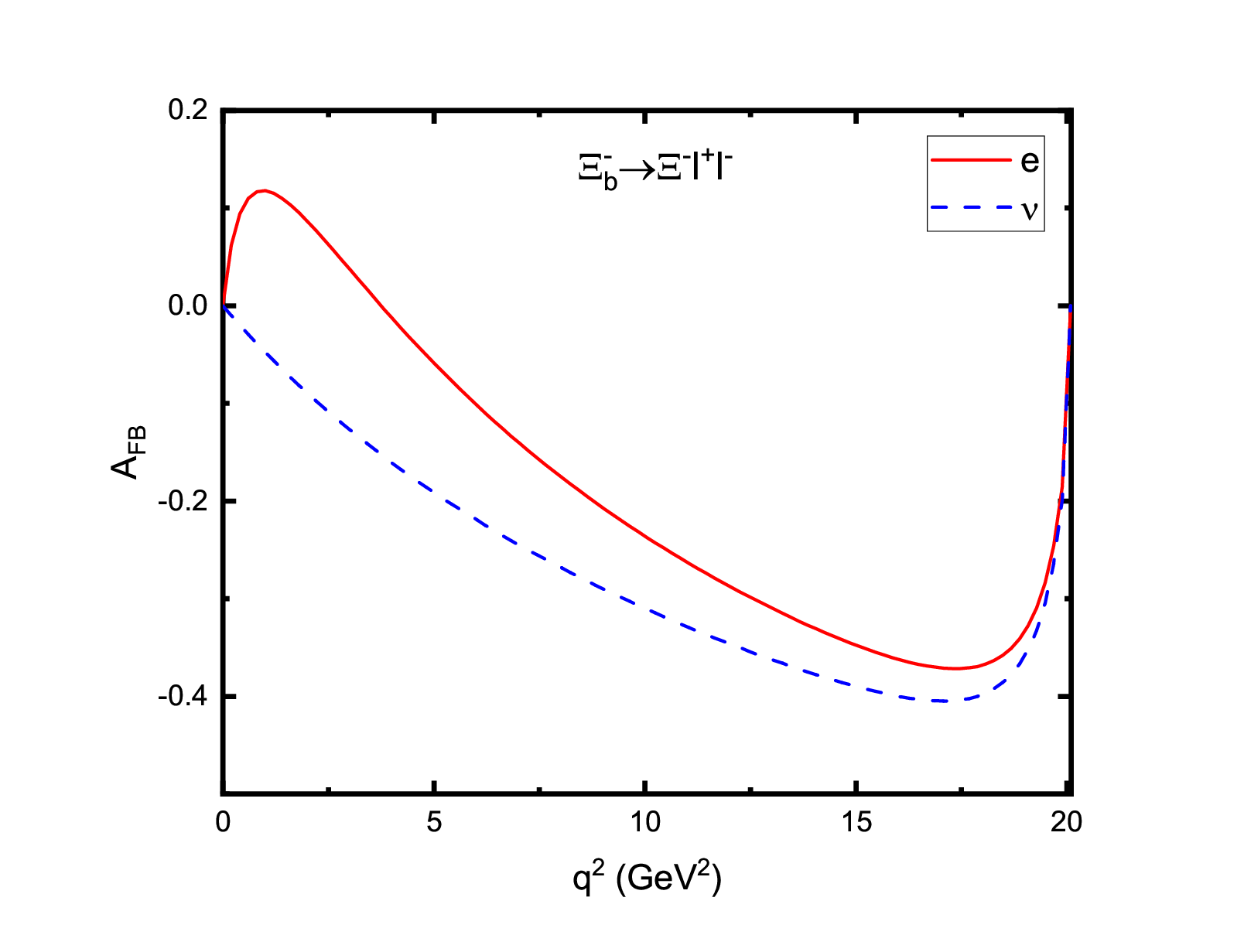} }
\hspace{-1.5cm}\subfigure{\epsfxsize=6.5cm \epsffile{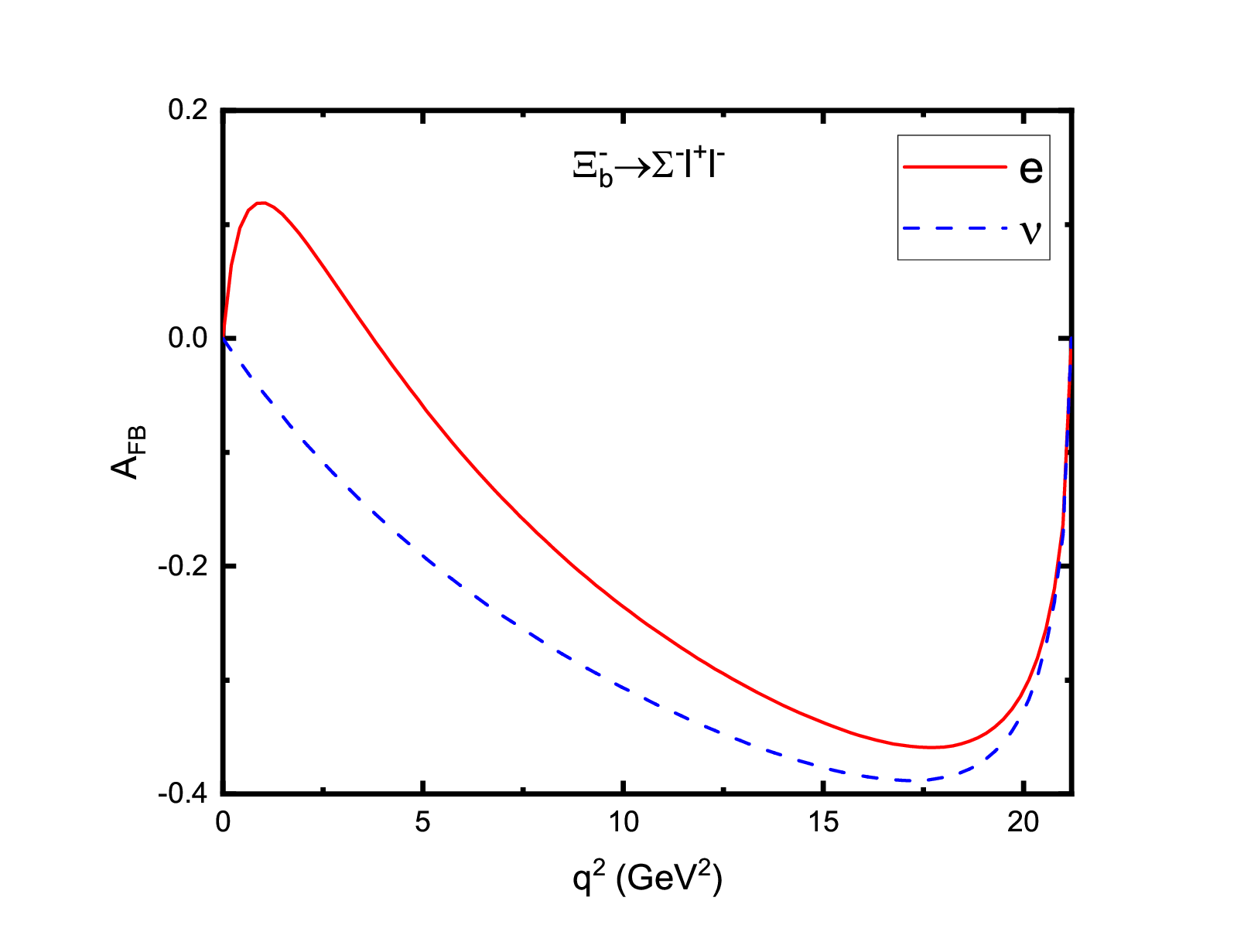} }
\hspace{-1.7cm}\subfigure{ \epsfxsize=6.5cm \epsffile{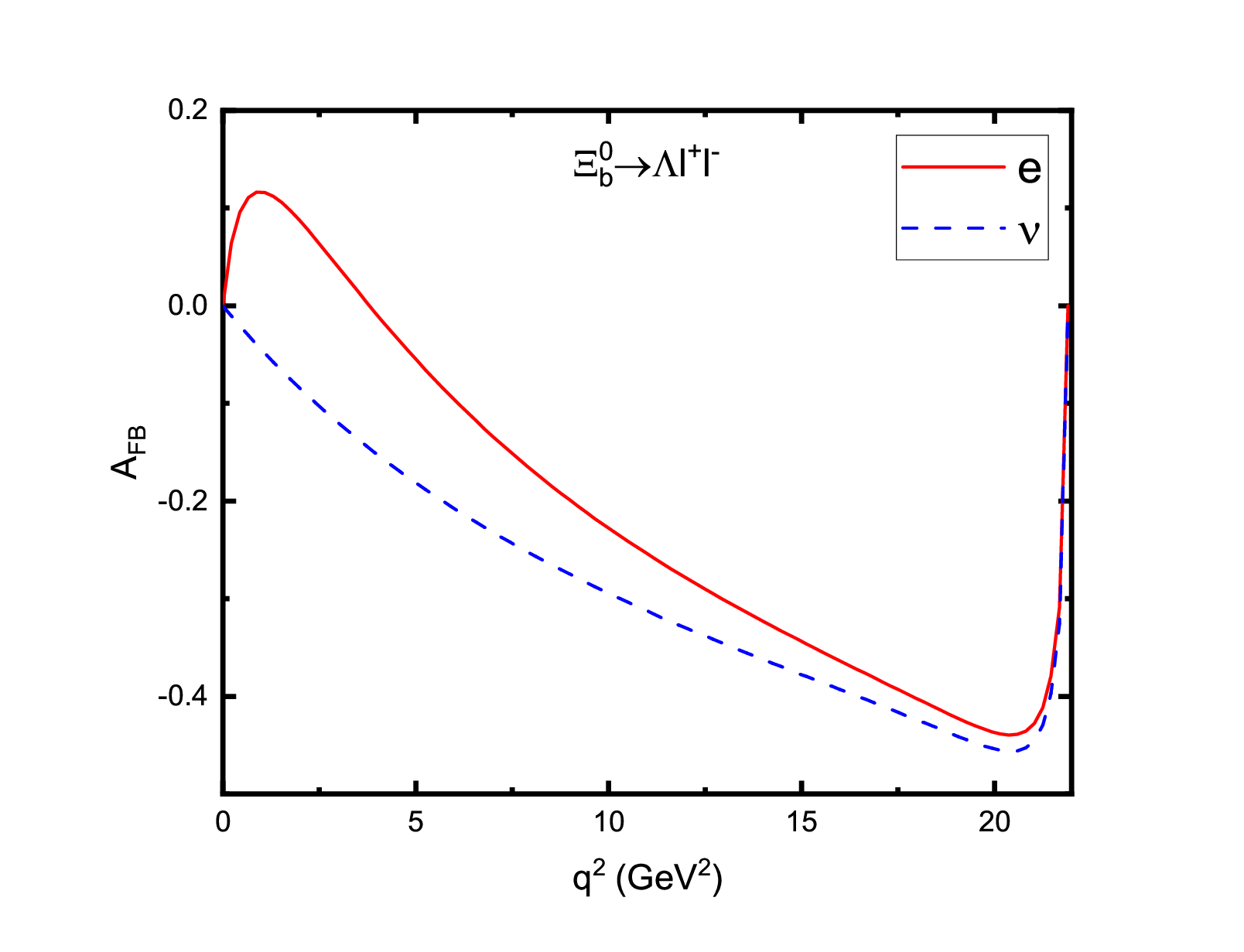}}}
\vspace{-0.8cm}
\centerline{
 \subfigure{\epsfxsize=6.5cm \epsffile{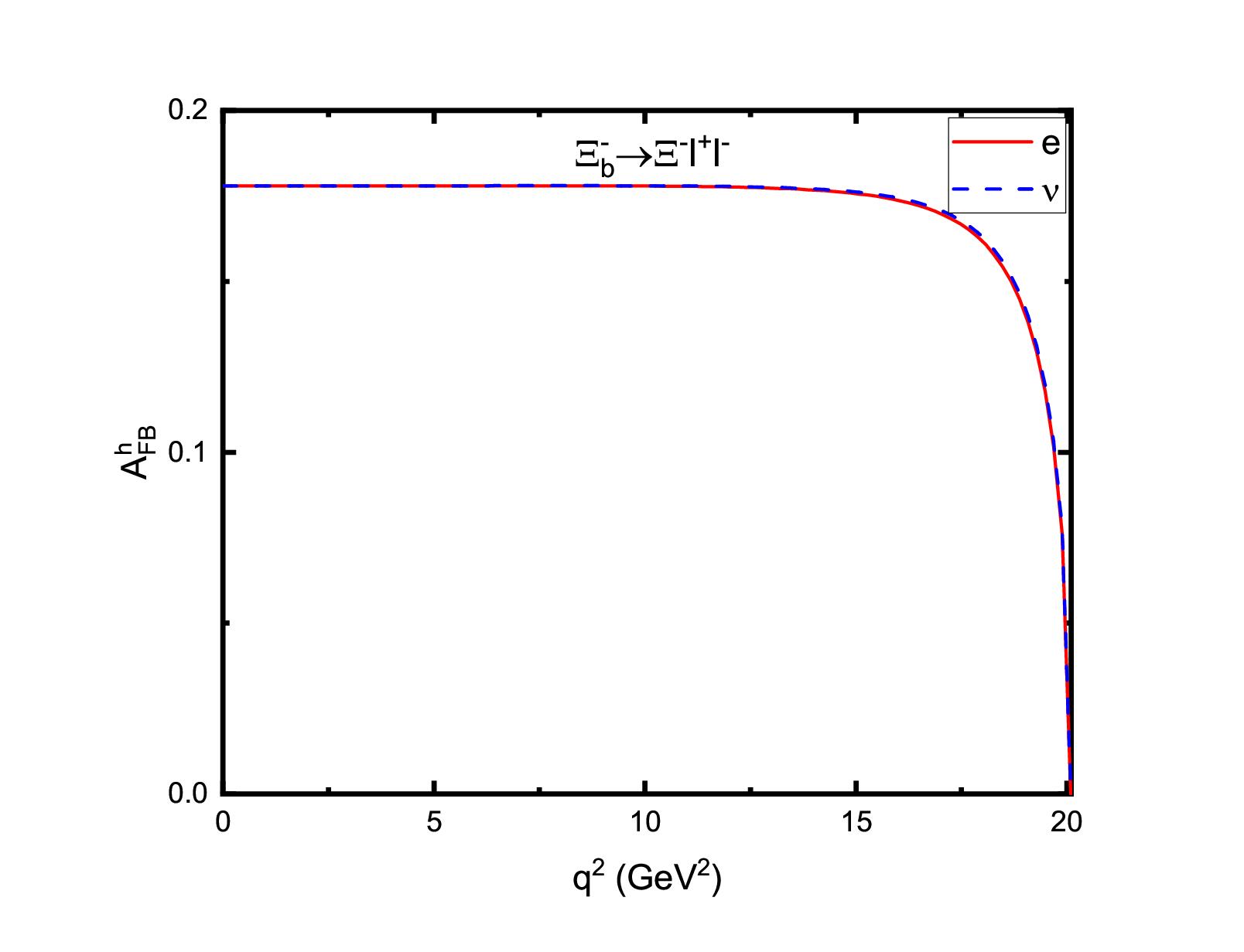} }
\hspace{-1.5cm}\subfigure{\epsfxsize=6.5cm \epsffile{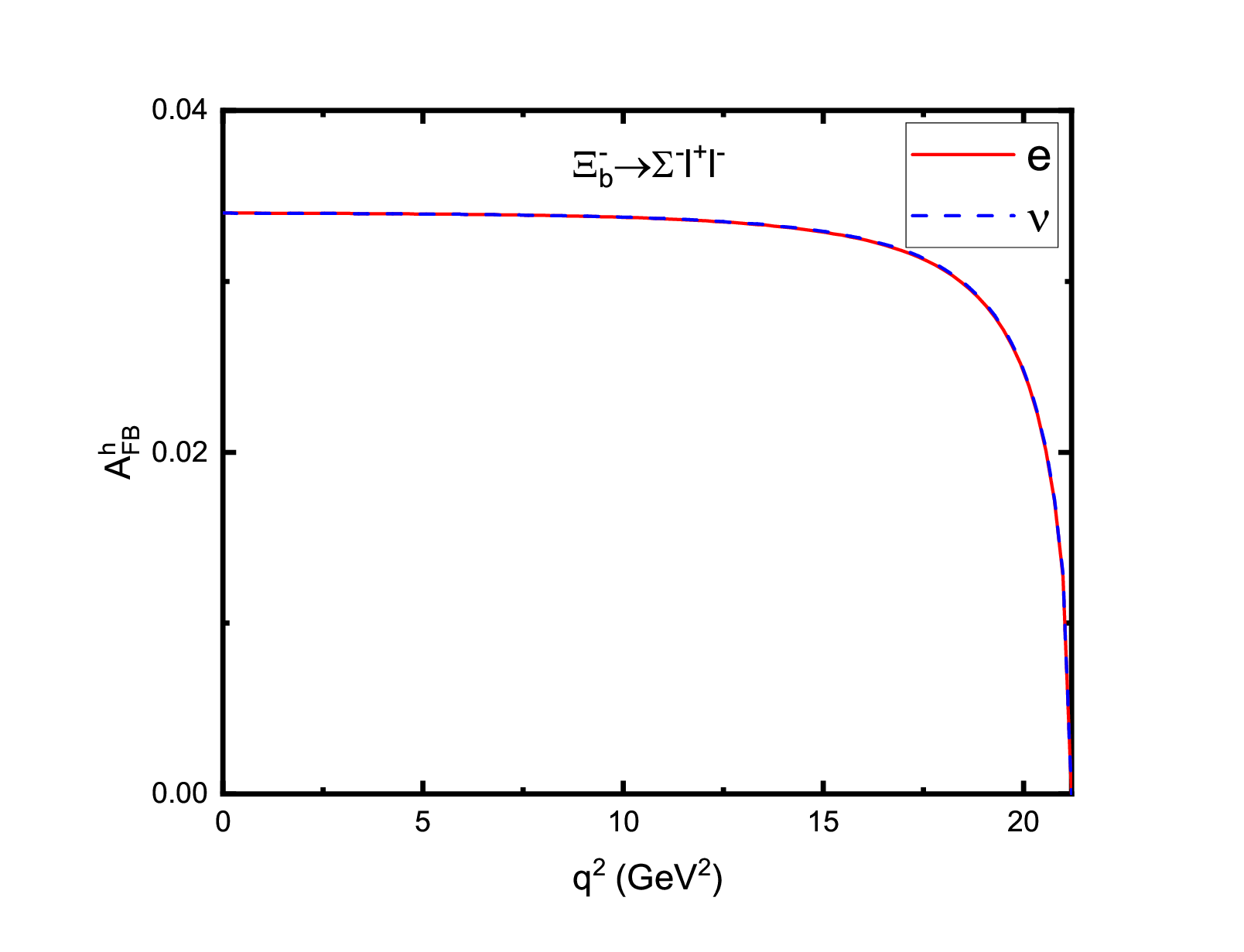} }
\hspace{-1.7cm}\subfigure{ \epsfxsize=6.5cm \epsffile{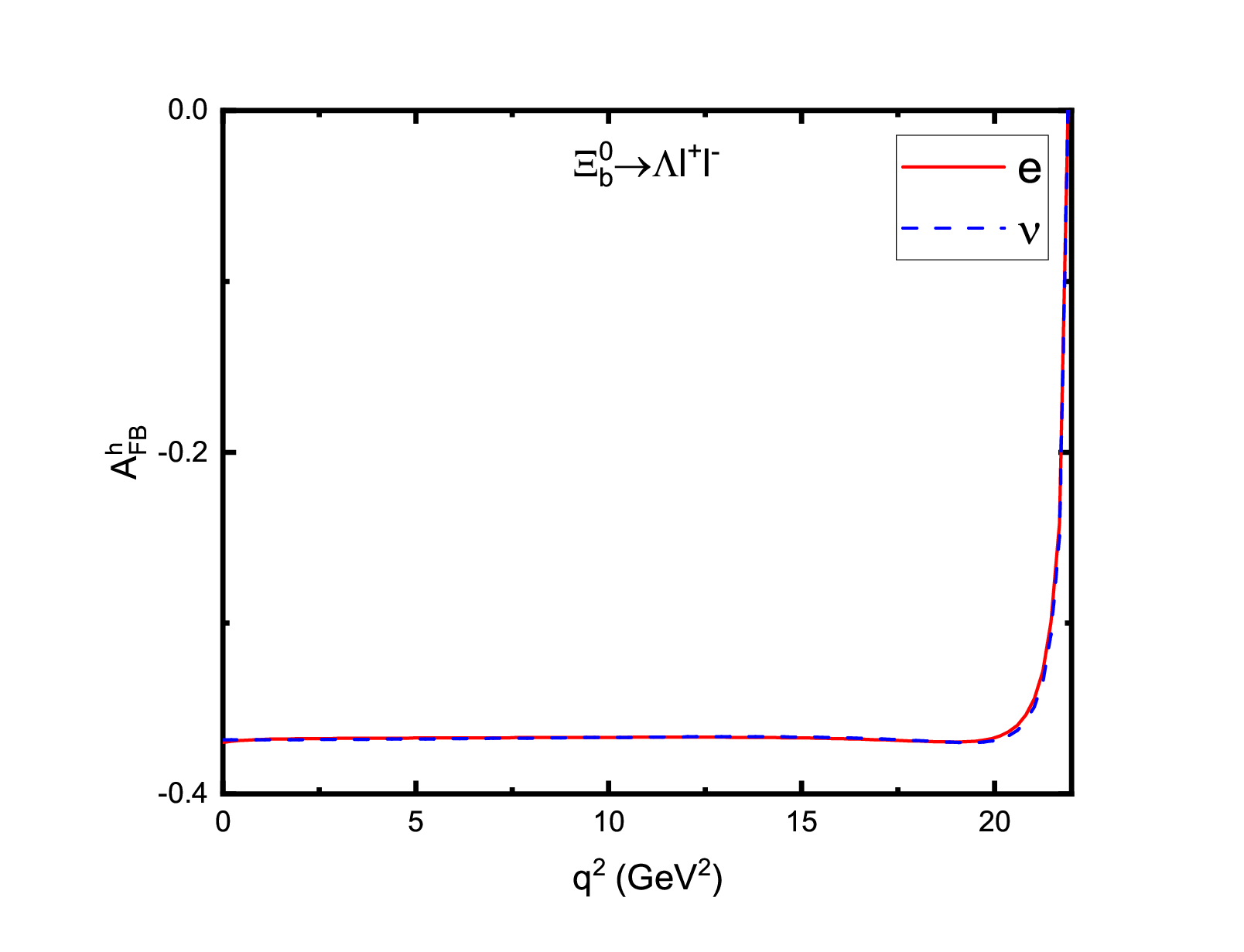}}}
\vspace{-0.8cm}
\centerline{
 \subfigure{\epsfxsize=6.5cm \epsffile{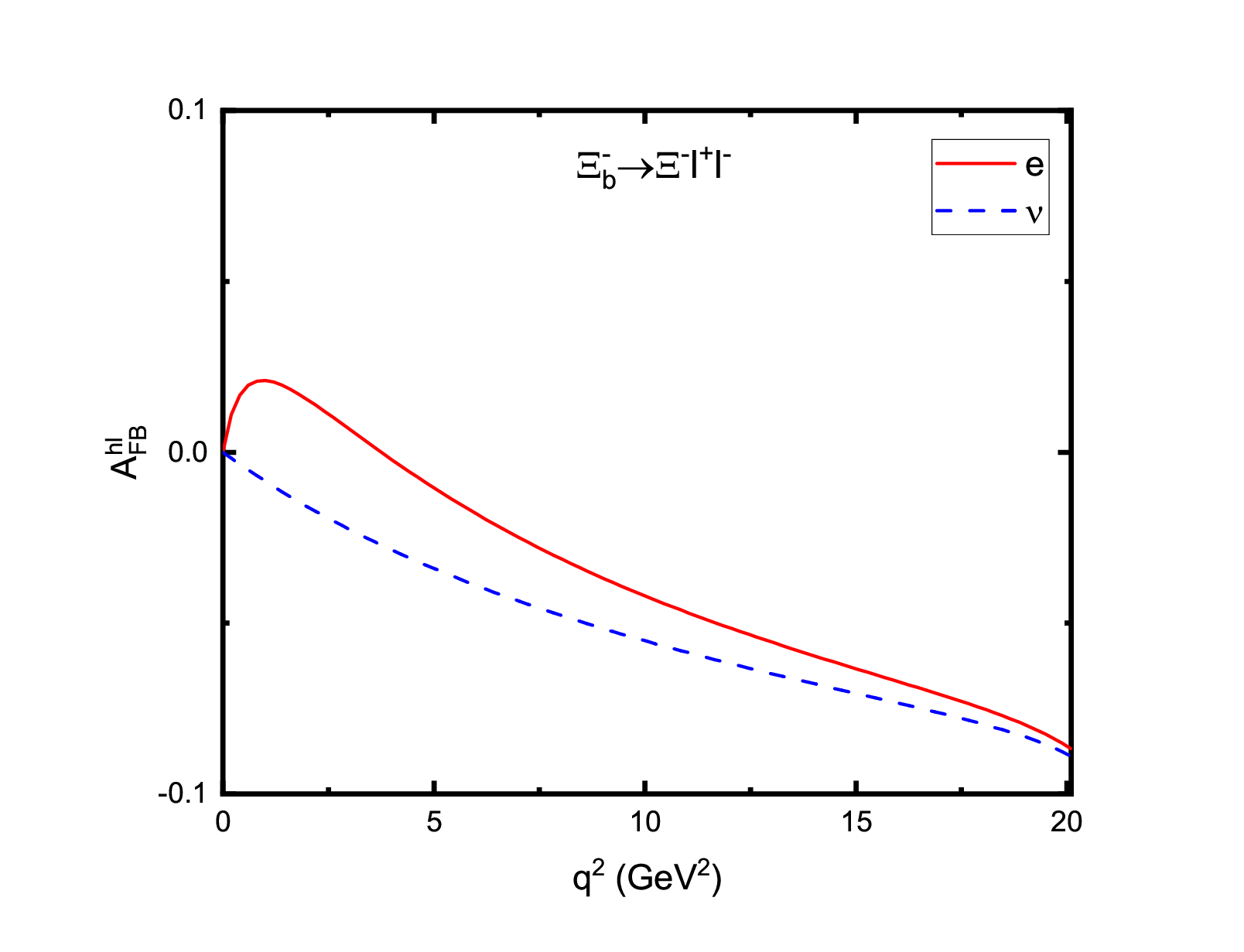} }
\hspace{-1.5cm}\subfigure{\epsfxsize=6.5cm \epsffile{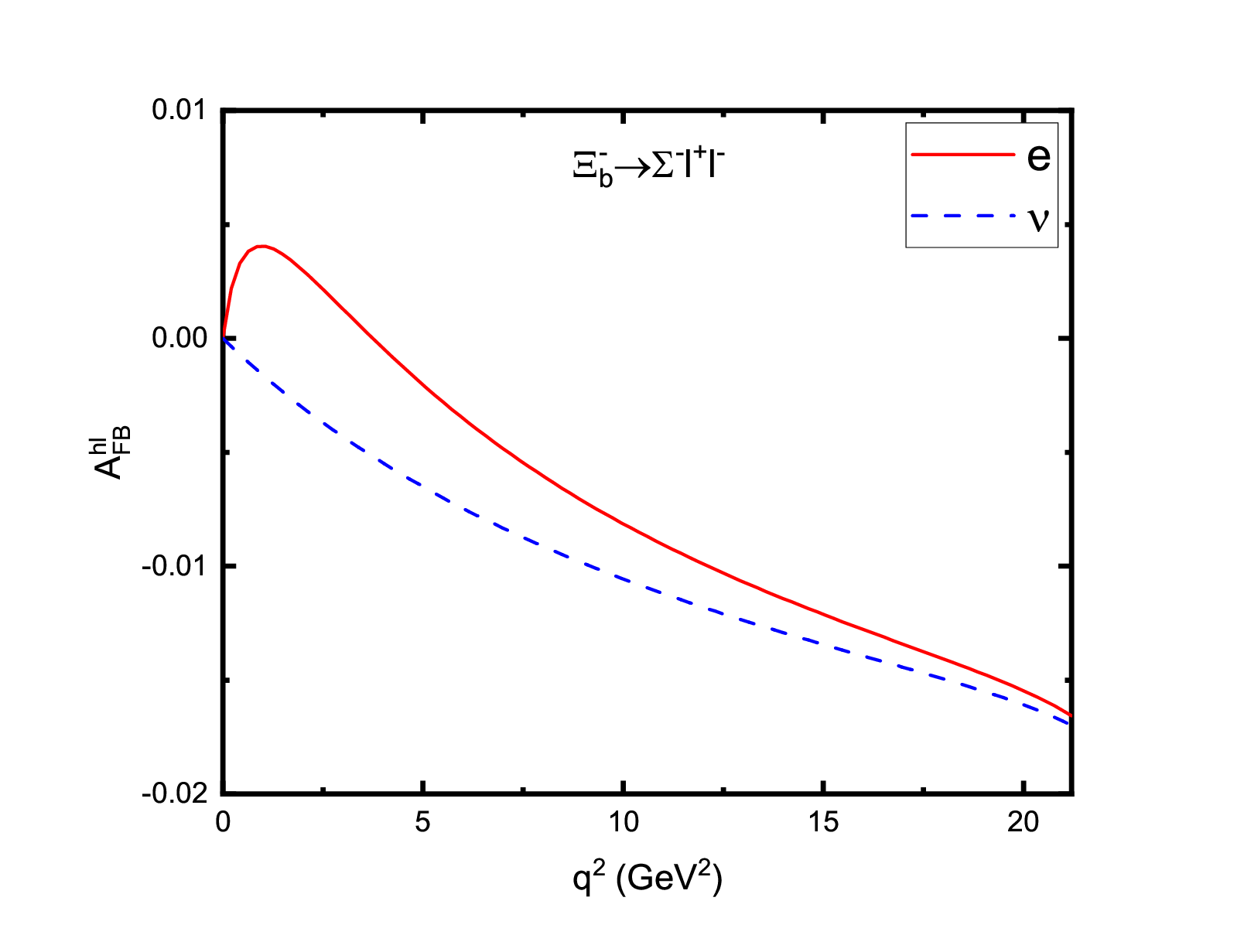} }
\hspace{-1.7cm}\subfigure{ \epsfxsize=6.5cm \epsffile{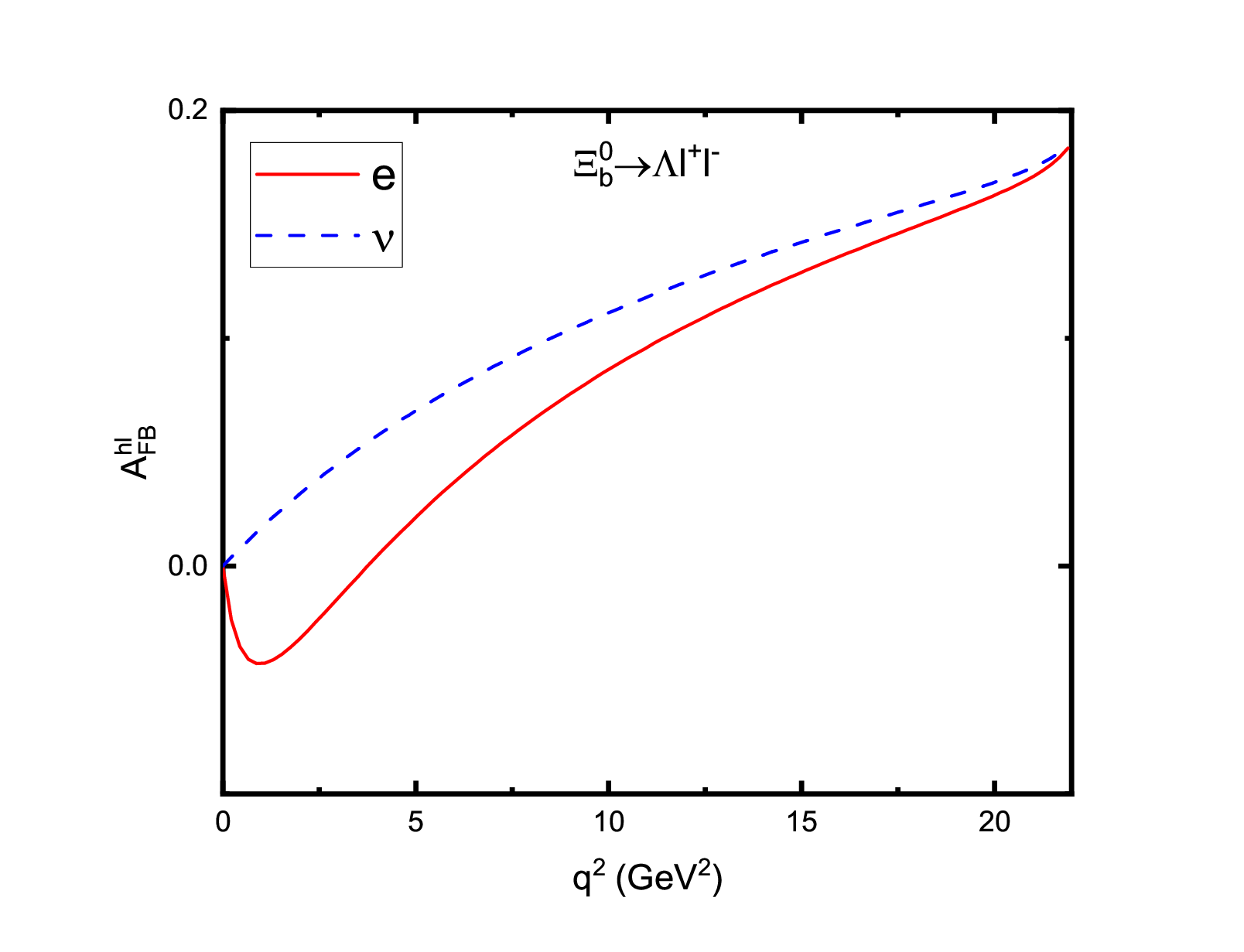}}}
\caption{ A comparison of the $q^2$ distributions of various observables between the electron (solid red) and neutrino (dashed blue) channels in the full $q^2$ kinematic region.}
 \label{fig:FLv}
\end{figure}

In this subsection, we turn to another class of rare FCNC decays mediated by the quark-level transitions $b\to s/d\nu\bar\nu$ within the PQCD framework. These channels offer complementary information on potential NP effects in comparison with the $b\to s/d\ell^+\ell^-$ modes. Importantly, the neutrino channels are theoretically ``clean'', as they are free from long-distance contaminations associated with intermediate charmonium resonances. The calculation of the dineutrino modes proceeds in close analogy to the dilepton case, relying on the same set of baryonic form factors. In Fig.~\ref{fig:FLv}, we show the $q^2$ distributions of several observables for the decays $\Xi_b\to(\Xi,\Sigma,\Lambda)\nu\bar\nu$, together with the corresponding electron-mode curves for comparison. The integrated values over the full kinematic range are summarized in Table~\ref{tab:Observables1}. Our main findings are outlined below.

\begin{itemize}

 \item
 The differential branching fractions of the dineutrino modes are generally larger than those of the corresponding dilepton decays.
 This enhancement is primarily due to the significantly larger magnitude of the Wilson coefficients involved in the dineutrino processes.
A similar pattern is also observed in the $B$ decays, for example, $\mathcal{B}(B\to K \nu \bar \nu) >\mathcal{B}(B\to Ke^+e^-)$~\cite{Parrott:2022zte}.
 Over the accessible $q^2$ region, the integrated branching fractions are predicted to lie in the range $10^{-8}$-$10^{-6}$. The branching fractions in the first row of Table~\ref{tab:Observables1} correspond to a single neutrino flavor. Summing over all three flavors increases these values by a factor of three, provided that long-distance effects from double charged-current interactions are neglected. In particular, the branching fraction $\mathcal{B}(\Xi_b\to \Xi\nu\bar\nu)$ can reach the $10^{-5}$ level, comparable to the Belle II measurement of $\mathcal{B}(B^+\to K^+ \nu\bar\nu)$~\cite{Belle-II:2023esi}.

  \item
  Significant differences arise between the longitudinal polarization fractions $F_L$ in the neutrino and electron channels. For the dineutrino modes, $F_L$ decreases monotonically with increasing $q^2$ over the entire kinematic range. By contrast, in the electron case, $F_L$ first rises to a maximum and then falls toward zero, as discussed previously. The two $F_L$ curves cross at approximately $q^2=2.0~\text{GeV}^2$. At zero recoil, both channels satisfy $F_L = 1/3$, in accordance with our earlier discussion. However, in the large-recoil limit, the neutrino longitudinal polarization tends toward unity, as shown in the second panel of Fig.~\ref{fig:FLv}. This behavior can be traced to the absence of the photon-penguin contribution in the dineutrino modes, which plays a substantial role in shaping the low-$q^2$ distribution in the dilepton case. Substituting the transversity amplitudes for the dineutrino channel from Eq.~(\ref{eq:ampv}) into the definition of $F_L$ in Eq.~(\ref{eq:o1}), one obtains
  \begin{eqnarray}\label{eq:FLv}
  F_L=\frac{(M+m)^2s_-f_0^2+(M-m)^2s_+g_0^2}
  {(M+m)^2s_-f_0^2+(M-m)^2s_+g_0^2+2q^2(s_-f_\perp^2+s_+g_\perp^2)},
  \end{eqnarray}
  where the overall Wilson coefficient $C_L$ cancels in the ratio. At small $q^2$, the last term in the denominator is strongly suppressed relative to the others and vanishes exactly at $q^2=0$, yielding $F_L=1$.

  \item
  The $q^2$ dependence of the forward-backward asymmetries also shows clear distinctions between the neutrino and electron channels, except for the hadron-side asymmetry. For the dineutrino modes, neither the lepton-side forward-backward asymmetry nor the combined asymmetry exhibits additional zero-crossing points. The lepton-side asymmetry remains negative throughout the full kinematic region for all three decay channels. By contrast, the sign of the combined forward-backward asymmetry is controlled by the parity-violating parameter $\alpha_h$, resulting in either positive or negative values depending on its sign.

  \item From Eq.~(\ref{eq:o3}), the hadron-side forward-backward asymmetry for the neutrino mode in the SM is given by
  \begin{eqnarray}\label{eq:AFBhv}
   A^h_{\text{FB}}=-\frac{\alpha_h\sqrt{s_+s_-}[(M^2-m^2)f_0g_0+2q^2f_\perp g_\perp]}{(M+m)^2s_-f_0^2+(M-m)^2s_+g_0^2+2q^2s_-f_\perp^2+2q^2s_+g_\perp^2}.
  \end{eqnarray}
  This observable is proportional to the factor $\sqrt{s_+ s_-}$ and therefore vanishes in the zero-recoil limit. Its magnitude is also controlled by the parity-violating parameter $\alpha_h$. The same analytic expression holds for the charged-lepton case when the lepton mass and photon-penguin contributions are neglected. Consequently, the resulting $q^2$ distributions for the electron and neutrino channels are identical in the SM and cannot be distinguished experimentally. This confirms our earlier statement that the ratio $\mathcal{R}(A^h_{\text{FB}})$ is essentially insensitive to the lepton flavor, making it a powerful probe for potential NP.

\end{itemize}

As a final observable, we consider the ratio of the branching fraction of the neutrino channel to that of the electron channel, defined as
\begin{eqnarray}\label{eq:RRR}
 \mathcal{R}^{e/\nu}_{\mathcal{B}_f}=\frac{\mathcal{B}_e}{3\mathcal{B}_\nu}.
\end{eqnarray}
Using the numerical results obtained in the preceding sections, we find that the integrated values of $\mathcal{R}^{e/\nu}_{\mathcal{B}_f}$ for the channels under study lie in the range 0.134-0.152. Further theoretical investigations, together with input from ongoing and future experimental programs, will be essential to  rigorously test these predictions.

\section{Conclusion}\label{sec:sum}
The observables associated with semileptonic baryonic transitions induced by the flavor-changing neutral current processes $b\to s$ and $b\to d$ offer phenomenological insights that are complementary to those obtained from mesonic and inclusive decays. In this work, we conduct a systematic study of the rare weak decays of the $\Xi_b$ baryon for all lepton species, including neutrinos, within the PQCD framework.

We compute ten independent transition form factors in the low-$q^2$ region and extrapolate them to the full kinematic range using a $z$-expansion parametrization. Their values at zero and maximum recoil are presented, and most of them satisfy the endpoint relations expected in the heavy-quark limit and in the large-recoil regime. Notably, our predictions differ substantially from those based on QCD sum rules, which tend to show inconsistencies with heavy-quark symmetry and endpoint behavior. The $q^2$ evolution of the form factors follows the anticipated pattern for weak decays, displaying smooth and steadily increasing magnitudes as $q^2$ grows.

We also analyze long-distance contributions arising from intermediate charmonium resonances in the differential branching ratios. As anticipated, narrow and pronounced peaks appear in the vicinity of the resonance regions once these effects are included. Interestingly, the integrated branching ratios with the resonance regions excluded remain slightly enhanced relative to results obtained without long-distance contributions. The largest branching fractions arise in the $b\to s$ channels, reaching the level of $10^{-6}$. When summed over all three neutrino flavors, the corresponding branching fractions can reach $10^{-5}$, making them promising targets for upcoming experimental searches. By contrast, typical $b\to d$ branching fractions are of order $10^{-7}$ or smaller, presenting a significant challenge for current detectors. 

We present predictions for several LFU ratios, highlighting their sensitivity to potential new physics. The ratios $\mathcal{R}_\mu$ are found to be very close to unity, reflecting the near universality between the electron and muon channels expected in the SM. Consequently, any deviation of $\mathcal{R}_\mu$ at the few-percent level or beyond would provide a clear indication of physics beyond the SM. The integrated values of $\mathcal{R}_\tau$ are typically around 0.6, consistent with analogous results in charged-current semileptonic decays. Precision measurements of these LFU ratios would therefore serve as important probes of lepton-flavor universality in FCNC baryonic transitions.

Employing the transversity basis and assuming unpolarized $\Xi_b$ baryons, we perform a comprehensive angular analysis beginning with the full fourfold differential decay distribution in $q^2$ and the three helicity angles. A broad array of angular observables are expressed in terms of short-distance Wilson coefficients and hadronic form factors. For the decays $\Xi_b\to (\Xi,\Sigma,\Lambda)(\ell^+\ell^-,\nu\bar\nu)$, we provide numerical predictions for the longitudinal polarization fraction and the forward-backward asymmetries defined on the lepton side, the hadron side, and their combination. These results are given both as functions of $q^2$ and as integrated quantities. A key observation is that many of these observables exhibit remarkably weak dependence on hadronic form factors, underscoring their potential as robust and relatively model-independent probes of the underlying Wilson coefficients. Conversely, they exhibit pronounced sensitivity to new-physics scenarios that modify the structure of the effective weak Hamiltonian. For instance, the zero-crossing points of the lepton-side and combined forward-backward asymmetries shift significantly in the presence of right-handed current contributions.

Finally, we emphasize that this work represents the first detailed investigation of semileptonic FCNC $\Xi_b$ decays involving neutrinos in the final state. A combined analysis of $b\to s\ell^+\ell^-$ and $b\to s\nu\bar\nu$ modes is theoretically well motivated because these channels are tightly constrained by $SU(2)_L$ gauge symmetry in both the SM and its extensions. Correlating observables across the two classes of baryonic transitions may prove valuable for understanding the anomalous behavior observed in $B$-meson decays. The predictions presented here will therefore provide a useful benchmark for future theoretical developments and experimental measurements.

\section*{Acknowledgments}
This work is supported in part by the National Science Foundation of China under the Grants No. 12375089, No.12435004, and No.12075086, and the Natural Science Foundation of Shandong province under the Grant No. ZR2022ZD26 and ZR2022MA035.

\begin{appendix}

\section{Factorization formulas }\label{sec:for}
\begin{figure}[htbp]
	\begin{center}
    \centerline{\epsfxsize=10cm \epsffile{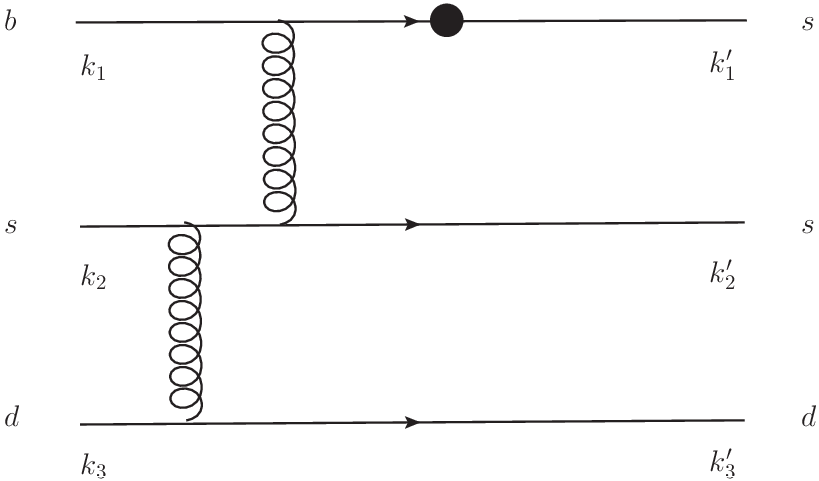}}
     \vspace{0.5cm}
     \caption{ Sample topological diagram for $\Xi^-_b\to \Xi^-$ transition form factors.
The symbol $\bullet$ denotes the electroweak vertex, from which the lepton pair emerges.}
		\label{fig:C}
	\end{center}
\end{figure}
In PQCD formulism, the formulas of $H_\xi(x_i,x'_i)$ in Eq.~(\ref{eq:FG})  are rather lengthy due to many higher twist LCDAs are included.
Here, we only give some details for a sample topological diagram in $\Xi^-_b\to \Xi^-$ transition as shown in Fig.~\ref{fig:C}. 
For a more complete set of Feynman diagrams, one refers to our previous work~\cite{Rui:2025bsu}.
The generalization to the calculation of other diagrams is straightforward.
The calculations are performed in the $\mathcal{B}_i$ rest frame  with the daughter baryon $\mathcal{B}_f$ moving in the plus direction.
The momentum $p$ and $p'$ are chosen, in the light-cone coordinates, as
\begin{eqnarray}
p=\frac{M}{\sqrt{2}}\left(1,1,\textbf{0}_{T}\right), \quad p'=\frac{M}{\sqrt{2}}\left(f^+,f^-,\textbf{0}_{T}\right),
\end{eqnarray}
where the factors $f^{\pm}=(f\pm \sqrt{f^2-1})r$ are defined in terms of the velocity transfer $f=v \cdot v'$ with $v=p/M$ and $v'=p'/m$.
The partonic momenta inside baryons are parametrized as
\begin{eqnarray}
k_{2,3}=\left(0,\frac{M}{\sqrt{2}}x_{2,3},\textbf{k}_{iT}\right), \quad k_1=p-k_2-k_3, \quad
  k'_i&=&\left(\frac{M}{\sqrt{2}}f^+x'_i,0,\textbf{k}'_{iT}\right).
\end{eqnarray}
where only $b$ quark is considered to be massive, and other light quarks are treated as massless.
Collected all above ingredients, we obtain the tensor form factors $ H^{F_i^T}_a(x_i,x'_i)$ at large recoil limit
\begin{eqnarray}
H^{F^T_1}_a(x_i,x'_i)&=&M^4 (r+1) (r^2 (-(x_2+x_3)) \phi _2 x'_1 (2 \mathcal{A}_3-4 \mathcal{A}_5-\mathcal{P}_1-2 \mathcal{P}_2+\mathcal{S}_1-2
   \mathcal{S}_2\nonumber\\&&+2 (\mathcal{T}_5-\mathcal{T}_6+\mathcal{V}_3+2 \mathcal{V}_5)+\mathcal{T}_1-\mathcal{T}_2-\mathcal{T}_4)\nonumber\\&&-2 r \phi _{3 s}
   ((x_2+x_3) x'_1 (2 \mathcal{A}_3+\mathcal{P}_1+\mathcal{S}_1-\mathcal{T}_1-\mathcal{T}_2+4 \mathcal{T}_3+\mathcal{T}_4-2
   \mathcal{V}_3)\nonumber\\&&-x'_3 (\mathcal{A}_1+2 \mathcal{A}_2+2 \mathcal{A}_3-2 \mathcal{T}_1+8 \mathcal{T}_3+4 \mathcal{T}_4+\mathcal{V}_1-2
   (\mathcal{V}_2+\mathcal{V}_3)))\nonumber\\&&+\phi _4 x'_3 (-\mathcal{A}_1+2 \mathcal{T}_1+\mathcal{V}_1)),
\end{eqnarray}
\begin{eqnarray}
H^{F^T_2}_a(x_i,x'_i)&=&\frac{M^4}{r+1} ((x_2+x_3) \phi _2 (2 (\mathcal{A}_3-2 \mathcal{A}_5+\mathcal{T}_1-\mathcal{T}_2-\mathcal{T}_4+2 \mathcal{T}_5-2
   \mathcal{T}_6+\mathcal{V}_3+2 \mathcal{V}_5)\nonumber\\&&+(-(2 r+1) \mathcal{P}_1-2 (2 r+1) \mathcal{P}_2+\mathcal{S}_1-2
   \mathcal{S}_2-\mathcal{T}_1+\mathcal{T}_2+\mathcal{T}_4-2 \mathcal{T}_5+2 \mathcal{T}_6\nonumber\\&&+r (\mathcal{A}_1+2 \mathcal{A}_2+4 \mathcal{A}_3+2
   \mathcal{A}_4-6 \mathcal{A}_5+4 \mathcal{A}_6+2 \mathcal{S}_1-4 \mathcal{S}_2-2 \mathcal{T}_2+2 \mathcal{T}_4\nonumber\\&&-4 \mathcal{T}_6+8
   \mathcal{T}_8-\mathcal{V}_1+2 \mathcal{V}_2+4 \mathcal{V}_3-2 \mathcal{V}_4+6 \mathcal{V}_5+4 \mathcal{V}_6)) x'_1) r^2\nonumber\\&&+2 \phi _{3 s}
   (-2 x_2 \mathcal{T}_1-2 x_3 \mathcal{T}_1+x_2 x'_1 \mathcal{T}_1+x_3 x'_1 \mathcal{T}_1-2 x_2 \mathcal{T}_2-2 x_3 \mathcal{T}_2\nonumber\\&&+8 x_2 \mathcal{T}_3+8
   x_3 \mathcal{T}_3+2 x_2 \mathcal{T}_4+2 x_3 \mathcal{T}_4-2 x_2 \mathcal{V}_3-2 x_3 \mathcal{V}_3-r x_2 \mathcal{A}_1 x'_1-r x_3 \mathcal{A}_1 x'_1\nonumber\\&&-2 r
   x_2 \mathcal{A}_4 x'_1-2 r x_3 \mathcal{A}_4 x'_1+2 r x_2 \mathcal{A}_5 x'_1+2 r x_3 \mathcal{A}_5 x'_1+2 r x_2 \mathcal{P}_1 x'_1+x_2 \mathcal{P}_1
   x'_1\nonumber\\&&+2 r x_3 \mathcal{P}_1 x'_1+x_3 \mathcal{P}_1 x'_1+2 r x_2 \mathcal{S}_1 x'_1+x_2 \mathcal{S}_1 x'_1+2 r x_3 \mathcal{S}_1 x'_1+x_3 \mathcal{S}_1
   x'_1\nonumber\\&&-2 r x_2 \mathcal{T}_2 x'_1+x_2 \mathcal{T}_2 x'_1-2 r x_3 \mathcal{T}_2 x'_1+x_3 \mathcal{T}_2 x'_1-4 x_2 \mathcal{T}_3 x'_1-4 x_3 \mathcal{T}_3
   x'_1\nonumber\\&&+2 r x_2 \mathcal{T}_4 x'_1-x_2 \mathcal{T}_4 x'_1+2 r x_3 \mathcal{T}_4 x'_1-x_3 \mathcal{T}_4 x'_1+4 r x_2 \mathcal{T}_5 x'_1+4 r x_3 \mathcal{T}_5
   x'_1\nonumber\\&&+16 r x_2 \mathcal{T}_7 x'_1+16 r x_3 \mathcal{T}_7 x'_1-r x_2 \mathcal{V}_1 x'_1-r x_3 \mathcal{V}_1 x'_1-2 r x_2 \mathcal{V}_3 x'_1-2 r x_3
   \mathcal{V}_3 x'_1\nonumber\\&&-2 r x_2 \mathcal{V}_4 x'_1-2 r x_3 \mathcal{V}_4 x'_1+2 r x_2 \mathcal{V}_5 x'_1+2 r x_3 \mathcal{V}_5 x'_1+2 (x_2+x_3)
   \mathcal{A}_3 (r x'_1+1)\nonumber\\&&-2 (r+1) \mathcal{A}_3 x'_3+(\mathcal{A}_1 x'_1 r^2+(-4 \mathcal{A}_5+\mathcal{P}_1+2
   \mathcal{P}_2+\mathcal{S}_1\nonumber\\&&-2 \mathcal{S}_2+3 \mathcal{T}_1+\mathcal{T}_2-12 \mathcal{T}_3-7 \mathcal{T}_4-2 (\mathcal{T}_5-\mathcal{T}_6+4
   \mathcal{T}_7+\mathcal{V}_1\nonumber\\&&-2 \mathcal{V}_2-\mathcal{V}_3+2 \mathcal{V}_5)+2 \mathcal{A}_2 (r (x'_1-1)-2)+r (2
   \mathcal{A}_4+2 \mathcal{A}_5\nonumber\\&&+4 \mathcal{A}_6+\mathcal{V}_1+2 (-\mathcal{V}_2+\mathcal{V}_4+\mathcal{V}_5-2 \mathcal{V}_6))
   (x'_1-1)) r-(r+1)^2 \mathcal{A}_1-2 \mathcal{A}_2\nonumber\\&&+2 \mathcal{T}_1-8 \mathcal{T}_3-4 \mathcal{T}_4-\mathcal{V}_1+2
   (\mathcal{V}_2+\mathcal{V}_3)) x'_3) r-\phi _4 (\mathcal{A}_1 x'_1 r^2+(\mathcal{P}_1-\mathcal{S}_1\nonumber\\&&+3
   \mathcal{T}_1-\mathcal{T}_2+\mathcal{T}_4+2 \mathcal{V}_1-2 \mathcal{V}_3+2 \mathcal{A}_3 (r (x'_1-1)-1)\nonumber\\&&+r (2 \mathcal{A}_4-2
   \mathcal{A}_5-\mathcal{V}_1+2 (\mathcal{V}_3-\mathcal{V}_4+\mathcal{V}_5)) (x'_1-1)) r\nonumber\\&&-(r+1)^2 \mathcal{A}_1+2
   \mathcal{T}_1+\mathcal{V}_1) x'_3),
\end{eqnarray}
\begin{eqnarray}
H^{F^T_3}_a(x_i,x'_i)&=&\frac{M^4}{r+1} ((x_2+x_3) \phi _2 (2 (\mathcal{A}_3-2 \mathcal{A}_5+\mathcal{T}_1-\mathcal{T}_2-\mathcal{T}_4+2 \mathcal{T}_5-2
   \mathcal{T}_6+\mathcal{V}_3+2 \mathcal{V}_5)\nonumber\\&&+(-(2 r+1) \mathcal{P}_1-2 (2 r+1) \mathcal{P}_2+\mathcal{S}_1-2
   \mathcal{S}_2-\mathcal{T}_1+\mathcal{T}_2+\mathcal{T}_4-2 \mathcal{T}_5+2 \mathcal{T}_6\nonumber\\&&+r (\mathcal{A}_1+2 \mathcal{A}_2+4 \mathcal{A}_3+2
   \mathcal{A}_4-6 \mathcal{A}_5+4 \mathcal{A}_6+2 \mathcal{S}_1-4 \mathcal{S}_2-2 \mathcal{T}_2+2 \mathcal{T}_4-4 \mathcal{T}_6\nonumber\\&&+8
   \mathcal{T}_8-\mathcal{V}_1+2 \mathcal{V}_2+4 \mathcal{V}_3-2 \mathcal{V}_4+6 \mathcal{V}_5+4 \mathcal{V}_6)) x'_1) r^2\nonumber\\&&+2 \phi _{3 s}
   (-2 x_2 \mathcal{T}_1-2 x_3 \mathcal{T}_1+x_2 x'_1 \mathcal{T}_1+x_3 x'_1 \mathcal{T}_1-2 x_2 \mathcal{T}_2-2 x_3 \mathcal{T}_2+8 x_2 \mathcal{T}_3\nonumber\\&&+8
   x_3 \mathcal{T}_3+2 x_2 \mathcal{T}_4+2 x_3 \mathcal{T}_4-2 x_2 \mathcal{V}_3-2 x_3 \mathcal{V}_3-r x_2 \mathcal{A}_1 x'_1-r x_3 \mathcal{A}_1 x'_1\nonumber\\&&-2 r
   x_2 \mathcal{A}_4 x'_1-2 r x_3 \mathcal{A}_4 x'_1+2 r x_2 \mathcal{A}_5 x'_1+2 r x_3 \mathcal{A}_5 x'_1+2 r x_2 \mathcal{P}_1 x'_1+x_2 \mathcal{P}_1
   x'_1\nonumber\\&&+2 r x_3 \mathcal{P}_1 x'_1+x_3 \mathcal{P}_1 x'_1+2 r x_2 \mathcal{S}_1 x'_1+x_2 \mathcal{S}_1 x'_1+2 r x_3 \mathcal{S}_1 x'_1+x_3 \mathcal{S}_1
   x'_1\nonumber\\&&-2 r x_2 \mathcal{T}_2 x'_1+x_2 \mathcal{T}_2 x'_1-2 r x_3 \mathcal{T}_2 x'_1+x_3 \mathcal{T}_2 x'_1-4 x_2 \mathcal{T}_3 x'_1-4 x_3 \mathcal{T}_3
   x'_1\nonumber\\&&+2 r x_2 \mathcal{T}_4 x'_1-x_2 \mathcal{T}_4 x'_1+2 r x_3 \mathcal{T}_4 x'_1-x_3 \mathcal{T}_4 x'_1+4 r x_2 \mathcal{T}_5 x'_1+4 r x_3 \mathcal{T}_5
   x'_1\nonumber\\&&+16 r x_2 \mathcal{T}_7 x'_1+16 r x_3 \mathcal{T}_7 x'_1-r x_2 \mathcal{V}_1 x'_1-r x_3 \mathcal{V}_1 x'_1-2 r x_2 \mathcal{V}_3 x'_1-2 r x_3
   \mathcal{V}_3 x'_1\nonumber\\&&-2 r x_2 \mathcal{V}_4 x'_1-2 r x_3 \mathcal{V}_4 x'_1+2 r x_2 \mathcal{V}_5 x'_1+2 r x_3 \mathcal{V}_5 x'_1+2 (x_2+x_3)
   \mathcal{A}_3 (r x'_1+1)\nonumber\\&&-2 (r+1) \mathcal{A}_3 x'_3+(\mathcal{A}_1 x'_1 r^2+(-4 \mathcal{A}_5+\mathcal{P}_1+2
   \mathcal{P}_2+\mathcal{S}_1\nonumber\\&&-2 \mathcal{S}_2+3 \mathcal{T}_1+\mathcal{T}_2-12 \mathcal{T}_3-7 \mathcal{T}_4-2 (\mathcal{T}_5-\mathcal{T}_6+4
   \mathcal{T}_7+\mathcal{V}_1\nonumber\\&&-2 \mathcal{V}_2-\mathcal{V}_3+2 \mathcal{V}_5)+2 \mathcal{A}_2 (r (x'_1-1)-2)+r (2
   \mathcal{A}_4+2 \mathcal{A}_5\nonumber\\&&+4 \mathcal{A}_6+\mathcal{V}_1+2 (-\mathcal{V}_2+\mathcal{V}_4+\mathcal{V}_5-2 \mathcal{V}_6))
   (x'_1-1)) r-(r+1)^2 \mathcal{A}_1\nonumber\\&&-2 \mathcal{A}_2+2 \mathcal{T}_1-8 \mathcal{T}_3-4 \mathcal{T}_4-\mathcal{V}_1+2
   (\mathcal{V}_2+\mathcal{V}_3)) x'_3) r\nonumber\\&&-\phi _4 (\mathcal{A}_1 x'_1 r^2+(\mathcal{P}_1-\mathcal{S}_1+3
   \mathcal{T}_1-\mathcal{T}_2+\mathcal{T}_4+2 \mathcal{V}_1-2 \mathcal{V}_3+2 \mathcal{A}_3 (r (x'_1-1)-1)\nonumber\\&&+r (2 \mathcal{A}_4-2
   \mathcal{A}_5-\mathcal{V}_1+2 (\mathcal{V}_3-\mathcal{V}_4+\mathcal{V}_5)) (x'_1-1)) r\nonumber\\&&-(r+1)^2 \mathcal{A}_1+2
   \mathcal{T}_1+\mathcal{V}_1) x'_3).
\end{eqnarray}

The corresponding formulas for the  pseudotensor  ones can be obtained by the following replacement:
\begin{eqnarray}
H^{G^T_i}_\xi&=&\mp H^{F^T_i}_\xi|_{r\to -r,\quad \phi _{3s}\to -\phi _{3s}},
\end{eqnarray}
where the minus and plus signs refer to  $i=1$ and $i=2,3$, respectively.

\section{Light-cone distribution amplitudes}\label{sec:LCDAsp}
Similar to $\Lambda_b$ baryon LCDAs~\cite{Ball:2008fw,Bell:2013tfa,Ali:2012pn,Braun:2014npa,Wang:2015ndk,Ali:2012zza},
the LCDAs of $\Xi_b$ baryon up to twist-4 accuracy in the momentum space can be written as~\cite{Rui:2025iwa,Rui:2023fiz}
\begin{eqnarray}\label{eq:LCDAs20}
 (\Psi_{\mathcal{B}_i})_{\alpha\beta\gamma}(x_i) &=&\frac{1}{8N_c}\{\frac{ f^{(1)}}{\sqrt{2}}
[(\slashed{ n}\gamma_5C)_{\alpha\beta}\phi_2(x_2,x_3)+(\slashed{ v}\gamma_5C)_{\alpha\beta}\phi_4(x_2,x_3)](u_{\mathcal{B}_i})_\gamma \nonumber\\
&&+f^{(2)}[(\gamma_5C)_{\alpha\beta}\phi_{3s}(x_2,x_3)-\frac{i}{2}(\sigma_{ nv}\gamma_5C)_{\alpha\beta}\phi_{3a}(x_2,x_3)](u_{\mathcal{B}_i})_\gamma\},
\end{eqnarray}
with two light-cone vectors $n=(1,0,\textbf{0}_T)$ and $v=(0,1,\textbf{0}_T)$ satisfying $n\cdot v=1$.
$C$ is the charge conjugation matrix and $N_c$ is the number of colors.
The decay constants for the $\Xi_b$ baryon take the values  $f^{(1,2)}_{\Xi_b}=(0.032\pm 0.009)$ GeV$^3$ from the QCD sum rules~\cite{Wang:2010fq}.
The functions $\phi_{ 2,4,3s,3a}$ distinguish four LCDAs with different asymptotic behaviors, and the numbers in the subscript denote the corresponding twist.
Their specific forms in Gegenbauer model are given by
\begin{eqnarray}\label{eq:g2}
\phi_2  (\omega,u)&=&   \omega^2u(1-u) \sum_{l=0}^2\frac{a_l}{\epsilon_l^4}\frac{c_l^{3/2}(2u-1)}{|c_l^{3/2}|^2}e^{-\frac{\omega}{\epsilon_l}},\nonumber\\
\phi_{3s,3a}(\omega,u)&=& \frac{\omega}{2}  \sum_{l=0}^2\frac{a_l}{\epsilon_l^3}\frac{c_l^{1/2}(2u-1)}{|c_l^{1/2}|^2}e^{-\frac{\omega}{\epsilon_l}},\nonumber\\
\phi_4  (\omega,u)&=&   \sum_{l=0}^2\frac{a_l}{\epsilon_l^2}\frac{c_l^{1/2}(2u-1)}{|c_l^{1/2}|^2}e^{-\frac{\omega}{\epsilon_l}},
\end{eqnarray}
with $\omega=(x_2+x_3)M$ and $u=x_2/(x_2+x_3)$.
The Gegenbauer polynomials $c_l$ read as~\cite{Ali:2012pn}
\begin{eqnarray}
[c_0^{z}(x),c_1^{z}(x), c_2^{z}(x)]&=&[1, \quad 2zx,\quad 2z(1+z)x^2-z],\nonumber\\
(|c_0^{1/2}|^2,|c_1^{1/2}|^2,|c_2^{1/2}|^2)&=&(1,\frac{1}{3},\frac{1}{5}),\nonumber\\
(|c_0^{3/2}|^2,|c_1^{3/2}|^2,|c_2^{3/2}|^2)&=&(1,3,6).
\end{eqnarray}	
The two shape parameters $a_l$ and $\epsilon_l$ in Eq.~(\ref{eq:g2})
dependence on a free parameter $A=0.5\pm 0.2$ have been determined in~\cite{Ali:2012pn},
and we collect them in Table.~\ref{tab:LCDAss}  to make the paper self-contained.
%
\begin{table}[htbp]
	\caption{Shape parameters entering the LCDAs of $\Xi_b$ baryons at the scale $\mu=1.0$~GeV.}
	\label{tab:LCDAss}
	\centering
	\begin{tabular}{lcccc}
		\hline\hline
		Twist & $\phi_2$ & $\phi_{3s}$ & $\phi_{3a}$ & $\phi_4$ \\
		\hline
		$a_0$
		& $1$
		& $1$
		& $\dfrac{-0.16(1-A)+0.16}{(1-A)-1.3}$
		& $1$ \\[0.6em]
		$a_1$
		& $\dfrac{0.25(1-A)+0.46}{(1-A)+0.68}$
		& $\dfrac{0.04(1-A)-0.14}{(1-A)-1.6}$
		& $1$
		& $\dfrac{0.03(1-A)+0.11}{(1-A)+0.16}$ \\[0.8em]
		$a_2$
		& $\dfrac{6.(1-A)+0.6}{(1-A)+0.68}$
		& $\dfrac{0.12(1-A)-0.09}{(1-A)-1.6}$
		& $\dfrac{0.17(1-A)-0.17}{(1-A)-1.3}$
		& $\dfrac{-0.1(1-A)-0.03}{(1-A)+0.61}$ \\[0.8em]
		$\epsilon_0$ [GeV]
		& $\dfrac{1.4(1-A)+1}{(1-A)+6.7}$
		& $\dfrac{0.56(1-A)-0.91}{(1-A)-2.9}$
		& $\dfrac{0.11(1-A)-0.11}{(1-A)-1}$
		& $\dfrac{0.63(1-A)+0.38}{(1-A)+1.3}$ \\[0.8em]
		$\epsilon_1$ [GeV]
		& $\dfrac{0.57(1-A)+1.1}{(1-A)+4}$
		& $\dfrac{-27(1-A)+92}{160}$
		& $\dfrac{0.39(1-A)-0.49}{(1-A)-1.3}$
		& $\dfrac{-0.82(1-A)-3.1}{(1-A)-3.9}$ \\[0.8em]
		$\epsilon_2$ [GeV]
		& $\dfrac{0.36(1-A)+0.03}{(1-A)-0.02}$
		& $\dfrac{0.3(1-A)-0.24}{(1-A)+0.54}$
		& $\dfrac{0.33(1-A)-0.33}{(1-A)-1}$
		& $\dfrac{1.2(1-A)+0.34}{(1-A)+4.1}$ \\
		\hline\hline
	\end{tabular}
\end{table}

The LCDAs for the octet have been updated to twist six in the QCDSR~\cite{Braun:2000kw}.
The corresponding momentum-space projector for an outgoing light baryon can be written as
 \begin{eqnarray}\label{eq:LCDAsp}
 (\bar{\Psi}_{\mathcal{B}_f})_{\alpha\beta\gamma}(x'_i) &=&  -\frac{1}{8\sqrt{2}N_c} [ \mathcal{S}_1m( \bar{u}_{\mathcal{B}_f}\gamma_5)_\gamma C_{\beta\alpha}+
\mathcal{S}_2m^2( \bar{u}_{\mathcal{B}_f}\gamma_5\rlap{/}{z})_\gamma C_{\beta\alpha}+  \mathcal{P}_1m( \bar{u}_{\mathcal{B}_f})_\gamma (C\gamma_5)_{\beta\alpha}   \nonumber\\&&
+ \mathcal{P}_2m^2( \bar{u}_{\mathcal{B}_f}\rlap{/}{z})_\gamma (C\gamma_5)_{\beta\alpha} +\mathcal{V}_1( \bar{u}_{\mathcal{B}_f}\gamma_5)_\gamma (C\rlap{/}{p'})_{\beta\alpha}
+\mathcal{V}_2m( \bar{u}_{\mathcal{B}_f}\gamma_5\rlap{/}{z})_\gamma (C\rlap{/}{p'})_{\beta\alpha} \nonumber\\&& +\mathcal{V}_3m( \bar{u}_{\mathcal{B}_f}\gamma_5 \gamma_\mu)_\gamma (C\gamma^\mu)_{\beta\alpha}+\mathcal{V}_4m^2( \bar{u}_{\mathcal{B}_f}\gamma_5)_\gamma (C\rlap{/}{z})_{\beta\alpha}
-i\mathcal{V}_5m^2(  \bar{u}_{\mathcal{B}_f}\gamma_5\sigma^{\mu\nu}z_{\nu})_\gamma( C\gamma_\mu)_{\beta\alpha} \nonumber\\&&
+\mathcal{V}_6m^3( \bar{u}_{\mathcal{B}_f}\gamma_5\rlap{/}{z})_\gamma (C\rlap{/}{z})_{\beta\alpha}  +\mathcal{A}_1( \bar{u}_{\mathcal{B}_f})_\gamma (C\gamma_5\rlap{/}{p'})_{\beta\alpha}
+\mathcal{A}_2m( \bar{u}_{\mathcal{B}_f}\rlap{/}{z})_\gamma (C\gamma_5\rlap{/}{p'})_{\beta\alpha} \nonumber\\&&+\mathcal{A}_3m( \bar{u}_{\mathcal{B}_f}\gamma_\mu)_\gamma (C\gamma_5\gamma^\mu)_{\beta\alpha}+\mathcal{A}_4m^2( \bar{u}_{\mathcal{B}_f})_\gamma (C\gamma_5\rlap{/}{z})_{\beta\alpha}
-i\mathcal{A}_5m^2(  \bar{u}_{\mathcal{B}_f}\sigma^{\mu\nu}z_{\nu})_\gamma( C\gamma_5\gamma_\mu)_{\beta\alpha} \nonumber\\&&
+\mathcal{A}_6m^3( \bar{u}_{\mathcal{B}_f}\rlap{/}{z})_\gamma (C\gamma_5\rlap{/}{z})_{\beta\alpha}
-i\mathcal{T}_1( \bar{u}_{\mathcal{B}_f} \gamma_5\gamma^{\mu})_\gamma( C \sigma_{\mu\nu}p'^\nu)_{\beta\alpha}
-i\mathcal{T}_2m( \bar{u}_{\mathcal{B}_f} \gamma_5)_\gamma( C \sigma_{\mu\nu}p'^\mu z^\nu)_{\beta\alpha} \nonumber\\&&
+\mathcal{T}_3m( \bar{u}_{\mathcal{B}_f} \gamma_5 \sigma^{\mu\nu} )_\gamma( C \sigma_{\mu\nu})_{\beta\alpha}
+\mathcal{T}_4m( \bar{u}_{\mathcal{B}_f} \gamma_5 \sigma^{\mu\rho}z_\rho )_\gamma( C \sigma_{\mu\nu}p'^\nu)_{\beta\alpha}\nonumber\\&&
-i\mathcal{T}_5m^2( \bar{u}_{\mathcal{B}_f} \gamma_5\gamma^{\mu})_\gamma( C \sigma_{\mu\nu}z^\nu)_{\beta\alpha}
-i\mathcal{T}_6m^2( \bar{u}_{\mathcal{B}_f} \gamma_5\rlap{/}{z})_\gamma( C \sigma_{\mu\nu}z^\mu p'^\nu)_{\beta\alpha} \nonumber\\&&
+\mathcal{T}_7m^2( \bar{u}_{\mathcal{B}_f} \gamma_5\rlap{/}{z} \sigma^{\mu\rho} )_\gamma( C \sigma_{\mu\nu})_{\beta\alpha}
+\mathcal{T}_8m^3( \bar{u}_{\mathcal{B}_f} \gamma_5 \sigma^{\mu\rho} z_\rho)_\gamma( C \sigma_{\mu\nu}z^\nu)_{\beta\alpha}],
\end{eqnarray}
where $z$ is an arbitrary lightlike vector with $z^2=0$.
We choose $z=\frac{\sqrt{2}}{Mf^+}v$ so that $z\cdot p'=1$~\cite{Braun:1999te}.
Note that the ``calligraphic" invariant functions  $\mathcal{S}_i$, $\mathcal{P}_i$, $\mathcal{V}_i$, $\mathcal{A}_i$,  and $\mathcal{T}_i$
in Eq.~(\ref{eq:LCDAsp}) do not have a definite twist.
They are related to the ones with definite twist and symmetry by five matrixes as follows~\cite{Braun:1999te,Liu:2008yg}

\begin{eqnarray}
\left( \begin{array}{c}  \mathcal{S}_1\\ 2 p'\cdot z\mathcal{S}_2 \\ \end{array} \right)=
\left( \begin{array}{cc} 1&0 \\ 1&-1 \\ \end{array} \right)
\left( \begin{array}{l} S_1\\ S_2 \\ \end{array} \right),
\end{eqnarray}
\begin{eqnarray}
\left( \begin{array}{c}  \mathcal{P}_1\\ 2 p'\cdot z\mathcal{P}_2 \\ \end{array} \right)=
\left( \begin{array}{cc} 1&0 \\ -1&1 \\ \end{array} \right)
\left( \begin{array}{l} P_1\\ P_2 \\ \end{array} \right),
\end{eqnarray}
\begin{eqnarray}
\left( \begin{array}{c}  \mathcal{V}_1\\ 2 p'\cdot z\mathcal{V}_2 \\ 2\mathcal{V}_3\\ 4 p'\cdot z\mathcal{V}_4\\4 p'\cdot z\mathcal{V}_5\\ (2 p'\cdot z)^2\mathcal{V}_6\\\end{array} \right)=
\left( \begin{array}{cccccc} 1&0&0&0&0&0 \\ 1&-1&-1&0&0&0 \\ 0&0&1&0&0&0 \\-2&0&1&1&2&0 \\0&0&-1&1&0&0 \\-1&1&1&1&1&-1 \\ \end{array} \right)
\left( \begin{array}{l} V_1\\ V_2 \\ V_3\\ V_4 \\V_5\\ V_6 \\\end{array} \right),
\end{eqnarray}
\begin{eqnarray}
\left( \begin{array}{c}  \mathcal{A}_1\\ 2 p'\cdot z\mathcal{A}_2 \\ 2\mathcal{A}_3\\ 4 p'\cdot z\mathcal{A}_4\\4 p'\cdot z\mathcal{A}_5\\ (2 p'\cdot z)^2\mathcal{A}_6\\\end{array} \right)=
\left( \begin{array}{cccccc} 1&0&0&0&0&0 \\ -1&1&-1&0&0&0 \\ 0&0&1&0&0&0 \\-2&0&-1&-1&2&0 \\0&0&1&-1&0&0 \\1&-1&1&1&-1&1 \\ \end{array} \right)
\left( \begin{array}{l} A_1\\ A_2 \\ A_3\\ A_4 \\A_5\\ A_6 \\\end{array} \right),
\end{eqnarray}
\begin{eqnarray}
\left( \begin{array}{c}  \mathcal{T}_1\\ 2 p'\cdot z\mathcal{T}_2 \\ 2\mathcal{T}_3\\ 2 p'\cdot z\mathcal{T}_4\\2 p'\cdot z\mathcal{T}_5\\
(2 p'\cdot z)^2\mathcal{T}_6\\4 p'\cdot z\mathcal{T}_7\\ (2 p'\cdot z)^2\mathcal{T}_8\\\end{array} \right)=
\left( \begin{array}{cccccccc} 1&0&0&0&0&0 &0&0\\ 1&1&-2&0&0&0&0&0\\0&0&0&0&0&0&1&0\\1&-1&0&0&0&0&-2&0\\-1&0&0&0&1&0&0&2\\0&2&-2&-2&2&0&2&2\\0&0&0&0&0&0&1&-1\\-1&1&0&0&1&-1&2&2\\ \end{array} \right)
\left( \begin{array}{l} T_1\\ T_2 \\ T_3\\ T_4 \\T_5\\ T_6 \\T_7\\ T_8 \\\end{array} \right).
\end{eqnarray}
The explicit expressions of various twist LCDAs $S_i$, $P_i$, $V_i$, $A_i$,  and $T_i$ to leading order conformal spin expansion for $\Xi$, $\Sigma$, and $\Lambda$ baryons can be found in~\cite{Liu:2009uc,Liu:2008yg}  which will not be duplicated here.

\section{ Form factor parametrization}\label{sec:LCDAs}
The PQCD predictions for the form factors are reliable up to a limited region of the momentum transfer.
In order to extend the PQCD predictions to the whole physical region,
we follow~\cite{Faustov:2017wbh} and parametrize the Weinberg form factors using a $z$-expansion
\begin{eqnarray}
\mathcal{F}(q^2)&=& \frac{a_0+a_1z(q^2)}{1-q^2/M^2_{\text{pole}}},\nonumber\\
z(q^2)&=&\frac{\sqrt{t^2_+-q^2}-\sqrt{t^2_+-t^2_0}}{\sqrt{t^2_+-q^2}+\sqrt{t^2_+-t^2_0}},
\end{eqnarray}
with $t^2_0=q^2_{\text{max}}=(M-m)^2$.   $t_+$ is set equal to the threshold of two-particle states $BK(\pi)$ for the $b\to s(d)$ transition.
Such choice maps the complex $q^2$ plane, cut along the real axis for $q^2\geq t_+$, onto the disk $z<1$.
$M_{\text{pole}}$ denotes the mass of the low-lying $B$ resonances and the relevant inputs with appropriate spin are taken as~\cite{ParticleDataGroup:2024cfk}
\begin{eqnarray}
M_{\text{pole}}= \left\{
            \begin{array}{ll}
               B_s (1^-):5.415~\text{GeV}, & \text{for}~ f_{1,2}^{(T)} \\
               B_s (1^+):5.829~\text{GeV}, & \text{for}~ g_{1,2}^{(T)} \\
               B_s (0^+):5.833~\text{GeV}, & \text{for}~ f_{3} \\
               B_s (0^-):5.367~\text{GeV}, & \text{for}~ g_{3} ,\\
            \end{array}
          \right.
\end{eqnarray}
for the $b\to s$ transition, and
\begin{eqnarray}
M_{\text{pole}}= \left\{
            \begin{array}{ll}
               B (1^-):5.325~\text{GeV}, & \text{for}~ f_{1,2}^{(T)} \\
               B (1^+):5.726~\text{GeV}, & \text{for}~ g_{1,2}^{(T)} \\
               B (0^+):5.749~\text{GeV}, & \text{for}~ f_{3} \\
               B (0^-):5.28~\text{GeV}, & \text{for}~ g_{3} ,\\
            \end{array}
          \right.
\end{eqnarray}
for $b\to d$ transition.
Note that the scalar $B$ and $B_s$ mesons have not been observed experimentally yet.
The mass of the former is taken from~\cite{Faustov:2018ahb} and the latter is estimated by using an approximate SU(3) symmetry relation
 $m_{B_s(0^+)}-m_{B_s(0^-)}=m_{B(0^+)}-m_{B(0^-)}$~\cite{Wang:2015ndk}.
The fitted coefficients $a_0$ and $a_1$   have been collected in Table~\ref{tab:para}.

\end{appendix}

\end{document}